\UseRawInputEncoding

\pdfoutput=1
\documentclass[twocolumn,aps,pra,superscriptaddress, 10pt]{revtex4-2}

\usepackage{graphicx}
\usepackage{bm}
\usepackage[dvipsnames]{xcolor}
\usepackage[mathlines]{lineno}
\usepackage{braket}
\usepackage{hyperref}
\usepackage{tabularx}
\usepackage[english]{babel}
\usepackage{amsmath}
\usepackage{mathtools}
\usepackage[noend]{algpseudocode}
\usepackage{algorithm}
\usepackage{ulem}
\usepackage{placeins}
\usepackage{mwe}

\usepackage{booktabs} 
\usepackage{makecell}

\usepackage{xcolor}

\begin{document}

\title{Scalable registration of single quantum emitters within solid immersion lenses through femtosecond laser writing}





\author{Alexander R. Jones}
\thanks{Both authors contributed equally to this work.}
\affiliation{ 
Institute of Photonics and Quantum Sciences, SUPA, Heriot-Watt University, Edinburgh EH14 4AS, UK
}

\author{Xingrui Cheng}
\thanks{Both authors contributed equally to this work.}
\affiliation{ 
Department of Engineering Science, University of Oxford, Parks Road, Oxford OX1 3PJ, UK
}
\affiliation{ 
Department of Materials, University of Oxford, Parks Road, Oxford OX1 3PH, UK
}

\author{{Shravan Kumar Parthasarathy}}
\affiliation{ 
Institute of Applied Quantum Technologies, Friedrich-Alexander-University Erlangen-Nürnberg, 91052 Erlangen, Germany 
}
\affiliation{ 
Fraunhofer Institute for Integrated Systems and Device Technology (IISB), 91058 Erlangen, Germany
}

\author{Muhammad Junaid Arshad}
\affiliation{ 
Institute of Photonics and Quantum Sciences, SUPA, Heriot-Watt University, Edinburgh EH14 4AS, UK
}

\author{Pasquale Cilibrizzi}
\affiliation{ 
Institute of Photonics and Quantum Sciences, SUPA, Heriot-Watt University, Edinburgh EH14 4AS, UK
}

\author{{Roland Nagy}}
\affiliation{ 
Institute of Applied Quantum Technologies, Friedrich-Alexander-University Erlangen-Nürnberg, 91052 Erlangen, Germany 
}

\author{Patrick Salter}
\affiliation{ 
Department of Engineering Science, University of Oxford, Parks Road, Oxford OX1 3PJ, UK
}

\author{Jason Smith}
\affiliation{ 
Department of Materials, University of Oxford, Parks Road, Oxford OX1 3PH, UK
}

\author{Cristian Bonato}
\email{c.bonato@hw.ac.uk}
\affiliation{ 
Institute of Photonics and Quantum Sciences, SUPA, Heriot-Watt University, Edinburgh EH14 4AS, UK
}

\author{Christiaan Bekker}
\email{c.bekker@hw.ac.uk}
\affiliation{ 
Institute of Photonics and Quantum Sciences, SUPA, Heriot-Watt University, Edinburgh EH14 4AS, UK
}


\begin{abstract}
The precise registration of solid-state quantum emitters to photonic structures is a major technological challenge for fundamental research (e.g. in cavity quantum electrodynamics) and applications to quantum technology. Standard approaches include the complex multi-step fabrication of photonic structures on pre-existing emitters, both registered within a grid of lithographically-defined markers.

Here, we demonstrate a marker-free, femtosecond laser writing technique to generate individual quantum emitters within photonic structures. Characterization of 28  defect centers, laser-written at the centers of pre-existing solid immersion lens structures, showed offsets relative to the photonic structure's center of 260~nm in the x-direction and 60~nm in the y-direction, with standard deviations of $\pm 170$~nm and $\pm 90$~nm, respectively, resulting in an average 4.5 times enhancement of the optical collection efficiency. This method is scalable for developing integrated quantum devices using spin-photon interfaces in silicon carbide and easily extendable to other materials. 
\end{abstract}

\maketitle

\vspace{1em}
\vspace{2em}


Color centers in the solid state, such as optically-active point defects and impurities, are among the most prominent systems for quantum technology~\cite{awschalom_quantum_2018}. Spin-photon interfaces associated to color centers in diamond~\cite{parker_diamond_2024, hermans_qubit_2022, stolk_metropolitan-scale_2024}, silicon carbide (SiC) ~\cite{christle_isolated_2017, nagy_high-fidelity_2019, anderson_five-second_2022, cilibrizzi_ultra-narrow_2023, ecker_quantum_2024} and silicon~\cite{gritsch_optical_2025, inc_distributed_2024}, along with single rare-earth ion dopants in crystals~\cite{wu_near-infrared_2023}, have been used in pioneering demonstrations of long-distance quantum networks. The nitrogen-vacancy center in diamond, as well as other defects in SiC~\cite{wolfowicz_electrometry_2018, niethammer_vector_2016, yan_room-temperature_2020, jiang_quantum_2023} and hexagonal boron nitride ~\cite{stern_room-temperature_2022, rizzato_extending_2023, zhou_sensing_2024, stern_quantum_2024}, are used in a variety of quantum sensing applications, ranging from fundamental physics to healthcare~\cite{miller_spin-enhanced_2020}.

Enhancing photon collection efficiency through tailored photonic structures is critical for optically-read-out spin qubits, as it directly impacts spin-photon interface efficiency in quantum networks and sensitivity in quantum sensing~\cite{degen_quantum_2017,bradac_quantum_2019}. This has been achieved, for example, with simple structures that minimize the effect of total internal reflection inside a high-index material, such as solid immersion lenses (SILs)~\cite{hadden_strongly_2010, rogers_multiple_2014,bekker_scalable_2023,wan_efficient_2018, sardi_scalable_2020}, or light guiding structures such as waveguides~\cite{babin_fabrication_2022, burek_fiber-coupled_2017,sotillo_diamond_2016,tiecke_efficient_2015,guo_laser-written_2024, krumrein_precise_2024}. The alternative is to employ a microcavity, such as a nanopillar~\cite{rugar_characterization_2019, radulaski_scalable_2017, losero_neuronal_2023,orphal-kobin_optically_2023,hedrich_parabolic_2020}, bullseye cavity~\cite{addhya_photonic-cavity-enhanced_2024} or photonic crystal~\cite{quan_photonic_2010,sipahigil_integrated_2016}, to enhance light-matter interaction, maximize the fraction of emission into the coherent zero-phonon line, and increase optical extraction efficiency~\cite{castelletto_deterministic_2019, hessenauer_cavity_2025}.

A significant challenge in scaling the integration of quantum emitters into photonic structures is registering them to the location that provides the greatest optical enhancement. Depending on the structure, this region can range from $\sim1~\mu\text{m}^3$ in SILs to $\sim10~ \text{nm}^3$ for photonic crystal cavities. One approach to registering single quantum emitters in photonic structures consists of mapping the position of pre-existing emitters against a marker array, which can be aligned to during photonic structure fabrication~\cite{thon_strong_2009, marseglia_nanofabricated_2011, copeland_traceable_2024}. This procedure could provide accuracy down to a few nanometers~\cite{thon_strong_2009} and enables pre-selection of emitters with optimal properties for integration. However, it is extremely time-consuming and not easily scalable to large arrays. 

A second possibility is to implant the ion species required to create the emitter into existing photonic structures, for example, through the use of focused ion beams~\cite{schroder_scalable_2017, addhya_photonic-cavity-enhanced_2024, lefaucher_bright_2025, he_maskless_2023}. This technique is very effective, but can only be used to create shallow emitters (depths of $\lesssim 1$~$\mu$m depth), as the lateral accuracy of implanted ion placement is similar to the implantation depth. Higher-energy ions for deep implantation can degrade emitter quality~\cite{fu_conversion_2010, orwa_engineering_2011,van_dam_optical_2019}. 

Recently, the generation of quantum emitters by laser-writing has received increasing attention. This technique exploits highly energetic carriers, created either by below-~\cite{chen_laser_2019, chen_laser_2017-1, hao_laser_2024, liu_confocal_2020, guo_laser-written_2024, abdedou_photoluminescence_2024} or above-bandgap~\cite{day_laser_2023} illumination with a high-power femtosecond laser to initiate an avalanche ionization process. Above-bandgap illumination has been used to create quantum emitters in SiC nanophotonic structures~\cite{day_laser_2023}, and femtosecond NIR laser pulses have been utilized to generate ensembles of quantum emitters in diamond nano-cavities~\cite{addhya_photonic-cavity-enhanced_2024}.   

Here, we create single quantum emitters registered to SILs by direct below-bandgap femtosecond laser writing. Femtosecond laser writing enables us to create intrinsic point defects based on vacancies directly in the focal region of the SILs. This method does not require alignment to markers and can be fast, making it compatible with wafer-scale processing. The photonic structure itself can enhance the laser-writing field, potentially removing the need for any registration effort. By using the focusing effects of the lens, weaker laser pulse energies can be used to generate emitters. In contrast to ion implantation and above-bandgap illumination, direct below-bandgap femtosecond laser writing enables the creation of single quantum emitters deeper within micron-scale structures, where they are less sensitive to surface noise and typically exhibit better quantum coherence properties. 

We focus on quantum emitters in SiC, a material that uniquely combines spin-photon interfaces~\cite{christle_isolated_2017, nagy_high-fidelity_2019, cilibrizzi_ultra-narrow_2023} possessing long spin coherence times~\cite{christle_isolated_2015, anderson_five-second_2022} with integrated photonic~\cite{lukin_4h-silicon-carbide--insulator_2020, lukin_integrated_2020, castelletto_silicon_2022, babin_fabrication_2022} and microelectronic~\cite{scheller_quantum_2024, steidl_single_2024} functionalities. 
Our approach readily adapts to alternative materials, such as diamond \cite{hadden_strongly_2010, marseglia_nanofabricated_2011,jamali_microscopic_2014}.

\begin{figure*}[ht]
\centering
\includegraphics[width= 1\textwidth]{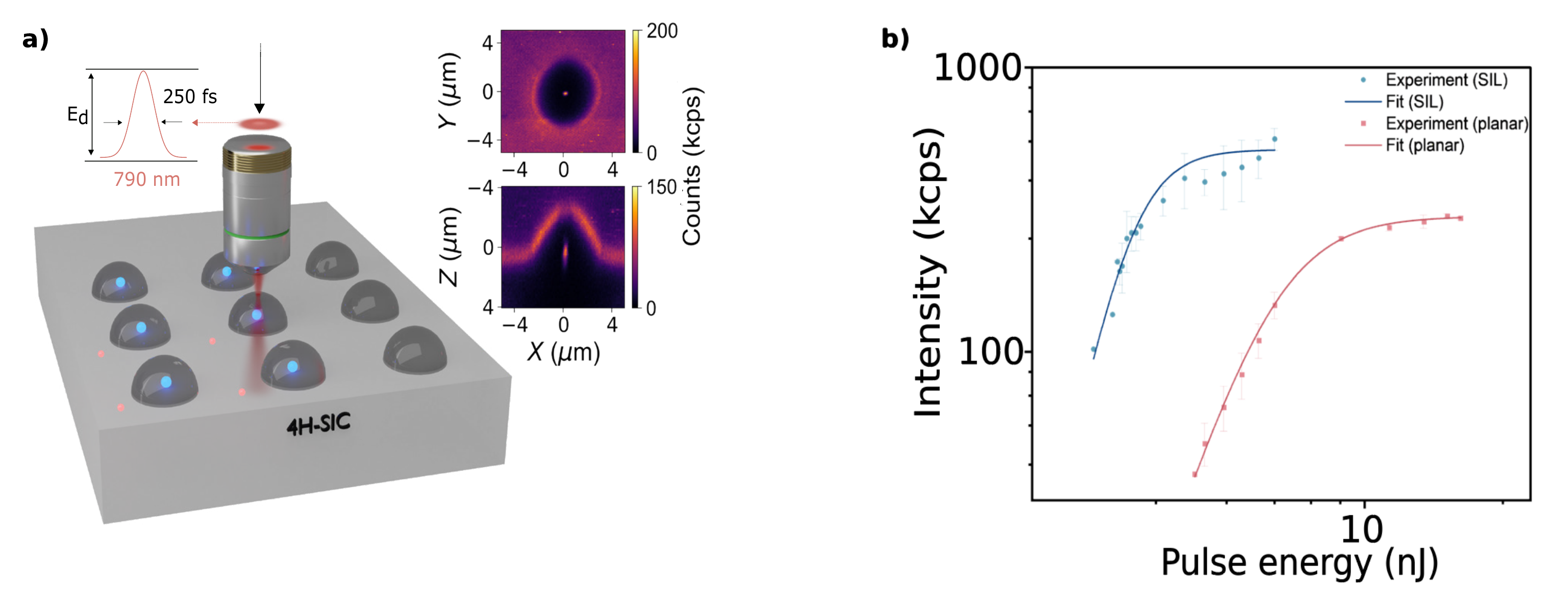}
\caption{(a) Schematic illustration of the formation of V$_\text{Si}$ centers in 4H-SiC via femtosecond laser writing using a single 250 fs FWHM pulse at 790~nm. Writing through the bulk (planar) interface results in lower efficiency compared to direct writing within SIL structures, as indicated by the contrast between dim red and bright blue spots, respectively. The inset presents photoluminescence (PL) maps of a defect ensemble written inside a SIL. The top panel shows an XY scan, and the bottom panel shows an XZ scan, demonstrating the precise spatial registration of laser-written defects in both axial and lateral dimensions relative to the SIL position. (b) Plot of the intensity of photoluminescence observed in laser-written defects as a function of writing pulse energy for bulk (red) and SIL (blue) interfaces, showing the effect of the SIL in reducing the waist of the focused beam and enhancing the local intensity of the pulse, resulting in an approximately 5-fold reduction in the required writing power for observing single emitters and higher light extraction efficiency. The experimental data were fitted using the saturation model described in Eqn~\ref{eq:Isat}.}
\label{fig:LW}
\end{figure*}

The experiment was conducted using commercial 4H-SiC material (Xiamen PowerWay\textsuperscript{\copyright}) diced into $5\times 5 ~\text{mm}$ chips (see SI section 1 for details). Arrays of hemispherical SILs with nominal radius $5\mu\text{m}$ were fabricated on these chips using the grayscale hard-mask lithography process set out in our previous work~\cite{bekker_scalable_2023}.

The laser writing process for generating quantum defects in the center of prefabricated SILs in 4H-SiC is illustrated in Fig.~\ref{fig:LW} a. The laser writing was performed using a regeneratively-amplified Ti:Sapphire laser source, delivering 250~fs pulses at a wavelength of 790~nm and a repetition rate of 1~kHz (shown in Fig. S1 and S2 in SI Section 2). The laser was focused through a high-NA oil immersion objective lens (Olympus 60$\times$, 1.4~NA).

A single femtosecond laser pulse with adjustable pulse energy was focused within the 4H-SiC sample. The degree of induced lattice damage can be controlled by varying the pulse energy, both when writing in the planar region and within SILs. For planar interfaces (red spots in Fig.~\ref{fig:LW} a), the focal depth was set to 5~$\mu$m, matching the radius of the hemispherical monolithic SILs. Focusing into high-index materials like 4H-SiC induces strong spherical aberrations, which were corrected using a spatial light modulator (SLM). 

During writing, the laser was focused through the SIL along its central axis at a depth corresponding to the sample's planar surface (blue spots in Fig.~\ref{fig:LW} a). This adjustment aimed to enable precise defect placement with minimal lattice perturbation and ensure optimal axial alignment between the SIL focus and laser-induced defects.

The laser writing process was initially characterized by photoluminescence (PL) under 532~nm CW optical excitation (1~mW) in a home-built confocal microscope with a $600-800$~nm collection window (details discussed in SI section 2). In both SIL and planar configurations, the PL intensity increases with higher pulse energy, as shown in Fig.~\ref{fig:LW}~b. For each pulse energy, data were collected from five spatially-separated sites (in the SIL case, 5 different neighboring SILs) written in a single column. All data points represent the arithmetic mean of intensities measured across these five equivalent fabrication points. Saturating behavior is observed in both cases, where the PL intensities detected from defect centers stabilize with respect to the writing power in the high pulse energy regime.

Notably, significantly lower pulse energies are required for laser writing through the SIL compared to the planar interface, evidenced by the lower pulse energy range of the blue (SIL) curve in Fig.~\ref{fig:LW}b (1.4 to 4.0~nJ) relative to the red (planar interface) curve (4.0 to 20~nJ). This highlights the additional focusing effect provided by the SIL. The larger error bars in PL counts through the SIL interface arises from minor deviations in SIL geometry and alignment uncertainties during focusing the writing laser, both of which introduce aberrations to the incident wavefront and associated variation in focal intensity. Furthermore, the saturation count rate for the SIL interface reaches 600~kcps, compared to 300~kcps for the planar interface, which indicates the light extraction efficiency achieved through the SIL interface is also enhanced.

\begin{figure}[b]
\centering
\includegraphics[width= \linewidth]{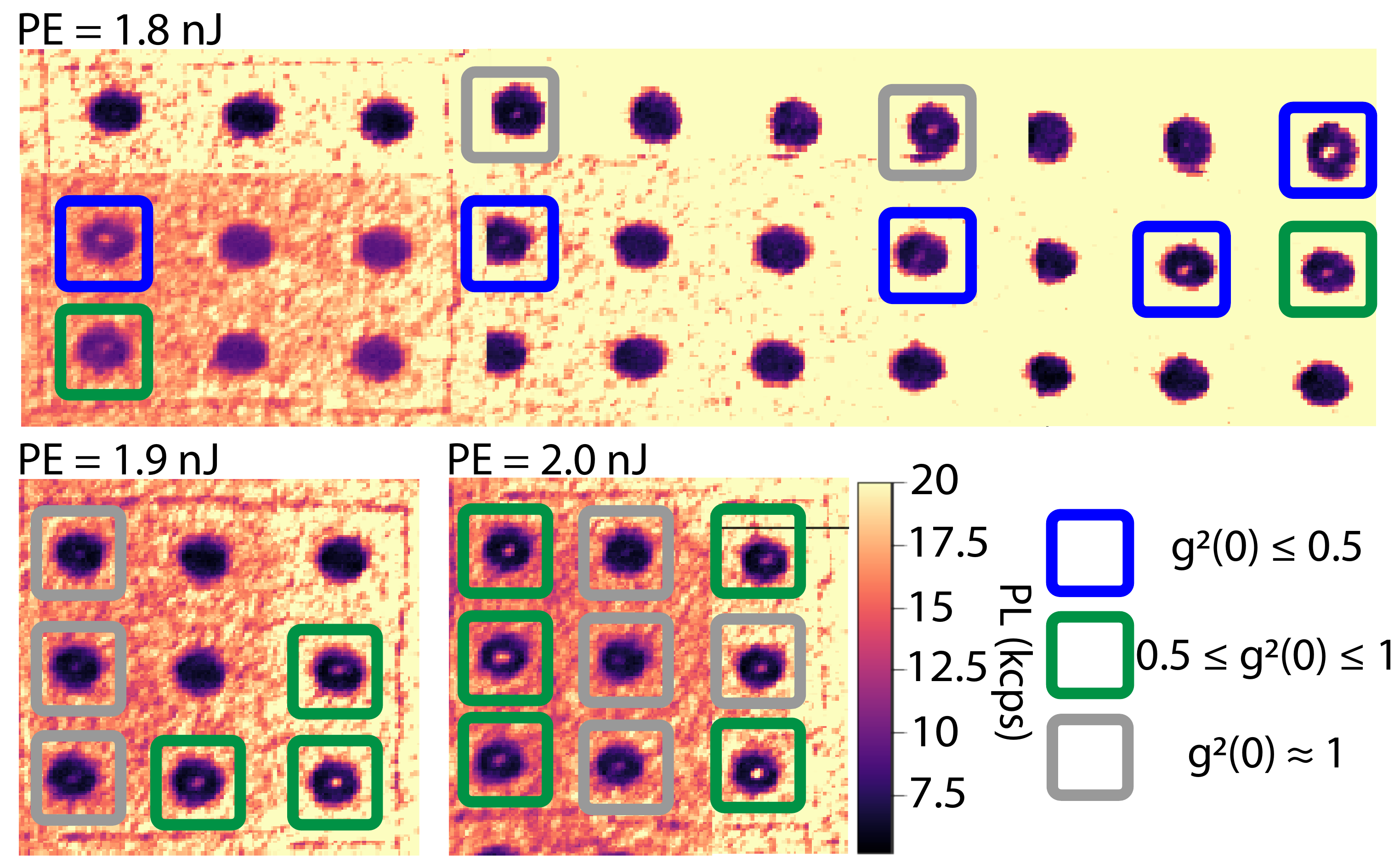}
\caption{PL maps of regions where SILs were written with single laser pulse energy (PE) range to create single V$_\text{Si}$ defects. This is shown by PE = 1.8~nJ, which has a statistically significant number of single photon sources distributed across the 30 SILs. Poissonian distributions in emitter creation are observed, progressing from 0.35 for PE=1.8~nJ to 1.0 for PE=2.0~nJ. When PE= 1.8~nJ, we also observe a higher yield of single color centers, namely 5 out of the 9 emitter spots observed in this region. This results in an expectation value of $\approx 0.35$ as, out of the 9 SILs where a photo-luminescent spot was created, 5 are confirmed as single photon sources. The other 21 SILs do not feature any photo-luminescence.}
\label{fig:regions}
\end{figure}

To elucidate the underlying mechanisms of laser-induced defect generation, we consider established models for ultrafast laser interactions in transparent materials, which typically involve multiphoton ionization (MPI), tunneling ionization (Zener breakdown), and avalanche ionization~\cite{keldysh_ionization_1965}. As a starting point, we adopt the framework developed for the creation of GR1 centers (neutral carbon vacancies) in diamond~\cite{chen_laser_2017}, which shares key similarities with the processes in 4H-SiC. The 250~fs laser pulses employed in this work are significantly shorter than the characteristic timescales of thermal diffusion (nanoseconds)~\cite{castelletto_femtosecond_2008}, allowing us to neglect thermal effects and focus on non-thermal ionization dynamics. The relative contributions of MPI and tunneling ionization are governed by the Keldysh parameter~\cite{keldysh_ionization_1965}, with MPI dominating when the laser intensity satisfies the condition:

\begin {equation}
I < \frac{mcn\epsilon_0E_g\omega^2}{e^2}
\label{eq:Keldysh}
\end {equation}
where $m = 0.37m_\text{e} = 3.370 \times 10^{-31}$~kg is the effective mass of lattice electrons, $n = 2.6$ and $E_g = 3.23$~eV are the refractive index and bandgap of 4H-SiC, and $\omega = 2.4 \times 10^{15}~\text{Hz}$ is the frequency of the photons. Substituting these values, the intensity threshold is calculated as $I < 1.38 \times 10^{17}\text{~W/m}^2$. The pulse energy $E$ is then determined as $E = I \times \pi \times (\text{beam waist})^2 \times (\text{pulse duration})$, where for the planar interface the beam waist is $350$~nm and pulse duration is $250$~fs. Substituting these parameters, $E_\text{planar} \approx 16.0\text{~nJ}$. For the SIL interface, the effective NA increases to $\sim2.6$, which reduces the beam waist to $190$~nm,  and the pulse energy is then calculated to be $E_{SIL} \approx 3\text{~nJ}$. 
At very high pulse energies, MPI loses dominance, resulting in a deviation from the expected power-law scaling, as shown in Fig.~\ref{fig:LW} b. This potentially leads to lattice breakdown which is associated with the onset of broadband PL emission and a saturation plateau in the photoluminescence count rate, as observed in Fig.~\ref{fig:LW}b. Assuming that the PL intensity relates to the number of defects generated by the laser pulse; we fit this saturation model to the PL intensity \textit{vs.} pulse energy data in Fig.~\ref{fig:LW}b (solid curves) to a saturation model:

\begin{equation}
I_{PL}(E) = \frac{aE^n}{1 + kE^n}
\label{eq:Isat}
\end{equation}
where $a$ is the amplitude coefficient, $n$ is the power law exponent, and $k$ is the saturation parameter (in slight contrast to the model presented in~\cite{chen_laser_2017} for the negatively charged carbon vacancy center in diamond). Fitting for the SIL interface yields $n = 5.75 \pm 0.15$ and $k = (12.5 \pm 1.26) \times 10^{-3}$, while for the planar interface, $n = 3.67 \pm 0.15$ and $k = (1.83 \pm 0.43) \times 10^{-3}$. The enhanced non-linearity in the SIL compared to the planar interface may arise from a stronger Zener breakdown contribution at lower pulse energies. The defect generation process involves photon energies that are multiples of the 790~nm laser photon energy. For the planar interface, this results in an energy of $5.8 \pm 0.24$~eV, while for the SIL interface, it yields $9.1 \pm 0.24$~eV. These energies exceed the SiC bandgap, enabling the generation of hot charge carriers rather than direct lattice disruption, given the non-polar covalent bonding in SiC. Instead, multiphoton absorption excites carriers which transfer energy to the lattice, leading to defect formation. This carrier-mediated mechanism resembles Frenkel defect formation in diamond~\cite{griffiths_microscopic_2021}, highlighting a shared pathway for laser-induced structural modification in wide-bandgap materials.
PL measurements were performed at room temperature using a 780nm laser, with 800nm dichroic and shortpass filters (Thorlabs FES800) placed in the excitation arm to suppress residual laser light, as described in previous work~\cite{bekker_scalable_2023}. By spatially mapping this PL emission over laser-written SILs, we could identify laser-induced defects localized near the center of SILs (Fig. \ref{fig:regions}).

For each SIL where a laser-written spot was detected, we measured the excitation power and brightness at which the emission saturates (\textit{power saturation} measurement) and the normalized depth of the second-order correlation function at zero delay ($g^{(2)} (0)$), using low-jitter ($\sim40$~ps) superconducting nanowire single-photon detectors (Single Quantum EOS). There is noticeable background luminescence from the SiC/air interface but not within the SIL itself. This agrees with previous studies in 4H-SiC that report intrinsic defects related to native surface oxide~\cite{day_laser_2023}.

Stochastic distributions of emitting defects consistent with Poissonian statistics could be detected across the written region, with the number and brightness of defects increasing with increasing laser-writing power. To minimize the creation of multiple emitters, writing is ideally performed in the regime where the Poissonian expectation value of creating any emitters  is much less than one ($\langle n \rangle\ll1$). For example, for an expectation value of $\langle n \rangle = 0.1$, the probability of creating a single emitter is $P(1)_{\langle n \rangle=0.1} = 9\%$, while the probability of creating two or more emitters is $P(>1)_{\langle n \rangle=0.1} = 0.4\%$. So, even for such a low yield, two emitters would be created approximately 5\% of the time. In the case of the region with laser writing pulse energy PE=1.8~nJ shown in Fig.~\ref{fig:regions}, 9 written defects occur within a set of 30 SILs, including five confirmed single photon emitters and two defects consistent with two emitters, leading to a Poissonian expectation value of $\langle n \rangle_\text{PE=1.8~nJ} \approx 0.35$.

The resulting values are reported in Table \ref{tab:statistics_table}; we show the second-order correlation results before and after background subtraction. Fabrication using laser pulse energy 1.8~nJ created the largest number of emitters with $g^{(2)} (0) < 0.5$, see Fig. \ref{fig:regions}. The SIL ID in Table \ref{tab:statistics_table} denotes its grid location on the sample. We also show photoluminescence counts at saturation for each emitter.

We classify each PL spot based on its $g^{(2)}(0)$ value to be either a single emitter ($g^{(2)}(0) \leq 0.5$), multiple emitters $0.5 < g^{(2)} (\tau = 0) < 1$, or as a PL spot not in the single-emitter regime $g^{(2)} (0) = 1$. 

Throughout several months of measurements, we have seen no evidence of photo-bleaching, and the PL emission for all examined color centers have remained stable and reproducible.

\begin{table}[t]
\centering
\caption{Statistics of color centers generated in SILs via femtosecond laser writing. SIL ID: column and row label of SIL with respect to one corner of the array. PE (nJ): writing laser pulse energy. PL$_\text{sat}$: saturation photoluminescence counts of each emitter. g$^{(2)}(0)$: normalized second-order correlation at zero delay without background subtraction (raw) and with background subtraction applied (bck. subtr.). After background subtraction, the laser-written defects are confirmed to be single photon emitters, occurring most frequently at PE=1.8~nJ. SIL ID T30 is the emitter characterized in Fig.~\ref{fig:single}.}
\label{tab:statistics_table}
\renewcommand{\arraystretch}{0.9}  
\begin{tabular}{ccccc}
\toprule
\textbf{SIL ID} 
& \makecell{\textbf{PE (nJ)}} 
& \makecell {\textbf{PL$_\text{sat}$} \\ \textbf{(kcps)}} 
& \makecell {\textbf{g$^{(2)}(0)$} \\ \textbf{(raw)}} 
& \makecell {\textbf{g$^{(2)}(0)$} \\ \textbf{(bkg. subtr.)}}
\\  
\midrule
T30 & 1.9 & $38.1\pm{0.5}$ & $0.34\pm{0.02}$ & $0.27\pm{0.11}$\\  
X18 & 1.8 & $36.0\pm{0.6}$ & $0.54\pm{0.05}$ & $0.45\pm{0.05}$\\
X23 & 1.8 & $29.0\pm{1.7}$ & $0.41\pm{0.04}$ & $0.21\pm{0.04}$\\
X26 & 1.8 & $45.7\pm{2.3}$ & $0.44\pm{0.07}$ & $0.27\pm{0.11}$\\
AA32 & 1.6 & $23.3\pm{1.8}$ & $0.55\pm{0.06}$ & $0.43\pm{0.06}$\\
Y17 & 1.8 & $32.3\pm{0.8}$ & $0.61\pm{0.04}$ & $0.40\pm{0.05}$\\
X20 & 1.8 & $38.3\pm{1.4}$ & $0.62\pm{0.05}$ & $0.54\pm{0.06}$\\
\bottomrule
\end{tabular}
\vspace{2mm}
\end{table}

\begin{figure*}[ht!]
\centering\includegraphics[width= 0.8\textwidth]{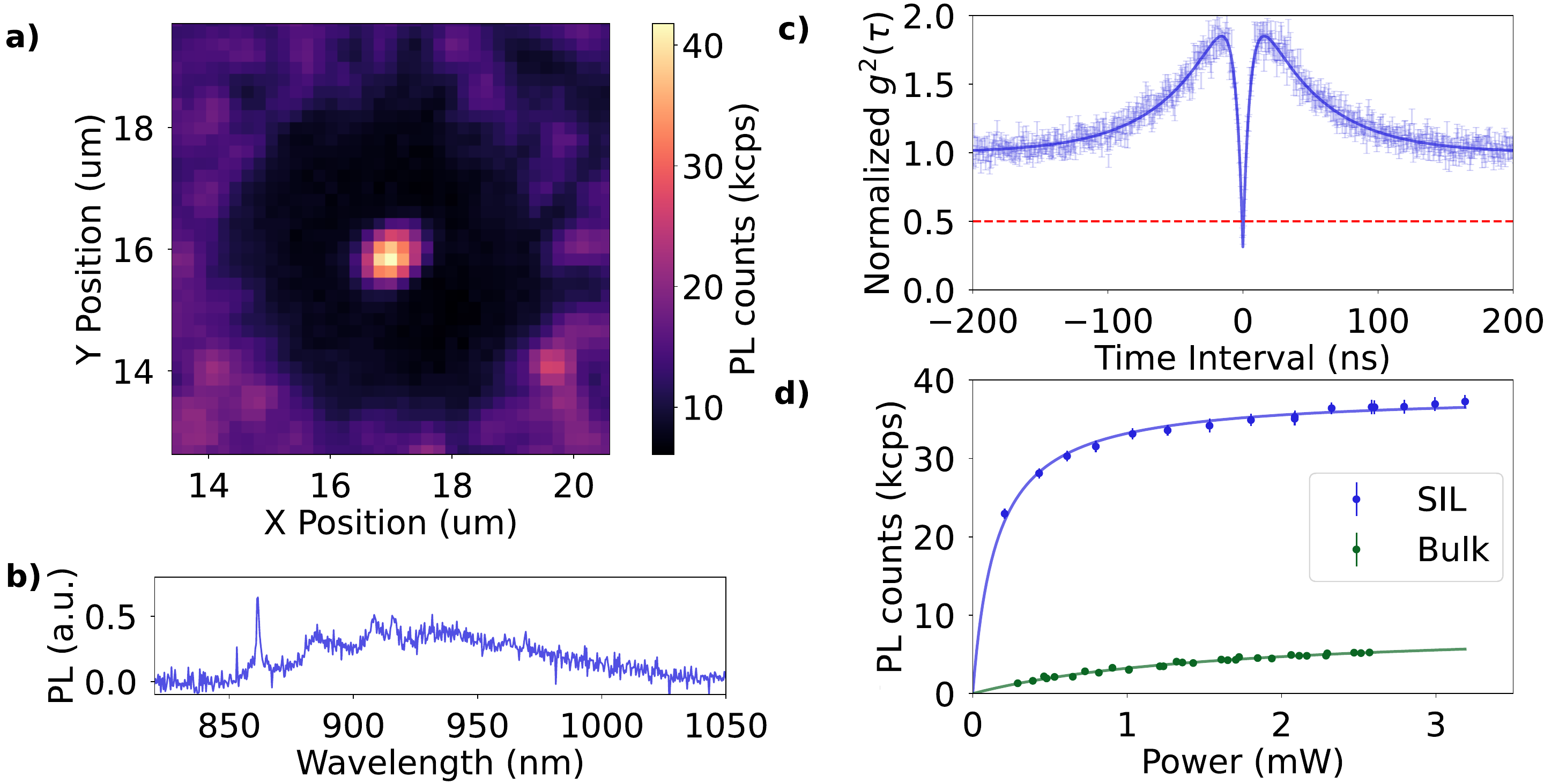}
\caption{Characterization of a single V$_\text{Si}$ written in SIL T30 with PE = 1.9~nJ. (a) Photoluminescence map taken at an excitation power of 4.2~mW. Note the background observed on the outskirts of the map is associated with SiC interfacial emitters~\cite{day_laser_2023} and does not occur within the SiC crystal. (b) An optical spectrum was taken at low temperature (4~K), showing a zero-phonon line at 861~nm characteristic of a V$_\text{Si}$ at a h lattice site. (c) Normalized autocorrelation plot of light from the emitter without background subtraction. The depth at zero time delay g$_2(0) = 0.35\pm 0.015$ confirms a single V$_\text{Si}$ has been created. (d) Power saturation curve of emitter T30 (blue), with a representative curve of a single emitter under a planar interface (green) for reference. The brightness of the emitter is enhanced by 4.47 times, and excitation power is intensified by a factor 9.2.} \label{fig:single}
\end{figure*}

We examined the PL spectra of generated quantum emitters at low temperature (T= 4~K, Montana s100 Cryostation) to identify characteristic zero-phonon lines (ZPLs). The emitters were excited with a 780~nm CW laser, and the PL spectra were measured with a grating spectrometer (OceanOptics QE Pro)). For further information on the setup and measured spectra, see SI section 5. 

Remarkably, we observe a wide spread of ZPL wavelengths in the spectral region 858~nm-985~nm. Out of 39 emission defects examined, 6 have spectral lines that can be classified as silicon-vacancy (V$_\text{Si}$) centers; these were 1 V$_1$' center (ZPL 858~nm), 4 V$_1$ centers (h-site, ZPL 861~nm), and 1 V$_2$ center (k-site, ZPL 916~nm). The remaining defects exhibit ZPLs at different wavelengths which do not appear to match other unidentified lines previously reported in this spectral range~\cite{fuchs_engineering_2015,bathen_resolving_2021, ruhl_controlled_2018}, but fall within the theoretically-predicted range of V$_\text{Si}$ centers modified by nearby carbon anti-sites~\cite{davidsson_exhaustive_2022} and experimentally-observed range for V$_\text{Si}$ centers in etched membranes \cite{heiler_spectral_2024}.

We further performed optically-detected magnetic resonance (ODMR) measurements at room temperature, in the frequency ranges associated with V$_1$ (zero-field splitting 4 MHz) and V$_2$ (zero-field splitting 70~MHz) centers. For as-written emission defects, no ODMR signals were observed in these ranges. However, after annealing the sample at 600$^\circ$C for 30 minutes in vacuum ($4\times10^-5$~mbar), a process known to improve V$_\text{Si}$ yield, we observed an ODMR signal associated with V$_2$ for one SIL (of 12 measured), though with a large linewidth of about $20$~MHz (see SI section 6 Fig. S6).

One possible explanation for the lack of ODMR signal might be that the femtosecond laser creates other defects that are not optically active, in addition to the V$_\text{Si}$. For example, carbon atoms feature much lower displacement energy than silicon atoms in SiC~\cite{li_threshold_2019, storasta_deep_2004, sullivan_investigation_2006,kaneko_formation_2011}, so that the creation of carbon vacancies (V$_\text{C}$) is favored. The presence of charge traps related to carbon vacancies and other defects~\cite{anderson_electrical_2019, scheller_quantum_2024, steidl_single_2024} is expected to create electromagnetic noise that may broaden observed optical and magnetic resonance linewidths. Further experiments, beyond the scope of this work, are needed to clarify the physics of femtosecond laser generation of point defects and to optimize laser writing parameters for improving the quality of generated quantum emitters.

\begin{figure*}[ht!]
\centering
\includegraphics[width= 1.0\textwidth]{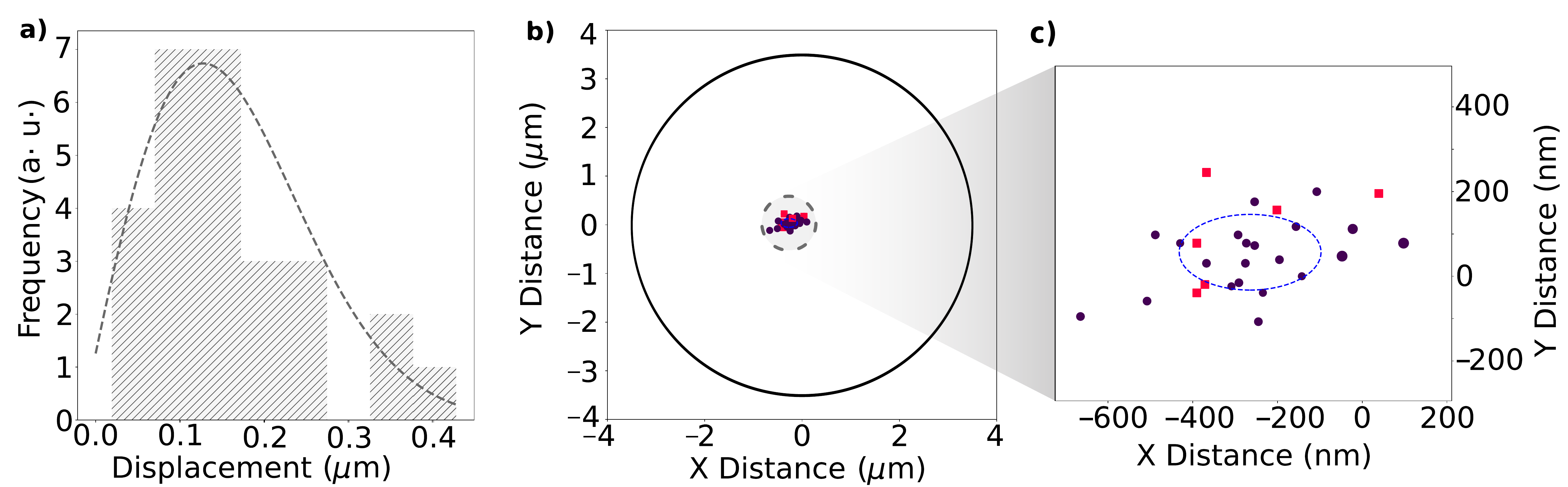}
\caption{Quantification of emitter distribution within SILs through registered laser writing. (a) Histogram of absolute displacement of the center of each observed emitter spot from the center of the SIL, determined using spatial PL maps. (b) Schematic of relative distributions in position space, with a SIL outline of radius 3.5$\mu$m as a guide for PL profiles. c) zoomed-in position map showing the distribution of emitter positions more clearly, with the dashed line denoting the mean displacement obtained from fitting the histogram data. Points in purple are registered positions with $g^2(\tau) > 0.5$; the red squares are single defects that are found in the Table \ref{tab:statistics_table}. 
\label{fig:distribution}
}
\end{figure*}

In Fig.~\ref{fig:single}, the key properties are given of a single V$_\text{Si}$ center generated in SIL T30 with a laser writing pulse energy $\text{PE}=1.9~\text{nJ}$. The photoluminescence map (Fig.~\ref{fig:single}~a) shows a spot centered well with respect to the SIL profile (dark circle). The photoluminescent background outside of the SIL region corresponds to surface defects on the etched surface of the SiC, postulated to arise at the SiC-surface oxide interface~\cite{day_laser_2023,babin_fabrication_2022, lohrmann_activation_2016, kaneko_impact_2023, onishi_generation_2024}. Spectroscopy of the emitted light at cryogenic temperatures (T = 4K) yielded a ZPL at 861~nm, consistent with a V$_1$-type V$_\text{Si}$ center. Furthermore, even without background subtraction, the normalized second-order correlation at zero delay  $g^{(2)} (0) = 0.34\pm0.018$, confirms that the emission comes from one single photon emitter. Taking power saturation measurements of the single V$_\text{Si}$ center in SIL T30 and a single electron-irradiated V$_\text{Si}$ center beneath a planar interface (Fig.~\ref{fig:single}, blue and green, respectively), the performance of the SIL-registered emitter could be benchmarked. Following our previous work~\cite{bekker_scalable_2023}, we determine an optical collection efficiency enhancement factor of 4.5 ($\text{PL}_\text{sat, SIL} = 38800\pm{500}$ \textit{vs.} $\text{PL}_\text{sat, bulk} = 8600\pm{600}$) and a power intensification factor of 9.1 ($\text{I}_\text{sat, SIL} = 0.181\pm{0.014}$ and $\text{I}_\text{sat, bulk} = 1.64\pm{0.21}$), consistent with the upper levels of performance observed for single V$_\text{Si}$ centers generated randomly throughout the SIL by electron irradiation.

Laser writing through the SIL interface was performed by microscope-aided alignment of the laser focus to the center of the SIL at a depth of $5~\mu$m, equivalent to the height of the lens structure. Contributions to the final emitter misalignment with respect to this target could arise both from the random probability of generating a defect within the volume of highest writing intensity and the error in writing laser alignment. 

To quantify these distributions, spatial PL maps of SILs with emitter spots were characterized using image analysis software (ImageJ) to extract the center of each SIL and emitter spot (see SI section 3). The absolute displacements of measured emitters from their collective mean position are given in Fig. \ref{fig:distribution} a. The statistics follow a Rayleigh distribution with scale parameter $\sigma = 0.14~\mu$m (dashed line).

When the emitters' absolute positions with respect to the SIL are plotted (Fig.~\ref{fig:distribution} b), clustering near the SIL center is evident, especially compared to emitter generation through electron irradiation in similar structures~\cite{bekker_scalable_2023}. Focusing on the SIL center (Fig.~\ref{fig:distribution} c), we find that the standard deviation of positions in the x-direction ($\pm{170}$~nm) is higher than in the y-direction ($\pm{90}$~nm). 
This is possibly due to the raster protocol for laser writing, where writing proceeds along all columns in a row before moving to the next row, so that more steps are taken in x than in y.

When aligning the writing laser to SILs, systematic offsets in positioning or beam angle can lead to the mean defect position deviating from the structure's center. This is observed in Fig.~\ref{fig:distribution} c, where the mean emitter position is offset by 260~nm in x and 60~nm in y from the SIL center.

In this work, we have demonstrated marker-free registration of single quantum emitters at the center of SILs by femtosecond laser writing. The SILs lower the laser dose required to generate emitters and aid in positioning them near the center of the structure, with an accuracy of $60\pm{90}~\text{nm}$ in the y-direction and $260\pm{170}~\text{nm}$ in the x-direction. Our approach, based on direct alignment of the laser-writing beam focus to the center of the SIL, achieves sub-diffraction-limit placement accuracy within the SIL to maximize optical collection enhancement. In future applications, better positioning (potentially $<100$~nm) could be achieved through use of alignment markers.

There remains an open question regarding the nature and diversity of created emitters and their performance for quantum technology applications. In our post$-$annealing study, controlled thermal treatment revealed opportunities to enhance the spin coherence of the generated defects; however, the yield of high$-$coherence centers remained low. Future investigations could systematically identify generated defect species and map their optical and spin$-$coherence properties as functions of laser irradiation parameters and thermal annealing protocols. Furthermore, deterministic generation of V$_\text{Si}$ centers can be further pursued by integrating laser$-$annealing sequences with in-situ photoluminescence monitoring.

\vspace{2cm}
\textbf{Supporting Information}: Experimental details of ODMR measurements, optical spectra, and XZ profiles, descriptions of the experimental setups, fitting procedure details, and sample preparation methods.

\section*{acknowledgments}
We express our gratitude to Nguyen Tien Son and Ivan G. Ivanov for valuable discussions. This work is funded by the Royal Academy of Engineering (RF2122-21-129), the Engineering and Physical Sciences Research Council (EP/S000550/1, EP/V053779/1, EP/Z533208/1, EP/Z533191/1, EP/W025256/1), the European Commission (QuanTELCO, grant agreement No 862721), the Leverhulme Trust (RPG-2019-388). It is further supported by the project 23NRM04 NoQTeS, which has received funding from the European Partnership on Metrology, co-financed from the European Union’s Horizon Europe Research and Innovation Programme and by the Participating State.

\clearpage

\renewcommand{\thepage}{S\arabic{page}}
\renewcommand{\thesection}{S\arabic{section}}
\renewcommand{\thesubsection}{S\arabic{subsection}}

\renewcommand{\thetable}{S\arabic{table}}
\renewcommand{\thefigure}{S\arabic{figure}}
\renewcommand{\figurename}{Supplementary Figure}

\setcounter{page}{1}
\setcounter{figure}{0}
\setcounter{section}{0}

\onecolumngrid

\section*{Supplementary Information}

\subsection{Sample Preparation}
\label{sec:sample_prep}

The experiment was conducted using commercial 4H-SiC material (Xiamen PowerWay\textsuperscript{\copyright}) diced into $5\times 5 ~\text{mm}$ chips.With substrate and epilayer thickness $500\;\mu\text{m}$ and $15\;\mu\text{m}$ respectively, and residual n-doping level of $<1\times 10^{14}\; \text{cm}^{-3}$. The material was diced into $5\times 5 \text{mm}$ chips. Arrays of hemispherical SILs with nominal radius $5\mu\text{m}$ were fabricated on these chips using the grayscale hard-mask lithography process set out in our previous work~\cite{bekker_scalable_2023}.

\subsection{Laser writing system}
\label{sec:LW_Setup}

The laser writing system employed in this work is consistent with that described in a previous study~\cite{chen_laser_2019}. The system comprises a Spectra Physics Mai Tai laser and a Spectra Physics Solstice amplifier, operating at a wavelength of 790~nm with a maximum repetition rate of 1~kHz. Pulse energy is regulated via a $\lambda/2$ waveplate in combination with a Glan-laser polarizer. The beam is subsequently expanded before reaching the spatial light modulator (SLM, Hamamatsu Photonics X10468-02). A dichroic mirror directs the fabrication laser toward the sample while transmitting both the excitation and PL signals from an integrated room-temperature confocal system, featuring a 532~nm CW laser as the excitation source. Laser fabrication and PL characterization employed an Olympus PlanApo 60× oil$-$immersion objective (NA = 1.4) with immersion oil of refractive index of n = 1.5. Emission under 532~nm excitation was spectrally filtered to the 600–800~nm window using a bandpass filter and a notch filter. Schematics of the laser writing system and confocal microscope are provided in Fig. S1 and Fig. S2, respectively.




\begin{figure*}[ht]
    \centering
    \begin{minipage}{0.75\textwidth}
        \centering
        \includegraphics[width= 0.9\linewidth]{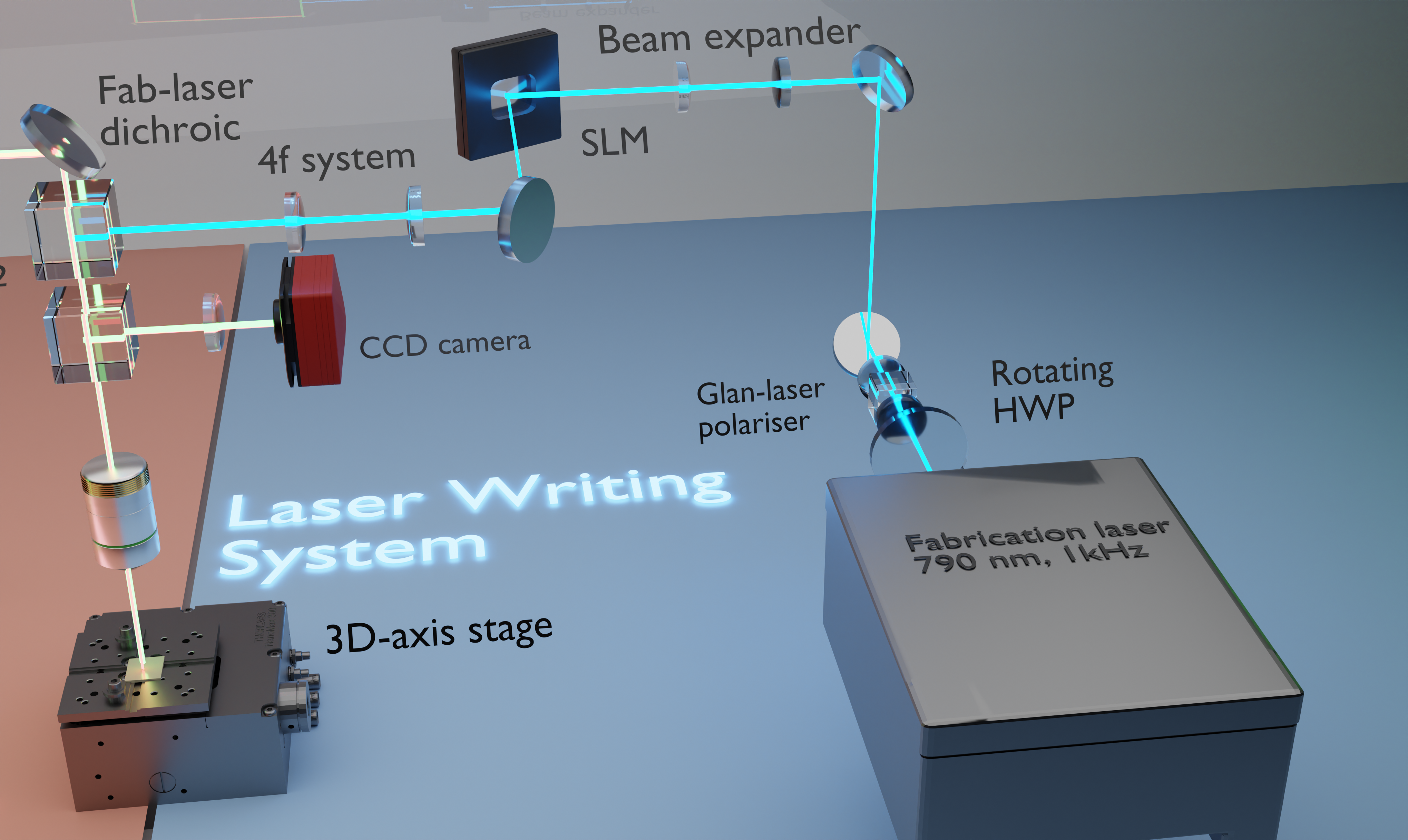}
        \caption{Schematic view of the Laser Writing System, including the fabrication laser for both kHz and MHz, SLM, ($\lambda/2$) waveplate, Glan-laser polarizer, CCD camera, 4f system, and translation stages for sample positioning.}
        \label{fig:Supp_writing_system}
    \end{minipage}
    \hfill
    \begin{minipage}{0.75\textwidth}
        \centering
        \includegraphics[width= 0.9\linewidth]{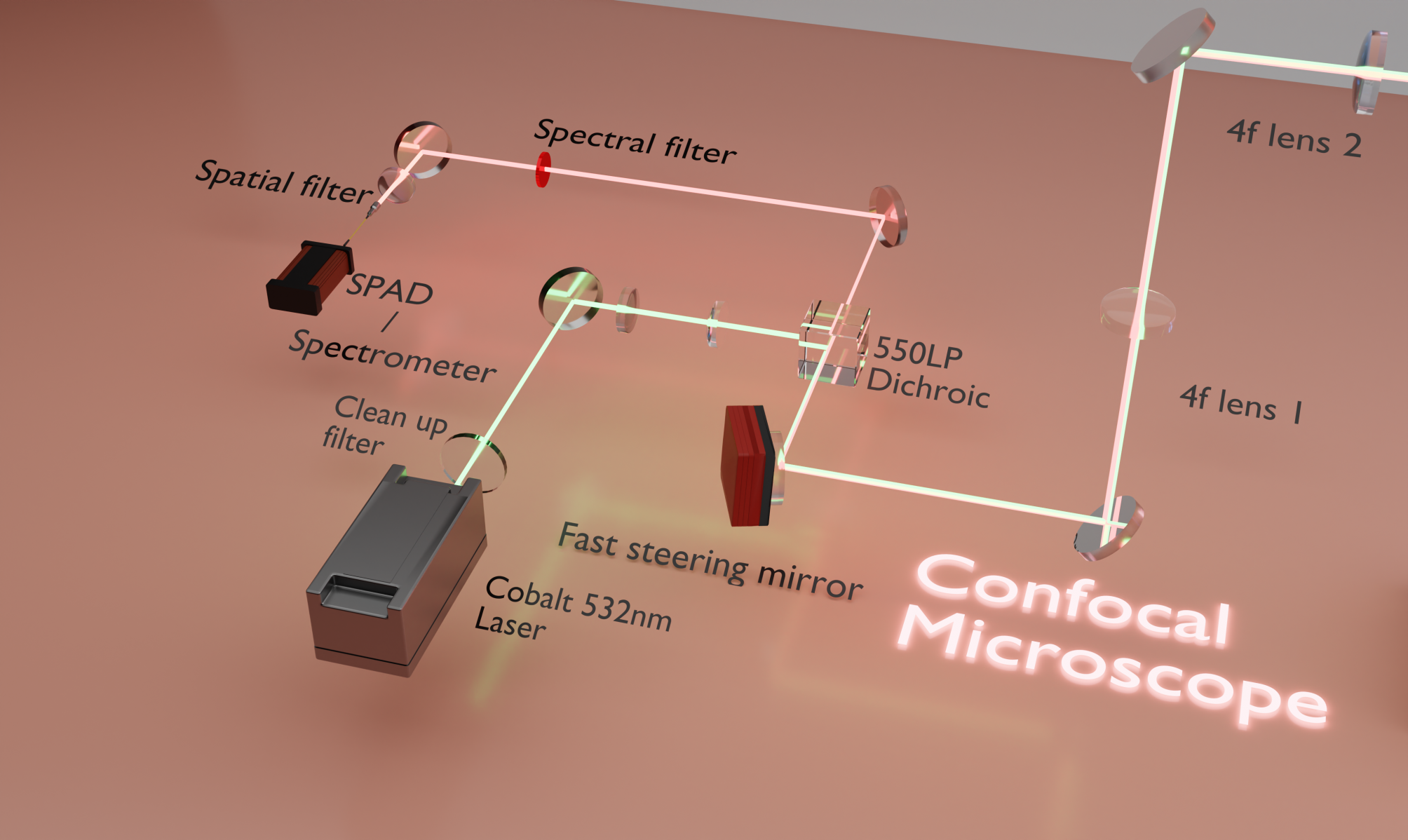}
        \caption{Schematic view of the confocal module, including the excitation laser, beam rastering module with FSM and 4f configuration relay lenses, and the collection system for SPAD or spectrometer detection.}
        \label{fig:Supp_writing_confocal}
    \end{minipage}
\end{figure*}

\subsection{Description of the circle-fitting procedure}
\label{sec:fit_procedure}

\begin{figure*}[ht]
\centering
\includegraphics[width= 0.65\textwidth]{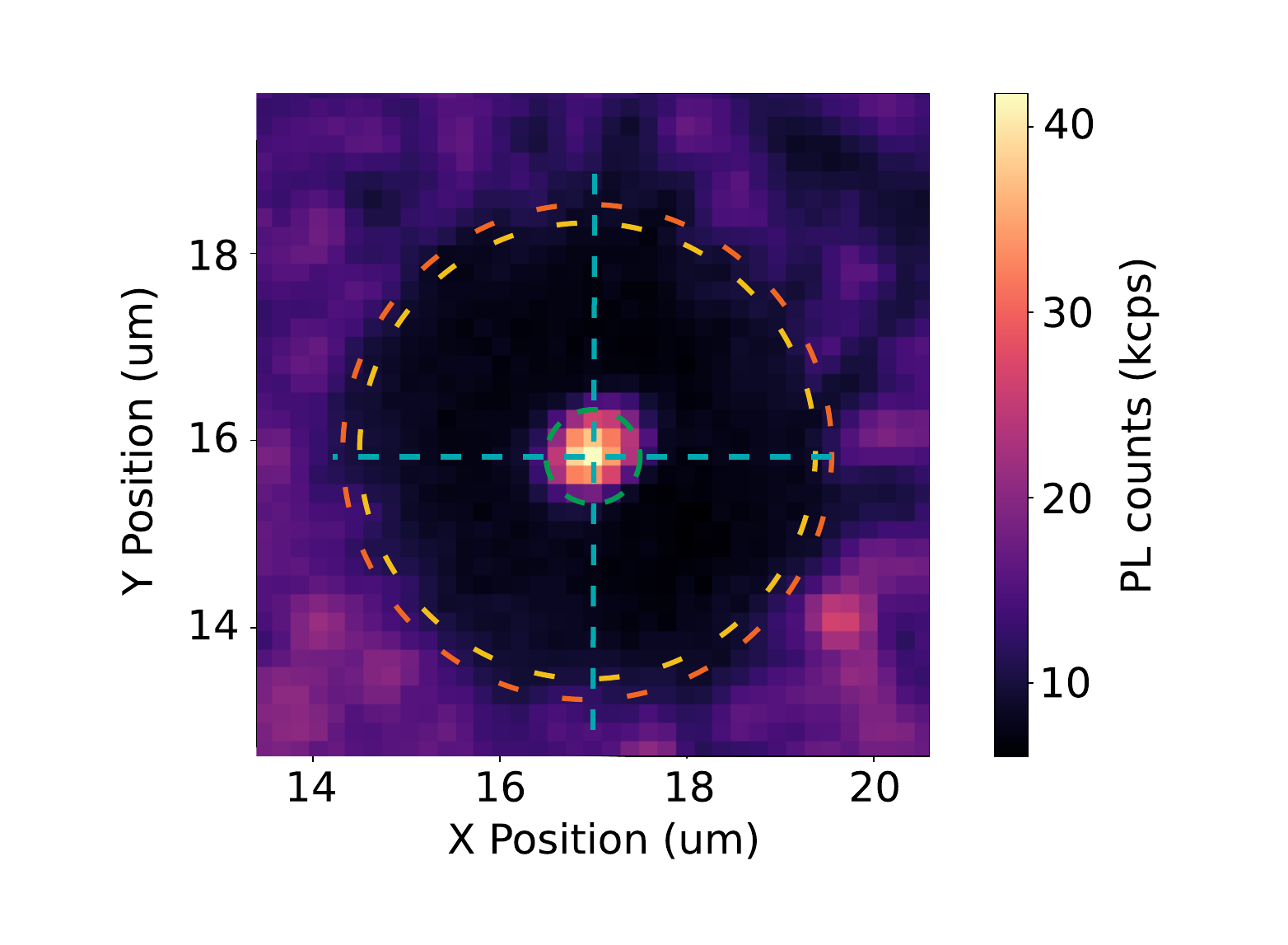}
\caption{Using SIL T30 from Fig.~\ref{fig:single}, we illustrate how we identified the positions of the registered defect centers within the SIL. The blue and green lines indicate the profile and circle methods, respectively. These were used to find the defect positions. The yellow, orange, and blue lines illustrate the circle, ellipse, and profile methods, respectively. These were used to find the SIL centers. } 
\label{fig:Supp_fitting_methods}
\end{figure*}

Here we describe the procedure to assess the position of the created emitter with respect to the SIL, based on PL maps with a stepsize of 0.13 $\mu$m. First of all, we determine the SIL circular edge to retrieve its center. We used a data analysis application (ImageJ) to determine the SIL circular edges, using three different methods to verify the reproducibility of the outcomes. First, we use the \textit{fit circle} tool to fit an exact circle around the SIL; second, we use the \textit{ellipse tool} to fit precisely the SIL edge. Lastly, we used the profile tool to measure the changes in PL, which are expected to be high at the SIL edge as the surface of SIL itself fluoresces. We then fit a Gaussian to each of these PL peaks to pinpoint a precise position for the edges. Fig.~S3 shows a comparison between the three methods for the same SIL. The \emph{ellipse} method in orange, the \emph{circle} method in yellow and the blue lines indicate where the profiles would be taken to find the SIL edge as the PL counts rise relative to the background within the SIL. 

Similarly, we compare results from two methods to determine the emitter position. First, we use the \emph{fit circle} tool in ImageJ, we fit a circle around the emitter confocal spot, and retrieve its center. Second, we fit a Gaussian to the emitter confocal spot. Fig.~S3 shows the typical positions for these methods, where the blue lines again show the profile positions and the green shows the circle that would fit around the defect to determine the central point to be taken as the defect position. 

We used the results for all SILs that generated a spot at 1.6~nJ writing power to benchmark the methods. The average SIL center using the ellipse and fitted circle methods were the same as $0.55\mu\text{m}$, $0.29\,\mu\text{m}$. The average calculated position relative to the SIL center using the profile method was $250\text{nm} \pm{0.09}$ in the x, and $78\text{nm} \pm{0.04}$ in the y. The average calculated position for the defect relative to the SIL center using the fitted circle's method was $190\text{nm}\pm{0.11}$ in x and $96\text{nm} \pm{0.05}$ in the y-direction. Due to the defects being within the lens, we performed a correction to account for the magnification effects on the measured distance between the center of the spot and the SIL.

\subsection{XZ profiles in SILs}

\begin{figure*}[h!]
\centering
\includegraphics[width= 0.9\textwidth]{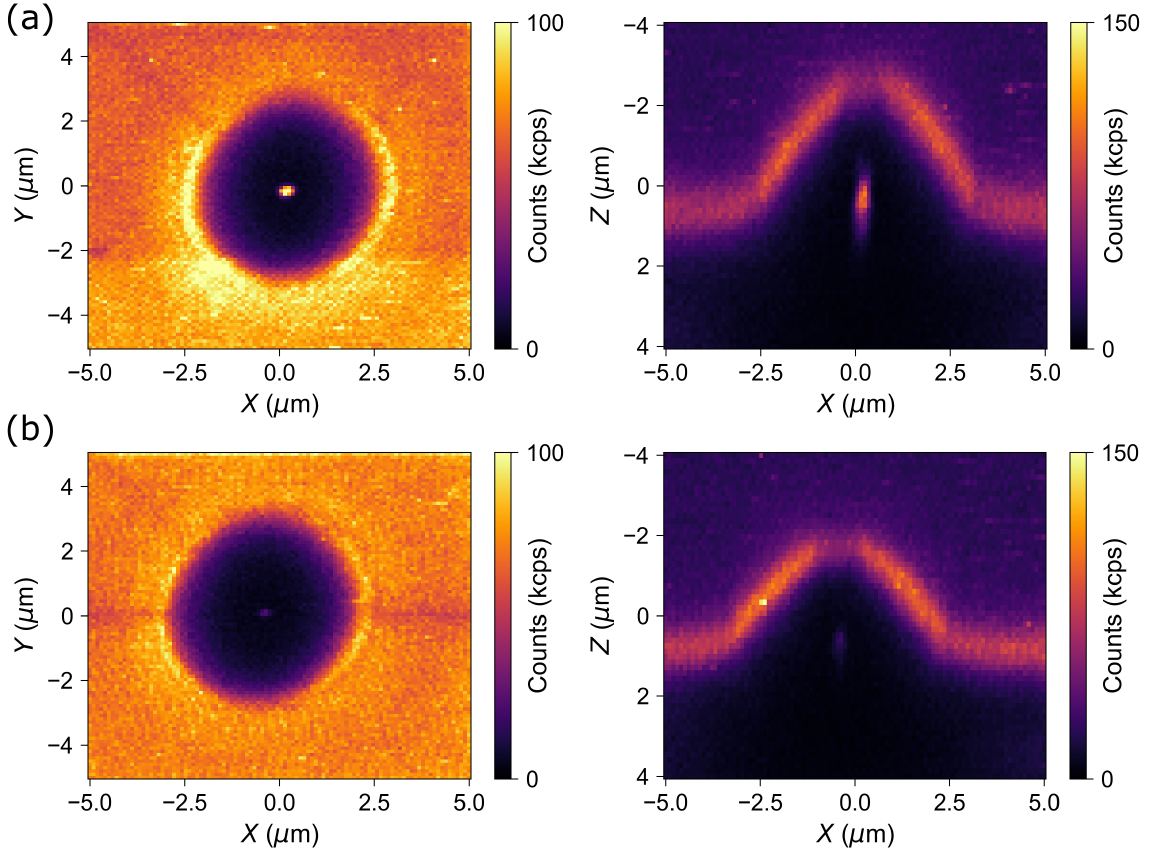}
\caption{XY and XZ 2D PL images of defect centers created with (a) high and (b) low pulse energies. High pulse energy results in an extended emission region due to lattice damage, while low pulse energy exhibits diffraction-limited behavior. Defect centers are aligned with the SIL focus, and the elongated XZ PSF reflects increased Abbe resolution.
\label{fig:Supp_XZ_profiles}
}
\end{figure*}

Fig.~\ref{fig:Supp_XZ_profiles} presents a series of XY and XZ 2D PL images for defects fabricated with high pulse energy of 3.5~nJ (a) and low pulse energy of 2~nJ (b). These PL images indicate that the defect centers are registered near the center of the SIL in both lateral and axial directions, aligning well with the SIL's focus. The elongated point spread function (PSF) observed in the axial PL profiles is attributed to the increased Abbe resolution in confocal microscopy. Quantitatively, the lateral and axial resolutions extracted from low pulse energy fabricated defect (110~nm and 170~nm, respectively) align with diffraction-limited expectations (100~nm lateral, 160~nm axial). In contrast, defect fabricated with higher pulse energy exhibits broadened emisssion profile (210~nm lateral and 985~nm axial, respectively), attributed to increased lattice damage induced by higher pulse energy that disrupts the point-source approximation.

\subsection{Optical Spectra of Emitters}
\label{sec:SI_Spectra}

To verify the characteristics of the quantum emitters produced, their photoluminescence (PL) spectrum was analyzed at low temperatures using a cryostat set to 4K (Montana s100 Cryostation) in combination with a custom confocal setup, as detailed by Cilibrizzi et al. \cite{cilibrizzi_ultra-narrow_2023}. The emitters were illuminated with a continuous wave (CW) laser at 780~nm, and their PL spectra were recorded with a grating spectrometer (OceanOptics QE Pro) with a 800nm longpass filter. The spectra obtained are displayed in Supplementary Figure S3.

The analysis uncovered a broad distribution of zero-phonon lines (ZPL) within the 858-985~nm spectral region. Out of 39 examined defect centers, six exhibited spectral lines typical of silicon-vacancy (V$_\text{Si}$) centers: one V$_1$' center (ZPL at 858~nm), four V$_1$ centres (h-site, ZPL at 861~nm), and one V$_2$ center (k-site, ZPL at 916~nm). Other defect centers showed ZPLs at various wavelengths within the 858-985~nm range, which do not correspond to previously reported unidentified lines in this range but align with the theoretically predicted range of V$_\text{Si}$ centers modified by nearby carbon anti-sites.

The broad line widths observed in the PL spectra and the lack of an ODMR signal might be attributed to the femtosecond laser's creation of additional non-optically active defects alongside the V$_{\text{Si}}$.

\begin{figure*}[h!]
\centering
\includegraphics[width= 1.0\textwidth]{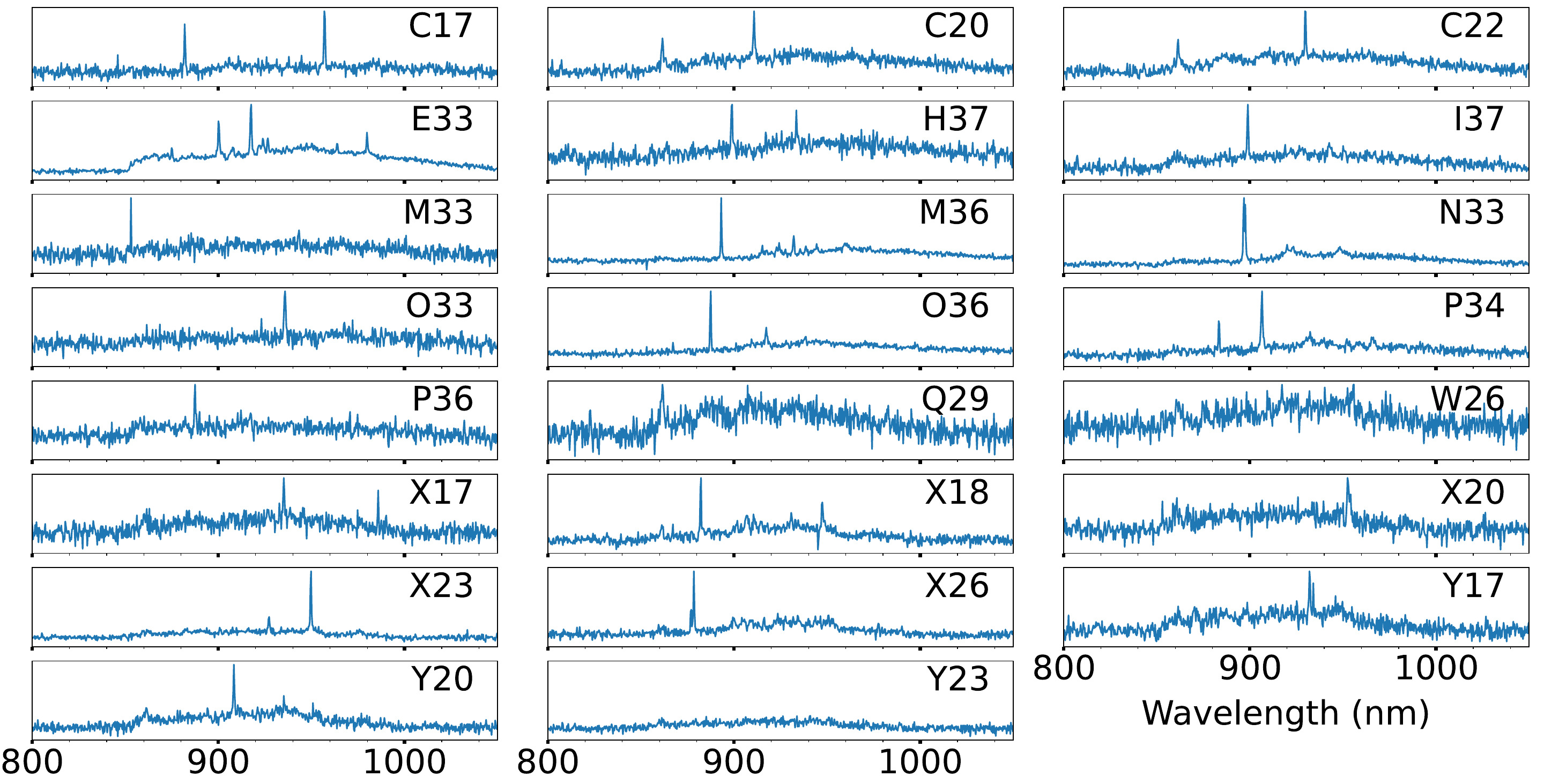}
\caption{Optical spectra were taken at low temperatures using the setup previously discussed in the Results Section for laser-written defects in SIL structures. They show a range of zero-phonon lines consistent with the predicted region for modified V$_\text{Si}$ centers. Labels correspond to SIL array position, as described in the main text. Zero-phonon lines are consistent with V$_\text{Si}$ centers observed for SILs C20, C22, Q29, T30, O36, and Y20.
\label{fig:Supp_spectra}
}
\end{figure*}

\subsection{ODMR after Anneal}

An anneal at 600 degrees was conducted for 30 minutes on the sample in vacuum at $4\times10^{-5}$ mbar to investigate the stability of the observed defects, and whether they could be converted to silicon vacancies. We re-characterised 5 SILs, which originally showed no ODMR. One evidenced an ODMR peak after annealing, with 0.48\% contrast and 22.36 MHz linewidth (Fig.~S6). 

The ODMR measurement is performed by focusing an off-resonant cw-laser (730~nm) on the a-plane side of the 4H-SiC sample inside the cryostat (Attocube Attodry 800) at 4K with a high-NA objective (Zeiss Epiplan-Neofluar 100x, NA 0.9). The spin state manipulation of the color center is orchestrated with the help of an Arbitrary Waveform Generator (QM-OPX) and a copper wire (50 $\mu m$ diameter) running on top of the sample.  The AWG feeds a radio-frequency signal with a power of -27dBm, amplified by a 44 dB amplifier (LZY-22+ Mini-circuits) and fed to the microwave antenna wire running over the sample. The manipulation of the spin states is realized once the frequency of microwave matches with the zero field splitting of ground state spin (~70 MHz for a $V_2$ color center at $B_0$ = 0G). The difference in spin state population while sweeping the MW frequency is seen through the photoluminescence emitted  by the V$_{2}$ V$_\text{Si}$ center which is filtered using a dichroic mirror (Semrock FF925-Di01) and along pass filter (FELH950 Thorlabs). The emission is readout through a super conducting nanowire single photon detector (Single Quantum). 

The filtering during spectral acquisition was carried out using a SEMROCK long-pass filter in the transmission direction, oriented at approximately $50^\circ$ to achieve an effective cutoff near 900~nm. This configuration was chosen to remove stray light from the 730~nm excitation laser, which has a mild band extending to 1000~nm. The 900~nm filter allows transmission of zero-phonon lines (ZPLs), including the $V_{2}$ center at 916~nm. To enhance clarity for the reader, cosmic ray artefacts were removed during post-processing.
  
\begin{figure}[h]
\centering
\includegraphics[width= 0.95\textwidth]{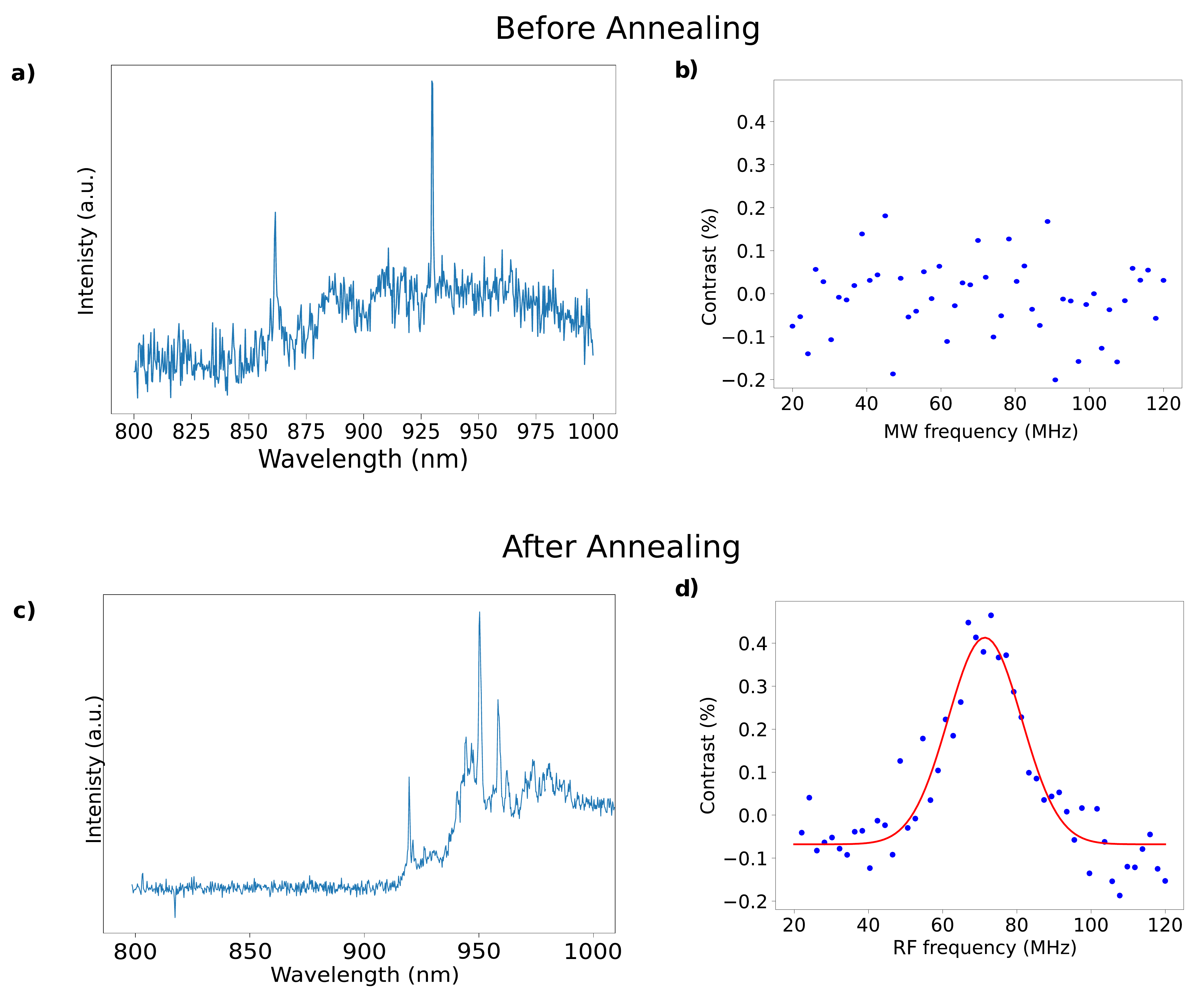}
\caption{This ODMR was conducted at room temperature on SIL C22 using a 730 nm later to excite the color center off resonantly while sweeping the microwave frequency (MW) from 20 MHz to 120 MHz. The spectrum is centered on $71.41 \pm{0.6}$MHz and Full-width half-maximum at $22.36 \pm{2.4}$MHz. The low-temperature spectra in a) is the same as previously shown in the SI section 5. 
\label{fig:Supp_ODMR}
}
\end{figure}

\clearpage

\bibliography{laser_writing_SILs_V2}

\begin{thebibliography}{83}%
\makeatletter
\providecommand \@ifxundefined [1]{%
 \@ifx{#1\undefined}
}%
\providecommand \@ifnum [1]{%
 \ifnum #1\expandafter \@firstoftwo
 \else \expandafter \@secondoftwo
 \fi
}%
\providecommand \@ifx [1]{%
 \ifx #1\expandafter \@firstoftwo
 \else \expandafter \@secondoftwo
 \fi
}%
\providecommand \natexlab [1]{#1}%
\providecommand \enquote  [1]{``#1''}%
\providecommand \bibnamefont  [1]{#1}%
\providecommand \bibfnamefont [1]{#1}%
\providecommand \citenamefont [1]{#1}%
\providecommand \href@noop [0]{\@secondoftwo}%
\providecommand \href [0]{\begingroup \@sanitize@url \@href}%
\providecommand \@href[1]{\@@startlink{#1}\@@href}%
\providecommand \@@href[1]{\endgroup#1\@@endlink}%
\providecommand \@sanitize@url [0]{\catcode `\\12\catcode `\$12\catcode `\&12\catcode `\#12\catcode `\^12\catcode `\_12\catcode `\%12\relax}%
\providecommand \@@startlink[1]{}%
\providecommand \@@endlink[0]{}%
\providecommand \url  [0]{\begingroup\@sanitize@url \@url }%
\providecommand \@url [1]{\endgroup\@href {#1}{\urlprefix }}%
\providecommand \urlprefix  [0]{URL }%
\providecommand \Eprint [0]{\href }%
\providecommand \doibase [0]{https://doi.org/}%
\providecommand \selectlanguage [0]{\@gobble}%
\providecommand \bibinfo  [0]{\@secondoftwo}%
\providecommand \bibfield  [0]{\@secondoftwo}%
\providecommand \translation [1]{[#1]}%
\providecommand \BibitemOpen [0]{}%
\providecommand \bibitemStop [0]{}%
\providecommand \bibitemNoStop [0]{.\EOS\space}%
\providecommand \EOS [0]{\spacefactor3000\relax}%
\providecommand \BibitemShut  [1]{\csname bibitem#1\endcsname}%
\let\auto@bib@innerbib\@empty
\bibitem [{\citenamefont {Awschalom}\ \emph {et~al.}(2018)\citenamefont {Awschalom}, \citenamefont {Hanson}, \citenamefont {Wrachtrup},\ and\ \citenamefont {Zhou}}]{awschalom_quantum_2018}%
  \BibitemOpen
  \bibfield  {author} {\bibinfo {author} {\bibfnamefont {D.~D.}\ \bibnamefont {Awschalom}}, \bibinfo {author} {\bibfnamefont {R.}~\bibnamefont {Hanson}}, \bibinfo {author} {\bibfnamefont {J.}~\bibnamefont {Wrachtrup}},\ and\ \bibinfo {author} {\bibfnamefont {B.~B.}\ \bibnamefont {Zhou}},\ }\href {https://doi.org/10.1038/s41566-018-0232-2} {\bibfield  {journal} {\bibinfo  {journal} {Nature Photonics}\ }\textbf {\bibinfo {volume} {12}},\ \bibinfo {pages} {516} (\bibinfo {year} {2018})}\BibitemShut {NoStop}%
\bibitem [{\citenamefont {Parker}\ \emph {et~al.}(2024)\citenamefont {Parker}, \citenamefont {Arjona~Martínez}, \citenamefont {Chen}, \citenamefont {Stramma}, \citenamefont {Harris}, \citenamefont {Michaels}, \citenamefont {Trusheim}, \citenamefont {Hayhurst~Appel}, \citenamefont {Purser}, \citenamefont {Roth}, \citenamefont {Englund},\ and\ \citenamefont {Atatüre}}]{parker_diamond_2024}%
  \BibitemOpen
  \bibfield  {author} {\bibinfo {author} {\bibfnamefont {R.~A.}\ \bibnamefont {Parker}}, \bibinfo {author} {\bibfnamefont {J.}~\bibnamefont {Arjona~Martínez}}, \bibinfo {author} {\bibfnamefont {K.~C.}\ \bibnamefont {Chen}}, \bibinfo {author} {\bibfnamefont {A.~M.}\ \bibnamefont {Stramma}}, \bibinfo {author} {\bibfnamefont {I.~B.}\ \bibnamefont {Harris}}, \bibinfo {author} {\bibfnamefont {C.~P.}\ \bibnamefont {Michaels}}, \bibinfo {author} {\bibfnamefont {M.~E.}\ \bibnamefont {Trusheim}}, \bibinfo {author} {\bibfnamefont {M.}~\bibnamefont {Hayhurst~Appel}}, \bibinfo {author} {\bibfnamefont {C.~M.}\ \bibnamefont {Purser}}, \bibinfo {author} {\bibfnamefont {W.~G.}\ \bibnamefont {Roth}}, \bibinfo {author} {\bibfnamefont {D.}~\bibnamefont {Englund}},\ and\ \bibinfo {author} {\bibfnamefont {M.}~\bibnamefont {Atatüre}},\ }\href {https://doi.org/10.1038/s41566-023-01332-8} {\bibfield  {journal} {\bibinfo  {journal} {Nature Photonics}\ }\textbf {\bibinfo {volume} {18}},\ \bibinfo {pages} {156} (\bibinfo {year}
  {2024})}\BibitemShut {NoStop}%
\bibitem [{\citenamefont {Hermans}\ \emph {et~al.}(2022)\citenamefont {Hermans}, \citenamefont {Pompili}, \citenamefont {Beukers}, \citenamefont {Baier}, \citenamefont {Borregaard},\ and\ \citenamefont {Hanson}}]{hermans_qubit_2022}%
  \BibitemOpen
  \bibfield  {author} {\bibinfo {author} {\bibfnamefont {S.~L.~N.}\ \bibnamefont {Hermans}}, \bibinfo {author} {\bibfnamefont {M.}~\bibnamefont {Pompili}}, \bibinfo {author} {\bibfnamefont {H.~K.~C.}\ \bibnamefont {Beukers}}, \bibinfo {author} {\bibfnamefont {S.}~\bibnamefont {Baier}}, \bibinfo {author} {\bibfnamefont {J.}~\bibnamefont {Borregaard}},\ and\ \bibinfo {author} {\bibfnamefont {R.}~\bibnamefont {Hanson}},\ }\href {https://doi.org/10.1038/s41586-022-04697-y} {\bibfield  {journal} {\bibinfo  {journal} {Nature}\ }\textbf {\bibinfo {volume} {605}},\ \bibinfo {pages} {663} (\bibinfo {year} {2022})}\BibitemShut {NoStop}%
\bibitem [{\citenamefont {Stolk}\ \emph {et~al.}(2024)\citenamefont {Stolk}, \citenamefont {van~der Enden}, \citenamefont {Slater}, \citenamefont {te~Raa-Derckx}, \citenamefont {Botma}, \citenamefont {van Rantwijk}, \citenamefont {Biemond}, \citenamefont {Hagen}, \citenamefont {Herfst}, \citenamefont {Koek}, \citenamefont {Meskers}, \citenamefont {Vollmer}, \citenamefont {van Zwet}, \citenamefont {Markham}, \citenamefont {Edmonds}, \citenamefont {Geus}, \citenamefont {Elsen}, \citenamefont {Jungbluth}, \citenamefont {Haefner}, \citenamefont {Tresp}, \citenamefont {Stuhler}, \citenamefont {Ritter},\ and\ \citenamefont {Hanson}}]{stolk_metropolitan-scale_2024}%
  \BibitemOpen
  \bibfield  {author} {\bibinfo {author} {\bibfnamefont {A.~J.}\ \bibnamefont {Stolk}}, \bibinfo {author} {\bibfnamefont {K.~L.}\ \bibnamefont {van~der Enden}}, \bibinfo {author} {\bibfnamefont {M.-C.}\ \bibnamefont {Slater}}, \bibinfo {author} {\bibfnamefont {I.}~\bibnamefont {te~Raa-Derckx}}, \bibinfo {author} {\bibfnamefont {P.}~\bibnamefont {Botma}}, \bibinfo {author} {\bibfnamefont {J.}~\bibnamefont {van Rantwijk}}, \bibinfo {author} {\bibfnamefont {J.~J.~B.}\ \bibnamefont {Biemond}}, \bibinfo {author} {\bibfnamefont {R.~A.~J.}\ \bibnamefont {Hagen}}, \bibinfo {author} {\bibfnamefont {R.~W.}\ \bibnamefont {Herfst}}, \bibinfo {author} {\bibfnamefont {W.~D.}\ \bibnamefont {Koek}}, \bibinfo {author} {\bibfnamefont {A.~J.~H.}\ \bibnamefont {Meskers}}, \bibinfo {author} {\bibfnamefont {R.}~\bibnamefont {Vollmer}}, \bibinfo {author} {\bibfnamefont {E.~J.}\ \bibnamefont {van Zwet}}, \bibinfo {author} {\bibfnamefont {M.}~\bibnamefont {Markham}}, \bibinfo {author} {\bibfnamefont {A.~M.}\ \bibnamefont {Edmonds}},
  \bibinfo {author} {\bibfnamefont {J.~F.}\ \bibnamefont {Geus}}, \bibinfo {author} {\bibfnamefont {F.}~\bibnamefont {Elsen}}, \bibinfo {author} {\bibfnamefont {B.}~\bibnamefont {Jungbluth}}, \bibinfo {author} {\bibfnamefont {C.}~\bibnamefont {Haefner}}, \bibinfo {author} {\bibfnamefont {C.}~\bibnamefont {Tresp}}, \bibinfo {author} {\bibfnamefont {J.}~\bibnamefont {Stuhler}}, \bibinfo {author} {\bibfnamefont {S.}~\bibnamefont {Ritter}},\ and\ \bibinfo {author} {\bibfnamefont {R.}~\bibnamefont {Hanson}},\ }\href {https://doi.org/10.1126/sciadv.adp6442} {\bibfield  {journal} {\bibinfo  {journal} {Science Advances}\ }\textbf {\bibinfo {volume} {10}},\ \bibinfo {pages} {eadp6442} (\bibinfo {year} {2024})},\ \bibinfo {note} {publisher: American Association for the Advancement of Science}\BibitemShut {NoStop}%
\bibitem [{\citenamefont {Christle}\ \emph {et~al.}(2017)\citenamefont {Christle}, \citenamefont {Klimov}, \citenamefont {de~las Casas}, \citenamefont {Szász}, \citenamefont {Ivády}, \citenamefont {Jokubavicius}, \citenamefont {Ul~Hassan}, \citenamefont {Syväjärvi}, \citenamefont {Koehl}, \citenamefont {Ohshima}, \citenamefont {Son}, \citenamefont {Janzén}, \citenamefont {Gali},\ and\ \citenamefont {Awschalom}}]{christle_isolated_2017}%
  \BibitemOpen
  \bibfield  {author} {\bibinfo {author} {\bibfnamefont {D.~J.}\ \bibnamefont {Christle}}, \bibinfo {author} {\bibfnamefont {P.~V.}\ \bibnamefont {Klimov}}, \bibinfo {author} {\bibfnamefont {C.~F.}\ \bibnamefont {de~las Casas}}, \bibinfo {author} {\bibfnamefont {K.}~\bibnamefont {Szász}}, \bibinfo {author} {\bibfnamefont {V.}~\bibnamefont {Ivády}}, \bibinfo {author} {\bibfnamefont {V.}~\bibnamefont {Jokubavicius}}, \bibinfo {author} {\bibfnamefont {J.}~\bibnamefont {Ul~Hassan}}, \bibinfo {author} {\bibfnamefont {M.}~\bibnamefont {Syväjärvi}}, \bibinfo {author} {\bibfnamefont {W.~F.}\ \bibnamefont {Koehl}}, \bibinfo {author} {\bibfnamefont {T.}~\bibnamefont {Ohshima}}, \bibinfo {author} {\bibfnamefont {N.~T.}\ \bibnamefont {Son}}, \bibinfo {author} {\bibfnamefont {E.}~\bibnamefont {Janzén}}, \bibinfo {author} {\bibfnamefont {Ã.}~\bibnamefont {Gali}},\ and\ \bibinfo {author} {\bibfnamefont {D.~D.}\ \bibnamefont {Awschalom}},\ }\href {https://doi.org/10.1103/PhysRevX.7.021046} {\bibfield  {journal}
  {\bibinfo  {journal} {Physical Review X}\ }\textbf {\bibinfo {volume} {7}},\ \bibinfo {pages} {021046} (\bibinfo {year} {2017})}\BibitemShut {NoStop}%
\bibitem [{\citenamefont {Nagy}\ \emph {et~al.}(2019)\citenamefont {Nagy}, \citenamefont {Niethammer}, \citenamefont {Widmann}, \citenamefont {Chen}, \citenamefont {Udvarhelyi}, \citenamefont {Bonato}, \citenamefont {Hassan}, \citenamefont {Karhu}, \citenamefont {Ivanov}, \citenamefont {Son}, \citenamefont {Maze}, \citenamefont {Ohshima}, \citenamefont {Soykal}, \citenamefont {Gali}, \citenamefont {Lee}, \citenamefont {Kaiser},\ and\ \citenamefont {Wrachtrup}}]{nagy_high-fidelity_2019}%
  \BibitemOpen
  \bibfield  {author} {\bibinfo {author} {\bibfnamefont {R.}~\bibnamefont {Nagy}}, \bibinfo {author} {\bibfnamefont {M.}~\bibnamefont {Niethammer}}, \bibinfo {author} {\bibfnamefont {M.}~\bibnamefont {Widmann}}, \bibinfo {author} {\bibfnamefont {Y.-C.}\ \bibnamefont {Chen}}, \bibinfo {author} {\bibfnamefont {P.}~\bibnamefont {Udvarhelyi}}, \bibinfo {author} {\bibfnamefont {C.}~\bibnamefont {Bonato}}, \bibinfo {author} {\bibfnamefont {J.~U.}\ \bibnamefont {Hassan}}, \bibinfo {author} {\bibfnamefont {R.}~\bibnamefont {Karhu}}, \bibinfo {author} {\bibfnamefont {I.~G.}\ \bibnamefont {Ivanov}}, \bibinfo {author} {\bibfnamefont {N.~T.}\ \bibnamefont {Son}}, \bibinfo {author} {\bibfnamefont {J.~R.}\ \bibnamefont {Maze}}, \bibinfo {author} {\bibfnamefont {T.}~\bibnamefont {Ohshima}}, \bibinfo {author} {\bibfnamefont {Ã.~O.}\ \bibnamefont {Soykal}}, \bibinfo {author} {\bibfnamefont {Ã.}~\bibnamefont {Gali}}, \bibinfo {author} {\bibfnamefont {S.-Y.}\ \bibnamefont {Lee}}, \bibinfo {author} {\bibfnamefont
  {F.}~\bibnamefont {Kaiser}},\ and\ \bibinfo {author} {\bibfnamefont {J.}~\bibnamefont {Wrachtrup}},\ }\href {https://doi.org/10.1038/s41467-019-09873-9} {\bibfield  {journal} {\bibinfo  {journal} {Nature Communications}\ }\textbf {\bibinfo {volume} {10}},\ \bibinfo {pages} {1954} (\bibinfo {year} {2019})}\BibitemShut {NoStop}%
\bibitem [{\citenamefont {Anderson}\ \emph {et~al.}(2022)\citenamefont {Anderson}, \citenamefont {Glen}, \citenamefont {Zeledon}, \citenamefont {Bourassa}, \citenamefont {Jin}, \citenamefont {Zhu}, \citenamefont {Vorwerk}, \citenamefont {Crook}, \citenamefont {Abe}, \citenamefont {Ul-Hassan}, \citenamefont {Ohshima}, \citenamefont {Son}, \citenamefont {Galli},\ and\ \citenamefont {Awschalom}}]{anderson_five-second_2022}%
  \BibitemOpen
  \bibfield  {author} {\bibinfo {author} {\bibfnamefont {C.~P.}\ \bibnamefont {Anderson}}, \bibinfo {author} {\bibfnamefont {E.~O.}\ \bibnamefont {Glen}}, \bibinfo {author} {\bibfnamefont {C.}~\bibnamefont {Zeledon}}, \bibinfo {author} {\bibfnamefont {A.}~\bibnamefont {Bourassa}}, \bibinfo {author} {\bibfnamefont {Y.}~\bibnamefont {Jin}}, \bibinfo {author} {\bibfnamefont {Y.}~\bibnamefont {Zhu}}, \bibinfo {author} {\bibfnamefont {C.}~\bibnamefont {Vorwerk}}, \bibinfo {author} {\bibfnamefont {A.~L.}\ \bibnamefont {Crook}}, \bibinfo {author} {\bibfnamefont {H.}~\bibnamefont {Abe}}, \bibinfo {author} {\bibfnamefont {J.}~\bibnamefont {Ul-Hassan}}, \bibinfo {author} {\bibfnamefont {T.}~\bibnamefont {Ohshima}}, \bibinfo {author} {\bibfnamefont {N.~T.}\ \bibnamefont {Son}}, \bibinfo {author} {\bibfnamefont {G.}~\bibnamefont {Galli}},\ and\ \bibinfo {author} {\bibfnamefont {D.~D.}\ \bibnamefont {Awschalom}},\ }\href {https://doi.org/10.1126/sciadv.abm5912} {\bibfield  {journal} {\bibinfo  {journal} {Science
  Advances}\ }\textbf {\bibinfo {volume} {8}},\ \bibinfo {pages} {eabm5912} (\bibinfo {year} {2022})}\BibitemShut {NoStop}%
\bibitem [{\citenamefont {Cilibrizzi}\ \emph {et~al.}(2023)\citenamefont {Cilibrizzi}, \citenamefont {Arshad}, \citenamefont {Tissot}, \citenamefont {Son}, \citenamefont {Ivanov}, \citenamefont {Astner}, \citenamefont {Koller}, \citenamefont {Ghezellou}, \citenamefont {Ul-Hassan}, \citenamefont {White}, \citenamefont {Bekker}, \citenamefont {Burkard}, \citenamefont {Trupke},\ and\ \citenamefont {Bonato}}]{cilibrizzi_ultra-narrow_2023}%
  \BibitemOpen
  \bibfield  {author} {\bibinfo {author} {\bibfnamefont {P.}~\bibnamefont {Cilibrizzi}}, \bibinfo {author} {\bibfnamefont {M.~J.}\ \bibnamefont {Arshad}}, \bibinfo {author} {\bibfnamefont {B.}~\bibnamefont {Tissot}}, \bibinfo {author} {\bibfnamefont {N.~T.}\ \bibnamefont {Son}}, \bibinfo {author} {\bibfnamefont {I.~G.}\ \bibnamefont {Ivanov}}, \bibinfo {author} {\bibfnamefont {T.}~\bibnamefont {Astner}}, \bibinfo {author} {\bibfnamefont {P.}~\bibnamefont {Koller}}, \bibinfo {author} {\bibfnamefont {M.}~\bibnamefont {Ghezellou}}, \bibinfo {author} {\bibfnamefont {J.}~\bibnamefont {Ul-Hassan}}, \bibinfo {author} {\bibfnamefont {D.}~\bibnamefont {White}}, \bibinfo {author} {\bibfnamefont {C.}~\bibnamefont {Bekker}}, \bibinfo {author} {\bibfnamefont {G.}~\bibnamefont {Burkard}}, \bibinfo {author} {\bibfnamefont {M.}~\bibnamefont {Trupke}},\ and\ \bibinfo {author} {\bibfnamefont {C.}~\bibnamefont {Bonato}},\ }\href {https://doi.org/10.1038/s41467-023-43923-7} {\bibfield  {journal} {\bibinfo  {journal} {Nature
  Communications}\ }\textbf {\bibinfo {volume} {14}},\ \bibinfo {pages} {8448} (\bibinfo {year} {2023})}\BibitemShut {NoStop}%
\bibitem [{\citenamefont {Ecker}\ \emph {et~al.}(2024)\citenamefont {Ecker}, \citenamefont {Fink}, \citenamefont {Scheidl}, \citenamefont {Sohr}, \citenamefont {Ursin}, \citenamefont {Arshad}, \citenamefont {Bonato}, \citenamefont {Cilibrizzi}, \citenamefont {Gali}, \citenamefont {Udvarhelyi}, \citenamefont {Politi}, \citenamefont {Trojak}, \citenamefont {Ghezellou}, \citenamefont {Hassan}, \citenamefont {Ivanov}, \citenamefont {Son}, \citenamefont {Burkard}, \citenamefont {Tissot}, \citenamefont {Hendriks}, \citenamefont {Gilardoni}, \citenamefont {Wal}, \citenamefont {David}, \citenamefont {Astner}, \citenamefont {Koller},\ and\ \citenamefont {Trupke}}]{ecker_quantum_2024}%
  \BibitemOpen
  \bibfield  {author} {\bibinfo {author} {\bibfnamefont {S.}~\bibnamefont {Ecker}}, \bibinfo {author} {\bibfnamefont {M.}~\bibnamefont {Fink}}, \bibinfo {author} {\bibfnamefont {T.}~\bibnamefont {Scheidl}}, \bibinfo {author} {\bibfnamefont {P.}~\bibnamefont {Sohr}}, \bibinfo {author} {\bibfnamefont {R.}~\bibnamefont {Ursin}}, \bibinfo {author} {\bibfnamefont {M.~J.}\ \bibnamefont {Arshad}}, \bibinfo {author} {\bibfnamefont {C.}~\bibnamefont {Bonato}}, \bibinfo {author} {\bibfnamefont {P.}~\bibnamefont {Cilibrizzi}}, \bibinfo {author} {\bibfnamefont {A.}~\bibnamefont {Gali}}, \bibinfo {author} {\bibfnamefont {P.}~\bibnamefont {Udvarhelyi}}, \bibinfo {author} {\bibfnamefont {A.}~\bibnamefont {Politi}}, \bibinfo {author} {\bibfnamefont {O.~J.}\ \bibnamefont {Trojak}}, \bibinfo {author} {\bibfnamefont {M.}~\bibnamefont {Ghezellou}}, \bibinfo {author} {\bibfnamefont {J.~U.}\ \bibnamefont {Hassan}}, \bibinfo {author} {\bibfnamefont {I.~G.}\ \bibnamefont {Ivanov}}, \bibinfo {author} {\bibfnamefont {N.~T.}\
  \bibnamefont {Son}}, \bibinfo {author} {\bibfnamefont {G.}~\bibnamefont {Burkard}}, \bibinfo {author} {\bibfnamefont {B.}~\bibnamefont {Tissot}}, \bibinfo {author} {\bibfnamefont {J.}~\bibnamefont {Hendriks}}, \bibinfo {author} {\bibfnamefont {C.~M.}\ \bibnamefont {Gilardoni}}, \bibinfo {author} {\bibfnamefont {C.~H. v.~d.}\ \bibnamefont {Wal}}, \bibinfo {author} {\bibfnamefont {C.}~\bibnamefont {David}}, \bibinfo {author} {\bibfnamefont {T.}~\bibnamefont {Astner}}, \bibinfo {author} {\bibfnamefont {P.}~\bibnamefont {Koller}},\ and\ \bibinfo {author} {\bibfnamefont {M.}~\bibnamefont {Trupke}},\ }\href {https://doi.org/10.48550/arXiv.2403.03284} {\bibinfo {title} {Quantum communication networks with defects in silicon carbide}} (\bibinfo {year} {2024}),\ \bibinfo {note} {arXiv:2403.03284 [quant-ph]}\BibitemShut {NoStop}%
\bibitem [{\citenamefont {Gritsch}\ \emph {et~al.}(2025)\citenamefont {Gritsch}, \citenamefont {Ulanowski}, \citenamefont {Pforr},\ and\ \citenamefont {Reiserer}}]{gritsch_optical_2025}%
  \BibitemOpen
  \bibfield  {author} {\bibinfo {author} {\bibfnamefont {A.}~\bibnamefont {Gritsch}}, \bibinfo {author} {\bibfnamefont {A.}~\bibnamefont {Ulanowski}}, \bibinfo {author} {\bibfnamefont {J.}~\bibnamefont {Pforr}},\ and\ \bibinfo {author} {\bibfnamefont {A.}~\bibnamefont {Reiserer}},\ }\href {https://doi.org/10.1038/s41467-024-55552-9} {\bibfield  {journal} {\bibinfo  {journal} {Nature Communications}\ }\textbf {\bibinfo {volume} {16}},\ \bibinfo {pages} {64} (\bibinfo {year} {2025})}\BibitemShut {NoStop}%
\bibitem [{\citenamefont {Inc}\ \emph {et~al.}(2024)\citenamefont {Inc}, \citenamefont {Afzal}, \citenamefont {Akhlaghi}, \citenamefont {Beale}, \citenamefont {Bedroya}, \citenamefont {Bell}, \citenamefont {Bergeron}, \citenamefont {Bonsma-Fisher}, \citenamefont {Bychkova}, \citenamefont {Chaisson}, \citenamefont {Chartrand}, \citenamefont {Clear}, \citenamefont {Darcie}, \citenamefont {DeAbreu}, \citenamefont {DeLisle}, \citenamefont {Duncan}, \citenamefont {Smith}, \citenamefont {Dunn}, \citenamefont {Ebrahimi}, \citenamefont {Evetts}, \citenamefont {Pinheiro}, \citenamefont {Fuentes}, \citenamefont {Georgiou}, \citenamefont {Guha}, \citenamefont {Haenel}, \citenamefont {Higginbottom}, \citenamefont {Jackson}, \citenamefont {Jahed}, \citenamefont {Khorshidahmad}, \citenamefont {Shandilya}, \citenamefont {Kurkjian}, \citenamefont {Lauk}, \citenamefont {Lee-Hone}, \citenamefont {Lin}, \citenamefont {Litynskyy}, \citenamefont {Lock}, \citenamefont {Ma}, \citenamefont {MacGilp}, \citenamefont {MacQuarrie},
  \citenamefont {Mar}, \citenamefont {Khah}, \citenamefont {Matiash}, \citenamefont {Meyer-Scott}, \citenamefont {Michaels}, \citenamefont {Motira}, \citenamefont {Noori}, \citenamefont {Ospadov}, \citenamefont {Patel}, \citenamefont {Patscheider}, \citenamefont {Paulson}, \citenamefont {Petruk}, \citenamefont {Ravindranath}, \citenamefont {Reznychenko}, \citenamefont {Ruether}, \citenamefont {Ruscica}, \citenamefont {Saxena}, \citenamefont {Schaller}, \citenamefont {Seidlitz}, \citenamefont {Senger}, \citenamefont {Lee}, \citenamefont {Sevoyan}, \citenamefont {Simmons}, \citenamefont {Soykal}, \citenamefont {Stott}, \citenamefont {Tran}, \citenamefont {Tserkis}, \citenamefont {Ulhaq}, \citenamefont {Vine}, \citenamefont {Weeks}, \citenamefont {Wolfowicz},\ and\ \citenamefont {Yoneda}}]{inc_distributed_2024}%
  \BibitemOpen
  \bibfield  {author} {\bibinfo {author} {\bibfnamefont {P.}~\bibnamefont {Inc}}, \bibinfo {author} {\bibfnamefont {F.}~\bibnamefont {Afzal}}, \bibinfo {author} {\bibfnamefont {M.}~\bibnamefont {Akhlaghi}}, \bibinfo {author} {\bibfnamefont {S.~J.}\ \bibnamefont {Beale}}, \bibinfo {author} {\bibfnamefont {O.}~\bibnamefont {Bedroya}}, \bibinfo {author} {\bibfnamefont {K.}~\bibnamefont {Bell}}, \bibinfo {author} {\bibfnamefont {L.}~\bibnamefont {Bergeron}}, \bibinfo {author} {\bibfnamefont {K.}~\bibnamefont {Bonsma-Fisher}}, \bibinfo {author} {\bibfnamefont {P.}~\bibnamefont {Bychkova}}, \bibinfo {author} {\bibfnamefont {Z.~M.~E.}\ \bibnamefont {Chaisson}}, \bibinfo {author} {\bibfnamefont {C.}~\bibnamefont {Chartrand}}, \bibinfo {author} {\bibfnamefont {C.}~\bibnamefont {Clear}}, \bibinfo {author} {\bibfnamefont {A.}~\bibnamefont {Darcie}}, \bibinfo {author} {\bibfnamefont {A.}~\bibnamefont {DeAbreu}}, \bibinfo {author} {\bibfnamefont {C.}~\bibnamefont {DeLisle}}, \bibinfo {author} {\bibfnamefont {L.~A.}\
  \bibnamefont {Duncan}}, \bibinfo {author} {\bibfnamefont {C.~D.}\ \bibnamefont {Smith}}, \bibinfo {author} {\bibfnamefont {J.}~\bibnamefont {Dunn}}, \bibinfo {author} {\bibfnamefont {A.}~\bibnamefont {Ebrahimi}}, \bibinfo {author} {\bibfnamefont {N.}~\bibnamefont {Evetts}}, \bibinfo {author} {\bibfnamefont {D.~F.}\ \bibnamefont {Pinheiro}}, \bibinfo {author} {\bibfnamefont {P.}~\bibnamefont {Fuentes}}, \bibinfo {author} {\bibfnamefont {T.}~\bibnamefont {Georgiou}}, \bibinfo {author} {\bibfnamefont {B.}~\bibnamefont {Guha}}, \bibinfo {author} {\bibfnamefont {R.}~\bibnamefont {Haenel}}, \bibinfo {author} {\bibfnamefont {D.}~\bibnamefont {Higginbottom}}, \bibinfo {author} {\bibfnamefont {D.~M.}\ \bibnamefont {Jackson}}, \bibinfo {author} {\bibfnamefont {N.}~\bibnamefont {Jahed}}, \bibinfo {author} {\bibfnamefont {A.}~\bibnamefont {Khorshidahmad}}, \bibinfo {author} {\bibfnamefont {P.~K.}\ \bibnamefont {Shandilya}}, \bibinfo {author} {\bibfnamefont {A.~T.~K.}\ \bibnamefont {Kurkjian}}, \bibinfo {author}
  {\bibfnamefont {N.}~\bibnamefont {Lauk}}, \bibinfo {author} {\bibfnamefont {N.~R.}\ \bibnamefont {Lee-Hone}}, \bibinfo {author} {\bibfnamefont {E.}~\bibnamefont {Lin}}, \bibinfo {author} {\bibfnamefont {R.}~\bibnamefont {Litynskyy}}, \bibinfo {author} {\bibfnamefont {D.}~\bibnamefont {Lock}}, \bibinfo {author} {\bibfnamefont {L.}~\bibnamefont {Ma}}, \bibinfo {author} {\bibfnamefont {I.}~\bibnamefont {MacGilp}}, \bibinfo {author} {\bibfnamefont {E.~R.}\ \bibnamefont {MacQuarrie}}, \bibinfo {author} {\bibfnamefont {A.}~\bibnamefont {Mar}}, \bibinfo {author} {\bibfnamefont {A.~M.}\ \bibnamefont {Khah}}, \bibinfo {author} {\bibfnamefont {A.}~\bibnamefont {Matiash}}, \bibinfo {author} {\bibfnamefont {E.}~\bibnamefont {Meyer-Scott}}, \bibinfo {author} {\bibfnamefont {C.~P.}\ \bibnamefont {Michaels}}, \bibinfo {author} {\bibfnamefont {J.}~\bibnamefont {Motira}}, \bibinfo {author} {\bibfnamefont {N.~K.}\ \bibnamefont {Noori}}, \bibinfo {author} {\bibfnamefont {E.}~\bibnamefont {Ospadov}}, \bibinfo {author}
  {\bibfnamefont {E.}~\bibnamefont {Patel}}, \bibinfo {author} {\bibfnamefont {A.}~\bibnamefont {Patscheider}}, \bibinfo {author} {\bibfnamefont {D.}~\bibnamefont {Paulson}}, \bibinfo {author} {\bibfnamefont {A.}~\bibnamefont {Petruk}}, \bibinfo {author} {\bibfnamefont {A.~L.}\ \bibnamefont {Ravindranath}}, \bibinfo {author} {\bibfnamefont {B.}~\bibnamefont {Reznychenko}}, \bibinfo {author} {\bibfnamefont {M.}~\bibnamefont {Ruether}}, \bibinfo {author} {\bibfnamefont {J.}~\bibnamefont {Ruscica}}, \bibinfo {author} {\bibfnamefont {K.}~\bibnamefont {Saxena}}, \bibinfo {author} {\bibfnamefont {Z.}~\bibnamefont {Schaller}}, \bibinfo {author} {\bibfnamefont {A.}~\bibnamefont {Seidlitz}}, \bibinfo {author} {\bibfnamefont {J.}~\bibnamefont {Senger}}, \bibinfo {author} {\bibfnamefont {Y.~S.}\ \bibnamefont {Lee}}, \bibinfo {author} {\bibfnamefont {O.}~\bibnamefont {Sevoyan}}, \bibinfo {author} {\bibfnamefont {S.}~\bibnamefont {Simmons}}, \bibinfo {author} {\bibfnamefont {O.}~\bibnamefont {Soykal}}, \bibinfo {author}
  {\bibfnamefont {L.}~\bibnamefont {Stott}}, \bibinfo {author} {\bibfnamefont {Q.}~\bibnamefont {Tran}}, \bibinfo {author} {\bibfnamefont {S.}~\bibnamefont {Tserkis}}, \bibinfo {author} {\bibfnamefont {A.}~\bibnamefont {Ulhaq}}, \bibinfo {author} {\bibfnamefont {W.}~\bibnamefont {Vine}}, \bibinfo {author} {\bibfnamefont {R.}~\bibnamefont {Weeks}}, \bibinfo {author} {\bibfnamefont {G.}~\bibnamefont {Wolfowicz}},\ and\ \bibinfo {author} {\bibfnamefont {I.}~\bibnamefont {Yoneda}},\ }\href {https://doi.org/10.48550/arXiv.2406.01704} {\bibinfo {title} {Distributed {Quantum} {Computing} in {Silicon}}} (\bibinfo {year} {2024}),\ \bibinfo {note} {arXiv:2406.01704 [quant-ph]}\BibitemShut {NoStop}%
\bibitem [{\citenamefont {Wu}\ \emph {et~al.}(2023)\citenamefont {Wu}, \citenamefont {Riedel}, \citenamefont {Ruskuc}, \citenamefont {Zhong}, \citenamefont {Kwon},\ and\ \citenamefont {Faraon}}]{wu_near-infrared_2023}%
  \BibitemOpen
  \bibfield  {author} {\bibinfo {author} {\bibfnamefont {C.-J.}\ \bibnamefont {Wu}}, \bibinfo {author} {\bibfnamefont {D.}~\bibnamefont {Riedel}}, \bibinfo {author} {\bibfnamefont {A.}~\bibnamefont {Ruskuc}}, \bibinfo {author} {\bibfnamefont {D.}~\bibnamefont {Zhong}}, \bibinfo {author} {\bibfnamefont {H.}~\bibnamefont {Kwon}},\ and\ \bibinfo {author} {\bibfnamefont {A.}~\bibnamefont {Faraon}},\ }\href {https://doi.org/10.1103/PhysRevApplied.20.044018} {\bibfield  {journal} {\bibinfo  {journal} {Physical Review Applied}\ }\textbf {\bibinfo {volume} {20}},\ \bibinfo {pages} {044018} (\bibinfo {year} {2023})}\BibitemShut {NoStop}%
\bibitem [{\citenamefont {Wolfowicz}\ \emph {et~al.}(2018)\citenamefont {Wolfowicz}, \citenamefont {Whiteley},\ and\ \citenamefont {Awschalom}}]{wolfowicz_electrometry_2018}%
  \BibitemOpen
  \bibfield  {author} {\bibinfo {author} {\bibfnamefont {G.}~\bibnamefont {Wolfowicz}}, \bibinfo {author} {\bibfnamefont {S.~J.}\ \bibnamefont {Whiteley}},\ and\ \bibinfo {author} {\bibfnamefont {D.~D.}\ \bibnamefont {Awschalom}},\ }\href {https://doi.org/10.1073/pnas.1806998115} {\bibfield  {journal} {\bibinfo  {journal} {Proceedings of the National Academy of Sciences}\ }\textbf {\bibinfo {volume} {115}},\ \bibinfo {pages} {7879} (\bibinfo {year} {2018})}\BibitemShut {NoStop}%
\bibitem [{\citenamefont {Niethammer}\ \emph {et~al.}(2016)\citenamefont {Niethammer}, \citenamefont {Widmann}, \citenamefont {Lee}, \citenamefont {Stenberg}, \citenamefont {Kordina}, \citenamefont {Ohshima}, \citenamefont {Son}, \citenamefont {Janzén},\ and\ \citenamefont {Wrachtrup}}]{niethammer_vector_2016}%
  \BibitemOpen
  \bibfield  {author} {\bibinfo {author} {\bibfnamefont {M.}~\bibnamefont {Niethammer}}, \bibinfo {author} {\bibfnamefont {M.}~\bibnamefont {Widmann}}, \bibinfo {author} {\bibfnamefont {S.-Y.}\ \bibnamefont {Lee}}, \bibinfo {author} {\bibfnamefont {P.}~\bibnamefont {Stenberg}}, \bibinfo {author} {\bibfnamefont {O.}~\bibnamefont {Kordina}}, \bibinfo {author} {\bibfnamefont {T.}~\bibnamefont {Ohshima}}, \bibinfo {author} {\bibfnamefont {N.~T.}\ \bibnamefont {Son}}, \bibinfo {author} {\bibfnamefont {E.}~\bibnamefont {Janzén}},\ and\ \bibinfo {author} {\bibfnamefont {J.}~\bibnamefont {Wrachtrup}},\ }\href {https://doi.org/10.1103/PhysRevApplied.6.034001} {\bibfield  {journal} {\bibinfo  {journal} {Physical Review Applied}\ }\textbf {\bibinfo {volume} {6}},\ \bibinfo {pages} {034001} (\bibinfo {year} {2016})}\BibitemShut {NoStop}%
\bibitem [{\citenamefont {Yan}\ \emph {et~al.}(2020)\citenamefont {Yan}, \citenamefont {Yi}, \citenamefont {Wang}, \citenamefont {Li}, \citenamefont {Yu}, \citenamefont {Zhang}, \citenamefont {Gali}, \citenamefont {Wang}, \citenamefont {Xu}, \citenamefont {Ou}, \citenamefont {Li},\ and\ \citenamefont {Guo}}]{yan_room-temperature_2020}%
  \BibitemOpen
  \bibfield  {author} {\bibinfo {author} {\bibfnamefont {F.-F.}\ \bibnamefont {Yan}}, \bibinfo {author} {\bibfnamefont {A.-L.}\ \bibnamefont {Yi}}, \bibinfo {author} {\bibfnamefont {J.-F.}\ \bibnamefont {Wang}}, \bibinfo {author} {\bibfnamefont {Q.}~\bibnamefont {Li}}, \bibinfo {author} {\bibfnamefont {P.}~\bibnamefont {Yu}}, \bibinfo {author} {\bibfnamefont {J.-X.}\ \bibnamefont {Zhang}}, \bibinfo {author} {\bibfnamefont {A.}~\bibnamefont {Gali}}, \bibinfo {author} {\bibfnamefont {Y.}~\bibnamefont {Wang}}, \bibinfo {author} {\bibfnamefont {J.-S.}\ \bibnamefont {Xu}}, \bibinfo {author} {\bibfnamefont {X.}~\bibnamefont {Ou}}, \bibinfo {author} {\bibfnamefont {C.-F.}\ \bibnamefont {Li}},\ and\ \bibinfo {author} {\bibfnamefont {G.-C.}\ \bibnamefont {Guo}},\ }\href {https://doi.org/10.1038/s41534-020-0270-8} {\bibfield  {journal} {\bibinfo  {journal} {npj Quantum Information}\ }\textbf {\bibinfo {volume} {6}},\ \bibinfo {pages} {38} (\bibinfo {year} {2020})}\BibitemShut {NoStop}%
\bibitem [{\citenamefont {Jiang}\ \emph {et~al.}(2023)\citenamefont {Jiang}, \citenamefont {Cai}, \citenamefont {Cernansky}, \citenamefont {Liu},\ and\ \citenamefont {Gao}}]{jiang_quantum_2023}%
  \BibitemOpen
  \bibfield  {author} {\bibinfo {author} {\bibfnamefont {Z.}~\bibnamefont {Jiang}}, \bibinfo {author} {\bibfnamefont {H.}~\bibnamefont {Cai}}, \bibinfo {author} {\bibfnamefont {R.}~\bibnamefont {Cernansky}}, \bibinfo {author} {\bibfnamefont {X.}~\bibnamefont {Liu}},\ and\ \bibinfo {author} {\bibfnamefont {W.}~\bibnamefont {Gao}},\ }\href {https://doi.org/10.1126/sciadv.adg2080} {\bibfield  {journal} {\bibinfo  {journal} {Science Advances}\ }\textbf {\bibinfo {volume} {9}},\ \bibinfo {pages} {eadg2080} (\bibinfo {year} {2023})}\BibitemShut {NoStop}%
\bibitem [{\citenamefont {Stern}\ \emph {et~al.}(2022)\citenamefont {Stern}, \citenamefont {Gu}, \citenamefont {Jarman}, \citenamefont {Eizagirre~Barker}, \citenamefont {Mendelson}, \citenamefont {Chugh}, \citenamefont {Schott}, \citenamefont {Tan}, \citenamefont {Sirringhaus}, \citenamefont {Aharonovich},\ and\ \citenamefont {Atatüre}}]{stern_room-temperature_2022}%
  \BibitemOpen
  \bibfield  {author} {\bibinfo {author} {\bibfnamefont {H.~L.}\ \bibnamefont {Stern}}, \bibinfo {author} {\bibfnamefont {Q.}~\bibnamefont {Gu}}, \bibinfo {author} {\bibfnamefont {J.}~\bibnamefont {Jarman}}, \bibinfo {author} {\bibfnamefont {S.}~\bibnamefont {Eizagirre~Barker}}, \bibinfo {author} {\bibfnamefont {N.}~\bibnamefont {Mendelson}}, \bibinfo {author} {\bibfnamefont {D.}~\bibnamefont {Chugh}}, \bibinfo {author} {\bibfnamefont {S.}~\bibnamefont {Schott}}, \bibinfo {author} {\bibfnamefont {H.~H.}\ \bibnamefont {Tan}}, \bibinfo {author} {\bibfnamefont {H.}~\bibnamefont {Sirringhaus}}, \bibinfo {author} {\bibfnamefont {I.}~\bibnamefont {Aharonovich}},\ and\ \bibinfo {author} {\bibfnamefont {M.}~\bibnamefont {Atatüre}},\ }\href {https://doi.org/10.1038/s41467-022-28169-z} {\bibfield  {journal} {\bibinfo  {journal} {Nature Communications}\ }\textbf {\bibinfo {volume} {13}},\ \bibinfo {pages} {618} (\bibinfo {year} {2022})}\BibitemShut {NoStop}%
\bibitem [{\citenamefont {Rizzato}\ \emph {et~al.}(2023)\citenamefont {Rizzato}, \citenamefont {Schalk}, \citenamefont {Mohr}, \citenamefont {Hermann}, \citenamefont {Leibold}, \citenamefont {Bruckmaier}, \citenamefont {Salvitti}, \citenamefont {Qian}, \citenamefont {Ji}, \citenamefont {Astakhov}, \citenamefont {Kentsch}, \citenamefont {Helm}, \citenamefont {Stier}, \citenamefont {Finley},\ and\ \citenamefont {Bucher}}]{rizzato_extending_2023}%
  \BibitemOpen
  \bibfield  {author} {\bibinfo {author} {\bibfnamefont {R.}~\bibnamefont {Rizzato}}, \bibinfo {author} {\bibfnamefont {M.}~\bibnamefont {Schalk}}, \bibinfo {author} {\bibfnamefont {S.}~\bibnamefont {Mohr}}, \bibinfo {author} {\bibfnamefont {J.~C.}\ \bibnamefont {Hermann}}, \bibinfo {author} {\bibfnamefont {J.~P.}\ \bibnamefont {Leibold}}, \bibinfo {author} {\bibfnamefont {F.}~\bibnamefont {Bruckmaier}}, \bibinfo {author} {\bibfnamefont {G.}~\bibnamefont {Salvitti}}, \bibinfo {author} {\bibfnamefont {C.}~\bibnamefont {Qian}}, \bibinfo {author} {\bibfnamefont {P.}~\bibnamefont {Ji}}, \bibinfo {author} {\bibfnamefont {G.~V.}\ \bibnamefont {Astakhov}}, \bibinfo {author} {\bibfnamefont {U.}~\bibnamefont {Kentsch}}, \bibinfo {author} {\bibfnamefont {M.}~\bibnamefont {Helm}}, \bibinfo {author} {\bibfnamefont {A.~V.}\ \bibnamefont {Stier}}, \bibinfo {author} {\bibfnamefont {J.~J.}\ \bibnamefont {Finley}},\ and\ \bibinfo {author} {\bibfnamefont {D.~B.}\ \bibnamefont {Bucher}},\ }\href
  {https://doi.org/10.1038/s41467-023-40473-w} {\bibfield  {journal} {\bibinfo  {journal} {Nature Communications}\ }\textbf {\bibinfo {volume} {14}},\ \bibinfo {pages} {5089} (\bibinfo {year} {2023})}\BibitemShut {NoStop}%
\bibitem [{\citenamefont {Zhou}\ \emph {et~al.}(2024)\citenamefont {Zhou}, \citenamefont {Lu}, \citenamefont {Chen}, \citenamefont {Huang}, \citenamefont {Yan}, \citenamefont {Al-matouq}, \citenamefont {Chang}, \citenamefont {Djugba}, \citenamefont {Jiang}, \citenamefont {Wang},\ and\ \citenamefont {Du}}]{zhou_sensing_2024}%
  \BibitemOpen
  \bibfield  {author} {\bibinfo {author} {\bibfnamefont {J.}~\bibnamefont {Zhou}}, \bibinfo {author} {\bibfnamefont {H.}~\bibnamefont {Lu}}, \bibinfo {author} {\bibfnamefont {D.}~\bibnamefont {Chen}}, \bibinfo {author} {\bibfnamefont {M.}~\bibnamefont {Huang}}, \bibinfo {author} {\bibfnamefont {G.~Q.}\ \bibnamefont {Yan}}, \bibinfo {author} {\bibfnamefont {F.}~\bibnamefont {Al-matouq}}, \bibinfo {author} {\bibfnamefont {J.}~\bibnamefont {Chang}}, \bibinfo {author} {\bibfnamefont {D.}~\bibnamefont {Djugba}}, \bibinfo {author} {\bibfnamefont {Z.}~\bibnamefont {Jiang}}, \bibinfo {author} {\bibfnamefont {H.}~\bibnamefont {Wang}},\ and\ \bibinfo {author} {\bibfnamefont {C.~R.}\ \bibnamefont {Du}},\ }\href {https://doi.org/10.1126/sciadv.adk8495} {\bibfield  {journal} {\bibinfo  {journal} {Science Advances}\ }\textbf {\bibinfo {volume} {10}},\ \bibinfo {pages} {eadk8495} (\bibinfo {year} {2024})}\BibitemShut {NoStop}%
\bibitem [{\citenamefont {Stern}\ \emph {et~al.}(2024)\citenamefont {Stern}, \citenamefont {M.~Gilardoni}, \citenamefont {Gu}, \citenamefont {Eizagirre~Barker}, \citenamefont {Powell}, \citenamefont {Deng}, \citenamefont {Fraser}, \citenamefont {Follet}, \citenamefont {Li}, \citenamefont {Ramsay}, \citenamefont {Tan}, \citenamefont {Aharonovich},\ and\ \citenamefont {Atatüre}}]{stern_quantum_2024}%
  \BibitemOpen
  \bibfield  {author} {\bibinfo {author} {\bibfnamefont {H.~L.}\ \bibnamefont {Stern}}, \bibinfo {author} {\bibfnamefont {C.}~\bibnamefont {M.~Gilardoni}}, \bibinfo {author} {\bibfnamefont {Q.}~\bibnamefont {Gu}}, \bibinfo {author} {\bibfnamefont {S.}~\bibnamefont {Eizagirre~Barker}}, \bibinfo {author} {\bibfnamefont {O.~F.~J.}\ \bibnamefont {Powell}}, \bibinfo {author} {\bibfnamefont {X.}~\bibnamefont {Deng}}, \bibinfo {author} {\bibfnamefont {S.~A.}\ \bibnamefont {Fraser}}, \bibinfo {author} {\bibfnamefont {L.}~\bibnamefont {Follet}}, \bibinfo {author} {\bibfnamefont {C.}~\bibnamefont {Li}}, \bibinfo {author} {\bibfnamefont {A.~J.}\ \bibnamefont {Ramsay}}, \bibinfo {author} {\bibfnamefont {H.~H.}\ \bibnamefont {Tan}}, \bibinfo {author} {\bibfnamefont {I.}~\bibnamefont {Aharonovich}},\ and\ \bibinfo {author} {\bibfnamefont {M.}~\bibnamefont {Atatüre}},\ }\href {https://doi.org/10.1038/s41563-024-01887-z} {\bibfield  {journal} {\bibinfo  {journal} {Nature Materials}\ }\textbf {\bibinfo {volume} {23}},\
  \bibinfo {pages} {1379} (\bibinfo {year} {2024})}\BibitemShut {NoStop}%
\bibitem [{\citenamefont {Miller}\ \emph {et~al.}(2020)\citenamefont {Miller}, \citenamefont {Bezinge}, \citenamefont {Gliddon}, \citenamefont {Huang}, \citenamefont {Dold}, \citenamefont {Gray}, \citenamefont {Heaney}, \citenamefont {Dobson}, \citenamefont {Nastouli}, \citenamefont {Morton},\ and\ \citenamefont {McKendry}}]{miller_spin-enhanced_2020}%
  \BibitemOpen
  \bibfield  {author} {\bibinfo {author} {\bibfnamefont {B.~S.}\ \bibnamefont {Miller}}, \bibinfo {author} {\bibfnamefont {L.}~\bibnamefont {Bezinge}}, \bibinfo {author} {\bibfnamefont {H.~D.}\ \bibnamefont {Gliddon}}, \bibinfo {author} {\bibfnamefont {D.}~\bibnamefont {Huang}}, \bibinfo {author} {\bibfnamefont {G.}~\bibnamefont {Dold}}, \bibinfo {author} {\bibfnamefont {E.~R.}\ \bibnamefont {Gray}}, \bibinfo {author} {\bibfnamefont {J.}~\bibnamefont {Heaney}}, \bibinfo {author} {\bibfnamefont {P.~J.}\ \bibnamefont {Dobson}}, \bibinfo {author} {\bibfnamefont {E.}~\bibnamefont {Nastouli}}, \bibinfo {author} {\bibfnamefont {J.~J.~L.}\ \bibnamefont {Morton}},\ and\ \bibinfo {author} {\bibfnamefont {R.~A.}\ \bibnamefont {McKendry}},\ }\href {https://doi.org/10.1038/s41586-020-2917-1} {\bibfield  {journal} {\bibinfo  {journal} {Nature}\ }\textbf {\bibinfo {volume} {587}},\ \bibinfo {pages} {588} (\bibinfo {year} {2020})}\BibitemShut {NoStop}%
\bibitem [{\citenamefont {Degen}\ \emph {et~al.}(2017)\citenamefont {Degen}, \citenamefont {Reinhard},\ and\ \citenamefont {Cappellaro}}]{degen_quantum_2017}%
  \BibitemOpen
  \bibfield  {author} {\bibinfo {author} {\bibfnamefont {C.}~\bibnamefont {Degen}}, \bibinfo {author} {\bibfnamefont {F.}~\bibnamefont {Reinhard}},\ and\ \bibinfo {author} {\bibfnamefont {P.}~\bibnamefont {Cappellaro}},\ }\href {https://doi.org/10.1103/RevModPhys.89.035002} {\bibfield  {journal} {\bibinfo  {journal} {Reviews of Modern Physics}\ }\textbf {\bibinfo {volume} {89}},\ \bibinfo {pages} {035002} (\bibinfo {year} {2017})},\ \bibinfo {note} {publisher: American Physical Society}\BibitemShut {NoStop}%
\bibitem [{\citenamefont {Bradac}\ \emph {et~al.}(2019)\citenamefont {Bradac}, \citenamefont {Gao}, \citenamefont {Forneris}, \citenamefont {Trusheim},\ and\ \citenamefont {Aharonovich}}]{bradac_quantum_2019}%
  \BibitemOpen
  \bibfield  {author} {\bibinfo {author} {\bibfnamefont {C.}~\bibnamefont {Bradac}}, \bibinfo {author} {\bibfnamefont {W.}~\bibnamefont {Gao}}, \bibinfo {author} {\bibfnamefont {J.}~\bibnamefont {Forneris}}, \bibinfo {author} {\bibfnamefont {M.~E.}\ \bibnamefont {Trusheim}},\ and\ \bibinfo {author} {\bibfnamefont {I.}~\bibnamefont {Aharonovich}},\ }\href {https://doi.org/10.1038/s41467-019-13332-w} {\bibfield  {journal} {\bibinfo  {journal} {Nature Communications}\ }\textbf {\bibinfo {volume} {10}},\ \bibinfo {pages} {5625} (\bibinfo {year} {2019})},\ \bibinfo {note} {publisher: Nature Publishing Group}\BibitemShut {NoStop}%
\bibitem [{\citenamefont {Hadden}\ \emph {et~al.}(2010)\citenamefont {Hadden}, \citenamefont {Harrison}, \citenamefont {Stanley-Clarke}, \citenamefont {Marseglia}, \citenamefont {Ho}, \citenamefont {Patton}, \citenamefont {O'Brien},\ and\ \citenamefont {Rarity}}]{hadden_strongly_2010}%
  \BibitemOpen
  \bibfield  {author} {\bibinfo {author} {\bibfnamefont {J.~P.}\ \bibnamefont {Hadden}}, \bibinfo {author} {\bibfnamefont {J.~P.}\ \bibnamefont {Harrison}}, \bibinfo {author} {\bibfnamefont {A.~C.}\ \bibnamefont {Stanley-Clarke}}, \bibinfo {author} {\bibfnamefont {L.}~\bibnamefont {Marseglia}}, \bibinfo {author} {\bibfnamefont {Y.-L.~D.}\ \bibnamefont {Ho}}, \bibinfo {author} {\bibfnamefont {B.~R.}\ \bibnamefont {Patton}}, \bibinfo {author} {\bibfnamefont {J.~L.}\ \bibnamefont {O'Brien}},\ and\ \bibinfo {author} {\bibfnamefont {J.~G.}\ \bibnamefont {Rarity}},\ }\href {https://doi.org/10.1063/1.3519847} {\bibfield  {journal} {\bibinfo  {journal} {Applied Physics Letters}\ }\textbf {\bibinfo {volume} {97}},\ \bibinfo {pages} {241901} (\bibinfo {year} {2010})}\BibitemShut {NoStop}%
\bibitem [{\citenamefont {Rogers}\ \emph {et~al.}(2014)\citenamefont {Rogers}, \citenamefont {Jahnke}, \citenamefont {Teraji}, \citenamefont {Marseglia}, \citenamefont {Müller}, \citenamefont {Naydenov}, \citenamefont {Schauffert}, \citenamefont {Kranz}, \citenamefont {Isoya}, \citenamefont {McGuinness},\ and\ \citenamefont {Jelezko}}]{rogers_multiple_2014}%
  \BibitemOpen
  \bibfield  {author} {\bibinfo {author} {\bibfnamefont {L.~J.}\ \bibnamefont {Rogers}}, \bibinfo {author} {\bibfnamefont {K.~D.}\ \bibnamefont {Jahnke}}, \bibinfo {author} {\bibfnamefont {T.}~\bibnamefont {Teraji}}, \bibinfo {author} {\bibfnamefont {L.}~\bibnamefont {Marseglia}}, \bibinfo {author} {\bibfnamefont {C.}~\bibnamefont {Müller}}, \bibinfo {author} {\bibfnamefont {B.}~\bibnamefont {Naydenov}}, \bibinfo {author} {\bibfnamefont {H.}~\bibnamefont {Schauffert}}, \bibinfo {author} {\bibfnamefont {C.}~\bibnamefont {Kranz}}, \bibinfo {author} {\bibfnamefont {J.}~\bibnamefont {Isoya}}, \bibinfo {author} {\bibfnamefont {L.~P.}\ \bibnamefont {McGuinness}},\ and\ \bibinfo {author} {\bibfnamefont {F.}~\bibnamefont {Jelezko}},\ }\href {https://doi.org/10.1038/ncomms5739} {\bibfield  {journal} {\bibinfo  {journal} {Nature Communications}\ }\textbf {\bibinfo {volume} {5}},\ \bibinfo {pages} {4739} (\bibinfo {year} {2014})}\BibitemShut {NoStop}%
\bibitem [{\citenamefont {Bekker}\ \emph {et~al.}(2023)\citenamefont {Bekker}, \citenamefont {Arshad}, \citenamefont {Cilibrizzi}, \citenamefont {Nikolatos}, \citenamefont {Lomax}, \citenamefont {Wood}, \citenamefont {Cheung}, \citenamefont {Knolle}, \citenamefont {Ross}, \citenamefont {Gerardot},\ and\ \citenamefont {Bonato}}]{bekker_scalable_2023}%
  \BibitemOpen
  \bibfield  {author} {\bibinfo {author} {\bibfnamefont {C.}~\bibnamefont {Bekker}}, \bibinfo {author} {\bibfnamefont {M.~J.}\ \bibnamefont {Arshad}}, \bibinfo {author} {\bibfnamefont {P.}~\bibnamefont {Cilibrizzi}}, \bibinfo {author} {\bibfnamefont {C.}~\bibnamefont {Nikolatos}}, \bibinfo {author} {\bibfnamefont {P.}~\bibnamefont {Lomax}}, \bibinfo {author} {\bibfnamefont {G.~S.}\ \bibnamefont {Wood}}, \bibinfo {author} {\bibfnamefont {R.}~\bibnamefont {Cheung}}, \bibinfo {author} {\bibfnamefont {W.}~\bibnamefont {Knolle}}, \bibinfo {author} {\bibfnamefont {N.}~\bibnamefont {Ross}}, \bibinfo {author} {\bibfnamefont {B.}~\bibnamefont {Gerardot}},\ and\ \bibinfo {author} {\bibfnamefont {C.}~\bibnamefont {Bonato}},\ }\href {https://doi.org/10.1063/5.0144684} {\bibfield  {journal} {\bibinfo  {journal} {Applied Physics Letters}\ }\textbf {\bibinfo {volume} {122}},\ \bibinfo {pages} {173507} (\bibinfo {year} {2023})}\BibitemShut {NoStop}%
\bibitem [{\citenamefont {Wan}\ \emph {et~al.}(2018)\citenamefont {Wan}, \citenamefont {Shields}, \citenamefont {Kim}, \citenamefont {Mouradian}, \citenamefont {Lienhard}, \citenamefont {Walsh}, \citenamefont {Bakhru}, \citenamefont {Schröder},\ and\ \citenamefont {Englund}}]{wan_efficient_2018}%
  \BibitemOpen
  \bibfield  {author} {\bibinfo {author} {\bibfnamefont {N.~H.}\ \bibnamefont {Wan}}, \bibinfo {author} {\bibfnamefont {B.~J.}\ \bibnamefont {Shields}}, \bibinfo {author} {\bibfnamefont {D.}~\bibnamefont {Kim}}, \bibinfo {author} {\bibfnamefont {S.}~\bibnamefont {Mouradian}}, \bibinfo {author} {\bibfnamefont {B.}~\bibnamefont {Lienhard}}, \bibinfo {author} {\bibfnamefont {M.}~\bibnamefont {Walsh}}, \bibinfo {author} {\bibfnamefont {H.}~\bibnamefont {Bakhru}}, \bibinfo {author} {\bibfnamefont {T.}~\bibnamefont {Schröder}},\ and\ \bibinfo {author} {\bibfnamefont {D.}~\bibnamefont {Englund}},\ }\href {https://doi.org/10.1021/acs.nanolett.7b04684} {\bibfield  {journal} {\bibinfo  {journal} {Nano Letters}\ }\textbf {\bibinfo {volume} {18}},\ \bibinfo {pages} {2787} (\bibinfo {year} {2018})}\BibitemShut {NoStop}%
\bibitem [{\citenamefont {Sardi}\ \emph {et~al.}(2020)\citenamefont {Sardi}, \citenamefont {Kornher}, \citenamefont {Widmann}, \citenamefont {Kolesov}, \citenamefont {Schiller}, \citenamefont {Reindl}, \citenamefont {Hagel},\ and\ \citenamefont {Wrachtrup}}]{sardi_scalable_2020}%
  \BibitemOpen
  \bibfield  {author} {\bibinfo {author} {\bibfnamefont {F.}~\bibnamefont {Sardi}}, \bibinfo {author} {\bibfnamefont {T.}~\bibnamefont {Kornher}}, \bibinfo {author} {\bibfnamefont {M.}~\bibnamefont {Widmann}}, \bibinfo {author} {\bibfnamefont {R.}~\bibnamefont {Kolesov}}, \bibinfo {author} {\bibfnamefont {F.}~\bibnamefont {Schiller}}, \bibinfo {author} {\bibfnamefont {T.}~\bibnamefont {Reindl}}, \bibinfo {author} {\bibfnamefont {M.}~\bibnamefont {Hagel}},\ and\ \bibinfo {author} {\bibfnamefont {J.}~\bibnamefont {Wrachtrup}},\ }\href {https://doi.org/10.1063/5.0011366} {\bibfield  {journal} {\bibinfo  {journal} {Applied Physics Letters}\ }\textbf {\bibinfo {volume} {117}},\ \bibinfo {pages} {022105} (\bibinfo {year} {2020})}\BibitemShut {NoStop}%
\bibitem [{\citenamefont {Babin}\ \emph {et~al.}(2022)\citenamefont {Babin}, \citenamefont {Stöhr}, \citenamefont {Morioka}, \citenamefont {Linkewitz}, \citenamefont {Steidl}, \citenamefont {Wörnle}, \citenamefont {Liu}, \citenamefont {Hesselmeier}, \citenamefont {Vorobyov}, \citenamefont {Denisenko}, \citenamefont {Hentschel}, \citenamefont {Gobert}, \citenamefont {Berwian}, \citenamefont {Astakhov}, \citenamefont {Knolle}, \citenamefont {Majety}, \citenamefont {Saha}, \citenamefont {Radulaski}, \citenamefont {Son}, \citenamefont {Ul-Hassan}, \citenamefont {Kaiser},\ and\ \citenamefont {Wrachtrup}}]{babin_fabrication_2022}%
  \BibitemOpen
  \bibfield  {author} {\bibinfo {author} {\bibfnamefont {C.}~\bibnamefont {Babin}}, \bibinfo {author} {\bibfnamefont {R.}~\bibnamefont {Stöhr}}, \bibinfo {author} {\bibfnamefont {N.}~\bibnamefont {Morioka}}, \bibinfo {author} {\bibfnamefont {T.}~\bibnamefont {Linkewitz}}, \bibinfo {author} {\bibfnamefont {T.}~\bibnamefont {Steidl}}, \bibinfo {author} {\bibfnamefont {R.}~\bibnamefont {Wörnle}}, \bibinfo {author} {\bibfnamefont {D.}~\bibnamefont {Liu}}, \bibinfo {author} {\bibfnamefont {E.}~\bibnamefont {Hesselmeier}}, \bibinfo {author} {\bibfnamefont {V.}~\bibnamefont {Vorobyov}}, \bibinfo {author} {\bibfnamefont {A.}~\bibnamefont {Denisenko}}, \bibinfo {author} {\bibfnamefont {M.}~\bibnamefont {Hentschel}}, \bibinfo {author} {\bibfnamefont {C.}~\bibnamefont {Gobert}}, \bibinfo {author} {\bibfnamefont {P.}~\bibnamefont {Berwian}}, \bibinfo {author} {\bibfnamefont {G.~V.}\ \bibnamefont {Astakhov}}, \bibinfo {author} {\bibfnamefont {W.}~\bibnamefont {Knolle}}, \bibinfo {author} {\bibfnamefont {S.}~\bibnamefont
  {Majety}}, \bibinfo {author} {\bibfnamefont {P.}~\bibnamefont {Saha}}, \bibinfo {author} {\bibfnamefont {M.}~\bibnamefont {Radulaski}}, \bibinfo {author} {\bibfnamefont {N.~T.}\ \bibnamefont {Son}}, \bibinfo {author} {\bibfnamefont {J.}~\bibnamefont {Ul-Hassan}}, \bibinfo {author} {\bibfnamefont {F.}~\bibnamefont {Kaiser}},\ and\ \bibinfo {author} {\bibfnamefont {J.}~\bibnamefont {Wrachtrup}},\ }\href {https://doi.org/10.1038/s41563-021-01148-3} {\bibfield  {journal} {\bibinfo  {journal} {Nature Materials}\ }\textbf {\bibinfo {volume} {21}},\ \bibinfo {pages} {67} (\bibinfo {year} {2022})}\BibitemShut {NoStop}%
\bibitem [{\citenamefont {Burek}\ \emph {et~al.}(2017)\citenamefont {Burek}, \citenamefont {Meuwly}, \citenamefont {Evans}, \citenamefont {Bhaskar}, \citenamefont {Sipahigil}, \citenamefont {Meesala}, \citenamefont {Machielse}, \citenamefont {Sukachev}, \citenamefont {Nguyen}, \citenamefont {Pacheco}, \citenamefont {Bielejec}, \citenamefont {Lukin},\ and\ \citenamefont {Lončar}}]{burek_fiber-coupled_2017}%
  \BibitemOpen
  \bibfield  {author} {\bibinfo {author} {\bibfnamefont {M.~J.}\ \bibnamefont {Burek}}, \bibinfo {author} {\bibfnamefont {C.}~\bibnamefont {Meuwly}}, \bibinfo {author} {\bibfnamefont {R.~E.}\ \bibnamefont {Evans}}, \bibinfo {author} {\bibfnamefont {M.~K.}\ \bibnamefont {Bhaskar}}, \bibinfo {author} {\bibfnamefont {A.}~\bibnamefont {Sipahigil}}, \bibinfo {author} {\bibfnamefont {S.}~\bibnamefont {Meesala}}, \bibinfo {author} {\bibfnamefont {B.}~\bibnamefont {Machielse}}, \bibinfo {author} {\bibfnamefont {D.~D.}\ \bibnamefont {Sukachev}}, \bibinfo {author} {\bibfnamefont {C.~T.}\ \bibnamefont {Nguyen}}, \bibinfo {author} {\bibfnamefont {J.~L.}\ \bibnamefont {Pacheco}}, \bibinfo {author} {\bibfnamefont {E.}~\bibnamefont {Bielejec}}, \bibinfo {author} {\bibfnamefont {M.~D.}\ \bibnamefont {Lukin}},\ and\ \bibinfo {author} {\bibfnamefont {M.}~\bibnamefont {Lončar}},\ }\href {https://doi.org/10.1103/PhysRevApplied.8.024026} {\bibfield  {journal} {\bibinfo  {journal} {Physical Review Applied}\ }\textbf {\bibinfo
  {volume} {8}},\ \bibinfo {pages} {024026} (\bibinfo {year} {2017})},\ \bibinfo {note} {publisher: American Physical Society}\BibitemShut {NoStop}%
\bibitem [{\citenamefont {Sotillo}\ \emph {et~al.}(2016)\citenamefont {Sotillo}, \citenamefont {Bharadwaj}, \citenamefont {Hadden}, \citenamefont {Sakakura}, \citenamefont {Chiappini}, \citenamefont {Fernandez}, \citenamefont {Longhi}, \citenamefont {Jedrkiewicz}, \citenamefont {Shimotsuma}, \citenamefont {Criante}, \citenamefont {Osellame}, \citenamefont {Galzerano}, \citenamefont {Ferrari}, \citenamefont {Miura}, \citenamefont {Ramponi}, \citenamefont {Barclay},\ and\ \citenamefont {Eaton}}]{sotillo_diamond_2016}%
  \BibitemOpen
  \bibfield  {author} {\bibinfo {author} {\bibfnamefont {B.}~\bibnamefont {Sotillo}}, \bibinfo {author} {\bibfnamefont {V.}~\bibnamefont {Bharadwaj}}, \bibinfo {author} {\bibfnamefont {J.~P.}\ \bibnamefont {Hadden}}, \bibinfo {author} {\bibfnamefont {M.}~\bibnamefont {Sakakura}}, \bibinfo {author} {\bibfnamefont {A.}~\bibnamefont {Chiappini}}, \bibinfo {author} {\bibfnamefont {T.~T.}\ \bibnamefont {Fernandez}}, \bibinfo {author} {\bibfnamefont {S.}~\bibnamefont {Longhi}}, \bibinfo {author} {\bibfnamefont {O.}~\bibnamefont {Jedrkiewicz}}, \bibinfo {author} {\bibfnamefont {Y.}~\bibnamefont {Shimotsuma}}, \bibinfo {author} {\bibfnamefont {L.}~\bibnamefont {Criante}}, \bibinfo {author} {\bibfnamefont {R.}~\bibnamefont {Osellame}}, \bibinfo {author} {\bibfnamefont {G.}~\bibnamefont {Galzerano}}, \bibinfo {author} {\bibfnamefont {M.}~\bibnamefont {Ferrari}}, \bibinfo {author} {\bibfnamefont {K.}~\bibnamefont {Miura}}, \bibinfo {author} {\bibfnamefont {R.}~\bibnamefont {Ramponi}}, \bibinfo {author} {\bibfnamefont
  {P.~E.}\ \bibnamefont {Barclay}},\ and\ \bibinfo {author} {\bibfnamefont {S.~M.}\ \bibnamefont {Eaton}},\ }\href {https://doi.org/10.1038/srep35566} {\bibfield  {journal} {\bibinfo  {journal} {Scientific Reports}\ }\textbf {\bibinfo {volume} {6}},\ \bibinfo {pages} {35566} (\bibinfo {year} {2016})},\ \bibinfo {note} {publisher: Nature Publishing Group}\BibitemShut {NoStop}%
\bibitem [{\citenamefont {Tiecke}\ \emph {et~al.}(2015)\citenamefont {Tiecke}, \citenamefont {Nayak}, \citenamefont {Thompson}, \citenamefont {Peyronel}, \citenamefont {Leon}, \citenamefont {Vuletić},\ and\ \citenamefont {Lukin}}]{tiecke_efficient_2015}%
  \BibitemOpen
  \bibfield  {author} {\bibinfo {author} {\bibfnamefont {T.~G.}\ \bibnamefont {Tiecke}}, \bibinfo {author} {\bibfnamefont {K.~P.}\ \bibnamefont {Nayak}}, \bibinfo {author} {\bibfnamefont {J.~D.}\ \bibnamefont {Thompson}}, \bibinfo {author} {\bibfnamefont {T.}~\bibnamefont {Peyronel}}, \bibinfo {author} {\bibfnamefont {N.~P.~d.}\ \bibnamefont {Leon}}, \bibinfo {author} {\bibfnamefont {V.}~\bibnamefont {Vuletić}},\ and\ \bibinfo {author} {\bibfnamefont {M.~D.}\ \bibnamefont {Lukin}},\ }\href {https://doi.org/10.1364/OPTICA.2.000070} {\bibfield  {journal} {\bibinfo  {journal} {Optica}\ }\textbf {\bibinfo {volume} {2}},\ \bibinfo {pages} {70} (\bibinfo {year} {2015})},\ \bibinfo {note} {publisher: Optica Publishing Group}\BibitemShut {NoStop}%
\bibitem [{\citenamefont {Guo}\ \emph {et~al.}(2024)\citenamefont {Guo}, \citenamefont {Hadden}, \citenamefont {Gorrini}, \citenamefont {Coccia}, \citenamefont {Bharadwaj}, \citenamefont {Kavatamane}, \citenamefont {Alam}, \citenamefont {Ramponi}, \citenamefont {Barclay}, \citenamefont {Chiappini}, \citenamefont {Ferrari}, \citenamefont {Kubanek}, \citenamefont {Bifone}, \citenamefont {Eaton},\ and\ \citenamefont {Bennett}}]{guo_laser-written_2024}%
  \BibitemOpen
  \bibfield  {author} {\bibinfo {author} {\bibfnamefont {Y.}~\bibnamefont {Guo}}, \bibinfo {author} {\bibfnamefont {J.~P.}\ \bibnamefont {Hadden}}, \bibinfo {author} {\bibfnamefont {F.}~\bibnamefont {Gorrini}}, \bibinfo {author} {\bibfnamefont {G.}~\bibnamefont {Coccia}}, \bibinfo {author} {\bibfnamefont {V.}~\bibnamefont {Bharadwaj}}, \bibinfo {author} {\bibfnamefont {V.~K.}\ \bibnamefont {Kavatamane}}, \bibinfo {author} {\bibfnamefont {M.~S.}\ \bibnamefont {Alam}}, \bibinfo {author} {\bibfnamefont {R.}~\bibnamefont {Ramponi}}, \bibinfo {author} {\bibfnamefont {P.~E.}\ \bibnamefont {Barclay}}, \bibinfo {author} {\bibfnamefont {A.}~\bibnamefont {Chiappini}}, \bibinfo {author} {\bibfnamefont {M.}~\bibnamefont {Ferrari}}, \bibinfo {author} {\bibfnamefont {A.}~\bibnamefont {Kubanek}}, \bibinfo {author} {\bibfnamefont {A.}~\bibnamefont {Bifone}}, \bibinfo {author} {\bibfnamefont {S.~M.}\ \bibnamefont {Eaton}},\ and\ \bibinfo {author} {\bibfnamefont {A.~J.}\ \bibnamefont {Bennett}},\ }\href
  {https://doi.org/10.1063/5.0209294} {\bibfield  {journal} {\bibinfo  {journal} {APL Photonics}\ }\textbf {\bibinfo {volume} {9}},\ \bibinfo {pages} {076103} (\bibinfo {year} {2024})}\BibitemShut {NoStop}%
\bibitem [{\citenamefont {Krumrein}\ \emph {et~al.}(2024)\citenamefont {Krumrein}, \citenamefont {Nold}, \citenamefont {Davidson-Marquis}, \citenamefont {Bouamra}, \citenamefont {Niechziol}, \citenamefont {Steidl}, \citenamefont {Peng}, \citenamefont {Körber}, \citenamefont {Stöhr}, \citenamefont {Gross}, \citenamefont {Smet}, \citenamefont {Ul-Hassan}, \citenamefont {Udvarhelyi}, \citenamefont {Gali}, \citenamefont {Kaiser},\ and\ \citenamefont {Wrachtrup}}]{krumrein_precise_2024}%
  \BibitemOpen
  \bibfield  {author} {\bibinfo {author} {\bibfnamefont {M.}~\bibnamefont {Krumrein}}, \bibinfo {author} {\bibfnamefont {R.}~\bibnamefont {Nold}}, \bibinfo {author} {\bibfnamefont {F.}~\bibnamefont {Davidson-Marquis}}, \bibinfo {author} {\bibfnamefont {A.}~\bibnamefont {Bouamra}}, \bibinfo {author} {\bibfnamefont {L.}~\bibnamefont {Niechziol}}, \bibinfo {author} {\bibfnamefont {T.}~\bibnamefont {Steidl}}, \bibinfo {author} {\bibfnamefont {R.}~\bibnamefont {Peng}}, \bibinfo {author} {\bibfnamefont {J.}~\bibnamefont {Körber}}, \bibinfo {author} {\bibfnamefont {R.}~\bibnamefont {Stöhr}}, \bibinfo {author} {\bibfnamefont {N.}~\bibnamefont {Gross}}, \bibinfo {author} {\bibfnamefont {J.~H.}\ \bibnamefont {Smet}}, \bibinfo {author} {\bibfnamefont {J.}~\bibnamefont {Ul-Hassan}}, \bibinfo {author} {\bibfnamefont {P.}~\bibnamefont {Udvarhelyi}}, \bibinfo {author} {\bibfnamefont {A.}~\bibnamefont {Gali}}, \bibinfo {author} {\bibfnamefont {F.}~\bibnamefont {Kaiser}},\ and\ \bibinfo {author} {\bibfnamefont
  {J.}~\bibnamefont {Wrachtrup}},\ }\href {https://doi.org/10.1021/acsphotonics.4c00538} {\bibfield  {journal} {\bibinfo  {journal} {ACS Photonics}\ }\textbf {\bibinfo {volume} {11}},\ \bibinfo {pages} {2160} (\bibinfo {year} {2024})},\ \bibinfo {note} {publisher: American Chemical Society}\BibitemShut {NoStop}%
\bibitem [{\citenamefont {Rugar}\ \emph {et~al.}(2019)\citenamefont {Rugar}, \citenamefont {Dory}, \citenamefont {Sun},\ and\ \citenamefont {Vučković}}]{rugar_characterization_2019}%
  \BibitemOpen
  \bibfield  {author} {\bibinfo {author} {\bibfnamefont {A.~E.}\ \bibnamefont {Rugar}}, \bibinfo {author} {\bibfnamefont {C.}~\bibnamefont {Dory}}, \bibinfo {author} {\bibfnamefont {S.}~\bibnamefont {Sun}},\ and\ \bibinfo {author} {\bibfnamefont {J.}~\bibnamefont {Vučković}},\ }\href {https://doi.org/10.1103/PhysRevB.99.205417} {\bibfield  {journal} {\bibinfo  {journal} {Physical Review B}\ }\textbf {\bibinfo {volume} {99}},\ \bibinfo {pages} {205417} (\bibinfo {year} {2019})},\ \bibinfo {note} {publisher: American Physical Society}\BibitemShut {NoStop}%
\bibitem [{\citenamefont {Radulaski}\ \emph {et~al.}(2017)\citenamefont {Radulaski}, \citenamefont {Widmann}, \citenamefont {Niethammer}, \citenamefont {Zhang}, \citenamefont {Lee}, \citenamefont {Rendler}, \citenamefont {Lagoudakis}, \citenamefont {Son}, \citenamefont {Janzén}, \citenamefont {Ohshima}, \citenamefont {Wrachtrup},\ and\ \citenamefont {Vučković}}]{radulaski_scalable_2017}%
  \BibitemOpen
  \bibfield  {author} {\bibinfo {author} {\bibfnamefont {M.}~\bibnamefont {Radulaski}}, \bibinfo {author} {\bibfnamefont {M.}~\bibnamefont {Widmann}}, \bibinfo {author} {\bibfnamefont {M.}~\bibnamefont {Niethammer}}, \bibinfo {author} {\bibfnamefont {J.~L.}\ \bibnamefont {Zhang}}, \bibinfo {author} {\bibfnamefont {S.-Y.}\ \bibnamefont {Lee}}, \bibinfo {author} {\bibfnamefont {T.}~\bibnamefont {Rendler}}, \bibinfo {author} {\bibfnamefont {K.~G.}\ \bibnamefont {Lagoudakis}}, \bibinfo {author} {\bibfnamefont {N.~T.}\ \bibnamefont {Son}}, \bibinfo {author} {\bibfnamefont {E.}~\bibnamefont {Janzén}}, \bibinfo {author} {\bibfnamefont {T.}~\bibnamefont {Ohshima}}, \bibinfo {author} {\bibfnamefont {J.}~\bibnamefont {Wrachtrup}},\ and\ \bibinfo {author} {\bibfnamefont {J.}~\bibnamefont {Vučković}},\ }\href {https://doi.org/10.1021/acs.nanolett.6b05102} {\bibfield  {journal} {\bibinfo  {journal} {Nano Letters}\ }\textbf {\bibinfo {volume} {17}},\ \bibinfo {pages} {1782} (\bibinfo {year} {2017})},\ \bibinfo {note}
  {publisher: American Chemical Society}\BibitemShut {NoStop}%
\bibitem [{\citenamefont {Losero}\ \emph {et~al.}(2023)\citenamefont {Losero}, \citenamefont {Jagannath}, \citenamefont {Pezzoli}, \citenamefont {Goblot}, \citenamefont {Babashah}, \citenamefont {Lashuel}, \citenamefont {Galland},\ and\ \citenamefont {Quack}}]{losero_neuronal_2023}%
  \BibitemOpen
  \bibfield  {author} {\bibinfo {author} {\bibfnamefont {E.}~\bibnamefont {Losero}}, \bibinfo {author} {\bibfnamefont {S.}~\bibnamefont {Jagannath}}, \bibinfo {author} {\bibfnamefont {M.}~\bibnamefont {Pezzoli}}, \bibinfo {author} {\bibfnamefont {V.}~\bibnamefont {Goblot}}, \bibinfo {author} {\bibfnamefont {H.}~\bibnamefont {Babashah}}, \bibinfo {author} {\bibfnamefont {H.~A.}\ \bibnamefont {Lashuel}}, \bibinfo {author} {\bibfnamefont {C.}~\bibnamefont {Galland}},\ and\ \bibinfo {author} {\bibfnamefont {N.}~\bibnamefont {Quack}},\ }\href {https://doi.org/10.1038/s41598-023-32235-x} {\bibfield  {journal} {\bibinfo  {journal} {Scientific Reports}\ }\textbf {\bibinfo {volume} {13}},\ \bibinfo {pages} {5909} (\bibinfo {year} {2023})},\ \bibinfo {note} {publisher: Nature Publishing Group}\BibitemShut {NoStop}%
\bibitem [{\citenamefont {Orphal-Kobin}\ \emph {et~al.}(2023)\citenamefont {Orphal-Kobin}, \citenamefont {Unterguggenberger}, \citenamefont {Pregnolato}, \citenamefont {Kemf}, \citenamefont {Matalla}, \citenamefont {Unger}, \citenamefont {Ostermay}, \citenamefont {Pieplow},\ and\ \citenamefont {Schröder}}]{orphal-kobin_optically_2023}%
  \BibitemOpen
  \bibfield  {author} {\bibinfo {author} {\bibfnamefont {L.}~\bibnamefont {Orphal-Kobin}}, \bibinfo {author} {\bibfnamefont {K.}~\bibnamefont {Unterguggenberger}}, \bibinfo {author} {\bibfnamefont {T.}~\bibnamefont {Pregnolato}}, \bibinfo {author} {\bibfnamefont {N.}~\bibnamefont {Kemf}}, \bibinfo {author} {\bibfnamefont {M.}~\bibnamefont {Matalla}}, \bibinfo {author} {\bibfnamefont {R.-S.}\ \bibnamefont {Unger}}, \bibinfo {author} {\bibfnamefont {I.}~\bibnamefont {Ostermay}}, \bibinfo {author} {\bibfnamefont {G.}~\bibnamefont {Pieplow}},\ and\ \bibinfo {author} {\bibfnamefont {T.}~\bibnamefont {Schröder}},\ }\href {https://doi.org/10.1103/PhysRevX.13.011042} {\bibfield  {journal} {\bibinfo  {journal} {Physical Review X}\ }\textbf {\bibinfo {volume} {13}},\ \bibinfo {pages} {011042} (\bibinfo {year} {2023})},\ \bibinfo {note} {publisher: American Physical Society}\BibitemShut {NoStop}%
\bibitem [{\citenamefont {Hedrich}\ \emph {et~al.}(2020)\citenamefont {Hedrich}, \citenamefont {Rohner}, \citenamefont {Batzer}, \citenamefont {Maletinsky},\ and\ \citenamefont {Shields}}]{hedrich_parabolic_2020}%
  \BibitemOpen
  \bibfield  {author} {\bibinfo {author} {\bibfnamefont {N.}~\bibnamefont {Hedrich}}, \bibinfo {author} {\bibfnamefont {D.}~\bibnamefont {Rohner}}, \bibinfo {author} {\bibfnamefont {M.}~\bibnamefont {Batzer}}, \bibinfo {author} {\bibfnamefont {P.}~\bibnamefont {Maletinsky}},\ and\ \bibinfo {author} {\bibfnamefont {B.~J.}\ \bibnamefont {Shields}},\ }\href {https://doi.org/10.1103/PhysRevApplied.14.064007} {\bibfield  {journal} {\bibinfo  {journal} {Physical Review Applied}\ }\textbf {\bibinfo {volume} {14}},\ \bibinfo {pages} {064007} (\bibinfo {year} {2020})},\ \bibinfo {note} {publisher: American Physical Society}\BibitemShut {NoStop}%
\bibitem [{\citenamefont {Addhya}\ \emph {et~al.}(2024)\citenamefont {Addhya}, \citenamefont {Tyne}, \citenamefont {Guo}, \citenamefont {Hammock}, \citenamefont {Li}, \citenamefont {Leung}, \citenamefont {DeVault}, \citenamefont {Awschalom}, \citenamefont {Delegan}, \citenamefont {Heremans},\ and\ \citenamefont {High}}]{addhya_photonic-cavity-enhanced_2024}%
  \BibitemOpen
  \bibfield  {author} {\bibinfo {author} {\bibfnamefont {A.}~\bibnamefont {Addhya}}, \bibinfo {author} {\bibfnamefont {V.}~\bibnamefont {Tyne}}, \bibinfo {author} {\bibfnamefont {X.}~\bibnamefont {Guo}}, \bibinfo {author} {\bibfnamefont {I.~N.}\ \bibnamefont {Hammock}}, \bibinfo {author} {\bibfnamefont {Z.}~\bibnamefont {Li}}, \bibinfo {author} {\bibfnamefont {M.}~\bibnamefont {Leung}}, \bibinfo {author} {\bibfnamefont {C.~T.}\ \bibnamefont {DeVault}}, \bibinfo {author} {\bibfnamefont {D.~D.}\ \bibnamefont {Awschalom}}, \bibinfo {author} {\bibfnamefont {N.}~\bibnamefont {Delegan}}, \bibinfo {author} {\bibfnamefont {F.~J.}\ \bibnamefont {Heremans}},\ and\ \bibinfo {author} {\bibfnamefont {A.~A.}\ \bibnamefont {High}},\ }\href {https://doi.org/10.1021/acs.nanolett.4c02639} {\bibfield  {journal} {\bibinfo  {journal} {Nano Letters}\ }\textbf {\bibinfo {volume} {24}},\ \bibinfo {pages} {11224} (\bibinfo {year} {2024})}\BibitemShut {NoStop}%
\bibitem [{\citenamefont {Quan}\ \emph {et~al.}(2010)\citenamefont {Quan}, \citenamefont {Deotare},\ and\ \citenamefont {Loncar}}]{quan_photonic_2010}%
  \BibitemOpen
  \bibfield  {author} {\bibinfo {author} {\bibfnamefont {Q.}~\bibnamefont {Quan}}, \bibinfo {author} {\bibfnamefont {P.~B.}\ \bibnamefont {Deotare}},\ and\ \bibinfo {author} {\bibfnamefont {M.}~\bibnamefont {Loncar}},\ }\href {https://doi.org/10.1063/1.3429125} {\bibfield  {journal} {\bibinfo  {journal} {Applied Physics Letters}\ }\textbf {\bibinfo {volume} {96}},\ \bibinfo {pages} {203102} (\bibinfo {year} {2010})}\BibitemShut {NoStop}%
\bibitem [{\citenamefont {Sipahigil}\ \emph {et~al.}(2016)\citenamefont {Sipahigil}, \citenamefont {Evans}, \citenamefont {Sukachev}, \citenamefont {Burek}, \citenamefont {Borregaard}, \citenamefont {Bhaskar}, \citenamefont {Nguyen}, \citenamefont {Pacheco}, \citenamefont {Atikian}, \citenamefont {Meuwly}, \citenamefont {Camacho}, \citenamefont {Jelezko}, \citenamefont {Bielejec}, \citenamefont {Park}, \citenamefont {Lončar},\ and\ \citenamefont {Lukin}}]{sipahigil_integrated_2016}%
  \BibitemOpen
  \bibfield  {author} {\bibinfo {author} {\bibfnamefont {A.}~\bibnamefont {Sipahigil}}, \bibinfo {author} {\bibfnamefont {R.~E.}\ \bibnamefont {Evans}}, \bibinfo {author} {\bibfnamefont {D.~D.}\ \bibnamefont {Sukachev}}, \bibinfo {author} {\bibfnamefont {M.~J.}\ \bibnamefont {Burek}}, \bibinfo {author} {\bibfnamefont {J.}~\bibnamefont {Borregaard}}, \bibinfo {author} {\bibfnamefont {M.~K.}\ \bibnamefont {Bhaskar}}, \bibinfo {author} {\bibfnamefont {C.~T.}\ \bibnamefont {Nguyen}}, \bibinfo {author} {\bibfnamefont {J.~L.}\ \bibnamefont {Pacheco}}, \bibinfo {author} {\bibfnamefont {H.~A.}\ \bibnamefont {Atikian}}, \bibinfo {author} {\bibfnamefont {C.}~\bibnamefont {Meuwly}}, \bibinfo {author} {\bibfnamefont {R.~M.}\ \bibnamefont {Camacho}}, \bibinfo {author} {\bibfnamefont {F.}~\bibnamefont {Jelezko}}, \bibinfo {author} {\bibfnamefont {E.}~\bibnamefont {Bielejec}}, \bibinfo {author} {\bibfnamefont {H.}~\bibnamefont {Park}}, \bibinfo {author} {\bibfnamefont {M.}~\bibnamefont {Lončar}},\ and\ \bibinfo {author}
  {\bibfnamefont {M.~D.}\ \bibnamefont {Lukin}},\ }\href {https://doi.org/10.1126/science.aah6875} {\bibfield  {journal} {\bibinfo  {journal} {Science}\ }\textbf {\bibinfo {volume} {354}},\ \bibinfo {pages} {847} (\bibinfo {year} {2016})},\ \bibinfo {note} {publisher: American Association for the Advancement of Science}\BibitemShut {NoStop}%
\bibitem [{\citenamefont {Castelletto}\ \emph {et~al.}(2019)\citenamefont {Castelletto}, \citenamefont {Atem}, \citenamefont {Inam}, \citenamefont {Bardeleben}, \citenamefont {Hameau}, \citenamefont {Almutairi}, \citenamefont {Guillot}, \citenamefont {Sato}, \citenamefont {Boretti},\ and\ \citenamefont {Bluet}}]{castelletto_deterministic_2019}%
  \BibitemOpen
  \bibfield  {author} {\bibinfo {author} {\bibfnamefont {S.}~\bibnamefont {Castelletto}}, \bibinfo {author} {\bibfnamefont {A.~S.~A.}\ \bibnamefont {Atem}}, \bibinfo {author} {\bibfnamefont {F.~A.}\ \bibnamefont {Inam}}, \bibinfo {author} {\bibfnamefont {H.~J.~v.}\ \bibnamefont {Bardeleben}}, \bibinfo {author} {\bibfnamefont {S.}~\bibnamefont {Hameau}}, \bibinfo {author} {\bibfnamefont {A.~F.}\ \bibnamefont {Almutairi}}, \bibinfo {author} {\bibfnamefont {G.}~\bibnamefont {Guillot}}, \bibinfo {author} {\bibfnamefont {S.-i.}\ \bibnamefont {Sato}}, \bibinfo {author} {\bibfnamefont {A.}~\bibnamefont {Boretti}},\ and\ \bibinfo {author} {\bibfnamefont {J.~M.}\ \bibnamefont {Bluet}},\ }\href {https://doi.org/10.3762/bjnano.10.229} {\bibfield  {journal} {\bibinfo  {journal} {Beilstein Journal of Nanotechnology}\ }\textbf {\bibinfo {volume} {10}},\ \bibinfo {pages} {2383} (\bibinfo {year} {2019})},\ \bibinfo {note} {publisher: Beilstein-Institut}\BibitemShut {NoStop}%
\bibitem [{\citenamefont {Hessenauer}\ \emph {et~al.}(2025)\citenamefont {Hessenauer}, \citenamefont {Körber}, \citenamefont {Ghezellou}, \citenamefont {Ul-Hassan}, \citenamefont {Astakhov}, \citenamefont {Knolle}, \citenamefont {Wrachtrup},\ and\ \citenamefont {Hunger}}]{hessenauer_cavity_2025}%
  \BibitemOpen
  \bibfield  {author} {\bibinfo {author} {\bibfnamefont {J.}~\bibnamefont {Hessenauer}}, \bibinfo {author} {\bibfnamefont {J.}~\bibnamefont {Körber}}, \bibinfo {author} {\bibfnamefont {M.}~\bibnamefont {Ghezellou}}, \bibinfo {author} {\bibfnamefont {J.}~\bibnamefont {Ul-Hassan}}, \bibinfo {author} {\bibfnamefont {G.~V.}\ \bibnamefont {Astakhov}}, \bibinfo {author} {\bibfnamefont {W.}~\bibnamefont {Knolle}}, \bibinfo {author} {\bibfnamefont {J.}~\bibnamefont {Wrachtrup}},\ and\ \bibinfo {author} {\bibfnamefont {D.}~\bibnamefont {Hunger}},\ }\href {https://doi.org/10.48550/arXiv.2501.04583} {\bibinfo {title} {Cavity enhancement of {V2} centers in {4H}-{SiC} with a fiber-based {Fabry}-{Pérot} microcavity}} (\bibinfo {year} {2025}),\ \bibinfo {note} {arXiv:2501.04583 [quant-ph]}\BibitemShut {NoStop}%
\bibitem [{\citenamefont {Thon}\ \emph {et~al.}(2009)\citenamefont {Thon}, \citenamefont {Rakher}, \citenamefont {Kim}, \citenamefont {Gudat}, \citenamefont {Irvine}, \citenamefont {Petroff},\ and\ \citenamefont {Bouwmeester}}]{thon_strong_2009}%
  \BibitemOpen
  \bibfield  {author} {\bibinfo {author} {\bibfnamefont {S.~M.}\ \bibnamefont {Thon}}, \bibinfo {author} {\bibfnamefont {M.~T.}\ \bibnamefont {Rakher}}, \bibinfo {author} {\bibfnamefont {H.}~\bibnamefont {Kim}}, \bibinfo {author} {\bibfnamefont {J.}~\bibnamefont {Gudat}}, \bibinfo {author} {\bibfnamefont {W.~T.~M.}\ \bibnamefont {Irvine}}, \bibinfo {author} {\bibfnamefont {P.~M.}\ \bibnamefont {Petroff}},\ and\ \bibinfo {author} {\bibfnamefont {D.}~\bibnamefont {Bouwmeester}},\ }\href {https://doi.org/10.1063/1.3103885} {\bibfield  {journal} {\bibinfo  {journal} {Applied Physics Letters}\ }\textbf {\bibinfo {volume} {94}},\ \bibinfo {pages} {111115} (\bibinfo {year} {2009})}\BibitemShut {NoStop}%
\bibitem [{\citenamefont {Marseglia}\ \emph {et~al.}(2011)\citenamefont {Marseglia}, \citenamefont {Hadden}, \citenamefont {Stanley-Clarke}, \citenamefont {Harrison}, \citenamefont {Patton}, \citenamefont {Ho}, \citenamefont {Naydenov}, \citenamefont {Jelezko}, \citenamefont {Meijer}, \citenamefont {Dolan}, \citenamefont {Smith}, \citenamefont {Rarity},\ and\ \citenamefont {O'Brien}}]{marseglia_nanofabricated_2011}%
  \BibitemOpen
  \bibfield  {author} {\bibinfo {author} {\bibfnamefont {L.}~\bibnamefont {Marseglia}}, \bibinfo {author} {\bibfnamefont {J.~P.}\ \bibnamefont {Hadden}}, \bibinfo {author} {\bibfnamefont {A.~C.}\ \bibnamefont {Stanley-Clarke}}, \bibinfo {author} {\bibfnamefont {J.~P.}\ \bibnamefont {Harrison}}, \bibinfo {author} {\bibfnamefont {B.}~\bibnamefont {Patton}}, \bibinfo {author} {\bibfnamefont {Y.-L.~D.}\ \bibnamefont {Ho}}, \bibinfo {author} {\bibfnamefont {B.}~\bibnamefont {Naydenov}}, \bibinfo {author} {\bibfnamefont {F.}~\bibnamefont {Jelezko}}, \bibinfo {author} {\bibfnamefont {J.}~\bibnamefont {Meijer}}, \bibinfo {author} {\bibfnamefont {P.~R.}\ \bibnamefont {Dolan}}, \bibinfo {author} {\bibfnamefont {J.~M.}\ \bibnamefont {Smith}}, \bibinfo {author} {\bibfnamefont {J.~G.}\ \bibnamefont {Rarity}},\ and\ \bibinfo {author} {\bibfnamefont {J.~L.}\ \bibnamefont {O'Brien}},\ }\href {https://doi.org/10.1063/1.3573870} {\bibfield  {journal} {\bibinfo  {journal} {Applied Physics Letters}\ }\textbf {\bibinfo {volume}
  {98}},\ \bibinfo {pages} {133107} (\bibinfo {year} {2011})}\BibitemShut {NoStop}%
\bibitem [{\citenamefont {Copeland}\ \emph {et~al.}(2024)\citenamefont {Copeland}, \citenamefont {Pintar}, \citenamefont {Dixson}, \citenamefont {Chanana}, \citenamefont {Srinivasan}, \citenamefont {Westly}, \citenamefont {Ilic}, \citenamefont {Davanco},\ and\ \citenamefont {Stavis}}]{copeland_traceable_2024}%
  \BibitemOpen
  \bibfield  {author} {\bibinfo {author} {\bibfnamefont {C.~R.}\ \bibnamefont {Copeland}}, \bibinfo {author} {\bibfnamefont {A.~L.}\ \bibnamefont {Pintar}}, \bibinfo {author} {\bibfnamefont {R.~G.}\ \bibnamefont {Dixson}}, \bibinfo {author} {\bibfnamefont {A.}~\bibnamefont {Chanana}}, \bibinfo {author} {\bibfnamefont {K.}~\bibnamefont {Srinivasan}}, \bibinfo {author} {\bibfnamefont {D.~A.}\ \bibnamefont {Westly}}, \bibinfo {author} {\bibfnamefont {B.~R.}\ \bibnamefont {Ilic}}, \bibinfo {author} {\bibfnamefont {M.~I.}\ \bibnamefont {Davanco}},\ and\ \bibinfo {author} {\bibfnamefont {S.~M.}\ \bibnamefont {Stavis}},\ }\href {https://doi.org/10.1364/OPTICAQ.502464} {\bibfield  {journal} {\bibinfo  {journal} {Optica Quantum}\ }\textbf {\bibinfo {volume} {2}},\ \bibinfo {pages} {72} (\bibinfo {year} {2024})}\BibitemShut {NoStop}%
\bibitem [{\citenamefont {Schröder}\ \emph {et~al.}(2017)\citenamefont {Schröder}, \citenamefont {Trusheim}, \citenamefont {Walsh}, \citenamefont {Li}, \citenamefont {Zheng}, \citenamefont {Schukraft}, \citenamefont {Sipahigil}, \citenamefont {Evans}, \citenamefont {Sukachev}, \citenamefont {Nguyen}, \citenamefont {Pacheco}, \citenamefont {Camacho}, \citenamefont {Bielejec}, \citenamefont {Lukin},\ and\ \citenamefont {Englund}}]{schroder_scalable_2017}%
  \BibitemOpen
  \bibfield  {author} {\bibinfo {author} {\bibfnamefont {T.}~\bibnamefont {Schröder}}, \bibinfo {author} {\bibfnamefont {M.~E.}\ \bibnamefont {Trusheim}}, \bibinfo {author} {\bibfnamefont {M.}~\bibnamefont {Walsh}}, \bibinfo {author} {\bibfnamefont {L.}~\bibnamefont {Li}}, \bibinfo {author} {\bibfnamefont {J.}~\bibnamefont {Zheng}}, \bibinfo {author} {\bibfnamefont {M.}~\bibnamefont {Schukraft}}, \bibinfo {author} {\bibfnamefont {A.}~\bibnamefont {Sipahigil}}, \bibinfo {author} {\bibfnamefont {R.~E.}\ \bibnamefont {Evans}}, \bibinfo {author} {\bibfnamefont {D.~D.}\ \bibnamefont {Sukachev}}, \bibinfo {author} {\bibfnamefont {C.~T.}\ \bibnamefont {Nguyen}}, \bibinfo {author} {\bibfnamefont {J.~L.}\ \bibnamefont {Pacheco}}, \bibinfo {author} {\bibfnamefont {R.~M.}\ \bibnamefont {Camacho}}, \bibinfo {author} {\bibfnamefont {E.~S.}\ \bibnamefont {Bielejec}}, \bibinfo {author} {\bibfnamefont {M.~D.}\ \bibnamefont {Lukin}},\ and\ \bibinfo {author} {\bibfnamefont {D.}~\bibnamefont {Englund}},\ }\href
  {https://doi.org/10.1038/ncomms15376} {\bibfield  {journal} {\bibinfo  {journal} {Nature Communications}\ }\textbf {\bibinfo {volume} {8}},\ \bibinfo {pages} {15376} (\bibinfo {year} {2017})}\BibitemShut {NoStop}%
\bibitem [{\citenamefont {Lefaucher}\ \emph {et~al.}(2025)\citenamefont {Lefaucher}, \citenamefont {Baron}, \citenamefont {Jager}, \citenamefont {Calvo}, \citenamefont {Elsässer}, \citenamefont {Coppola}, \citenamefont {Mazen}, \citenamefont {Kerdilès}, \citenamefont {Cache}, \citenamefont {Dréau},\ and\ \citenamefont {Gérard}}]{lefaucher_bright_2025}%
  \BibitemOpen
  \bibfield  {author} {\bibinfo {author} {\bibfnamefont {B.}~\bibnamefont {Lefaucher}}, \bibinfo {author} {\bibfnamefont {Y.}~\bibnamefont {Baron}}, \bibinfo {author} {\bibfnamefont {J.-B.}\ \bibnamefont {Jager}}, \bibinfo {author} {\bibfnamefont {V.}~\bibnamefont {Calvo}}, \bibinfo {author} {\bibfnamefont {C.}~\bibnamefont {Elsässer}}, \bibinfo {author} {\bibfnamefont {G.}~\bibnamefont {Coppola}}, \bibinfo {author} {\bibfnamefont {F.}~\bibnamefont {Mazen}}, \bibinfo {author} {\bibfnamefont {S.}~\bibnamefont {Kerdilès}}, \bibinfo {author} {\bibfnamefont {F.}~\bibnamefont {Cache}}, \bibinfo {author} {\bibfnamefont {A.}~\bibnamefont {Dréau}},\ and\ \bibinfo {author} {\bibfnamefont {J.-M.}\ \bibnamefont {Gérard}},\ }\href {https://doi.org/10.48550/arXiv.2501.12744} {\bibinfo {title} {Bright single-photon source in a silicon chip by nanoscale positioning of a color center in a microcavity}} (\bibinfo {year} {2025})\BibitemShut {NoStop}%
\bibitem [{\citenamefont {He}\ \emph {et~al.}(2023)\citenamefont {He}, \citenamefont {Li}, \citenamefont {Wen}, \citenamefont {Zhou}, \citenamefont {Lin}, \citenamefont {Hao}, \citenamefont {Xu}, \citenamefont {Li},\ and\ \citenamefont {Guo}}]{he_maskless_2023}%
  \BibitemOpen
  \bibfield  {author} {\bibinfo {author} {\bibfnamefont {Z.-X.}\ \bibnamefont {He}}, \bibinfo {author} {\bibfnamefont {Q.}~\bibnamefont {Li}}, \bibinfo {author} {\bibfnamefont {X.-L.}\ \bibnamefont {Wen}}, \bibinfo {author} {\bibfnamefont {J.-Y.}\ \bibnamefont {Zhou}}, \bibinfo {author} {\bibfnamefont {W.-X.}\ \bibnamefont {Lin}}, \bibinfo {author} {\bibfnamefont {Z.-H.}\ \bibnamefont {Hao}}, \bibinfo {author} {\bibfnamefont {J.-S.}\ \bibnamefont {Xu}}, \bibinfo {author} {\bibfnamefont {C.-F.}\ \bibnamefont {Li}},\ and\ \bibinfo {author} {\bibfnamefont {G.-C.}\ \bibnamefont {Guo}},\ }\href {https://doi.org/10.1021/acsphotonics.2c01209} {\bibfield  {journal} {\bibinfo  {journal} {ACS Photonics}\ }\textbf {\bibinfo {volume} {10}},\ \bibinfo {pages} {2234} (\bibinfo {year} {2023})},\ \bibinfo {note} {publisher: American Chemical Society}\BibitemShut {NoStop}%
\bibitem [{\citenamefont {Fu}\ \emph {et~al.}(2010)\citenamefont {Fu}, \citenamefont {Santori}, \citenamefont {Barclay},\ and\ \citenamefont {Beausoleil}}]{fu_conversion_2010}%
  \BibitemOpen
  \bibfield  {author} {\bibinfo {author} {\bibfnamefont {K.-M.~C.}\ \bibnamefont {Fu}}, \bibinfo {author} {\bibfnamefont {C.}~\bibnamefont {Santori}}, \bibinfo {author} {\bibfnamefont {P.~E.}\ \bibnamefont {Barclay}},\ and\ \bibinfo {author} {\bibfnamefont {R.~G.}\ \bibnamefont {Beausoleil}},\ }\href {https://doi.org/10.1063/1.3364135} {\bibfield  {journal} {\bibinfo  {journal} {Applied Physics Letters}\ }\textbf {\bibinfo {volume} {96}},\ \bibinfo {pages} {121907} (\bibinfo {year} {2010})}\BibitemShut {NoStop}%
\bibitem [{\citenamefont {Orwa}\ \emph {et~al.}(2011)\citenamefont {Orwa}, \citenamefont {Santori}, \citenamefont {Fu}, \citenamefont {Gibson}, \citenamefont {Simpson}, \citenamefont {Aharonovich}, \citenamefont {Stacey}, \citenamefont {Cimmino}, \citenamefont {Balog}, \citenamefont {Markham}, \citenamefont {Twitchen}, \citenamefont {Greentree}, \citenamefont {Beausoleil},\ and\ \citenamefont {Prawer}}]{orwa_engineering_2011}%
  \BibitemOpen
  \bibfield  {author} {\bibinfo {author} {\bibfnamefont {J.~O.}\ \bibnamefont {Orwa}}, \bibinfo {author} {\bibfnamefont {C.}~\bibnamefont {Santori}}, \bibinfo {author} {\bibfnamefont {K.~M.~C.}\ \bibnamefont {Fu}}, \bibinfo {author} {\bibfnamefont {B.}~\bibnamefont {Gibson}}, \bibinfo {author} {\bibfnamefont {D.}~\bibnamefont {Simpson}}, \bibinfo {author} {\bibfnamefont {I.}~\bibnamefont {Aharonovich}}, \bibinfo {author} {\bibfnamefont {A.}~\bibnamefont {Stacey}}, \bibinfo {author} {\bibfnamefont {A.}~\bibnamefont {Cimmino}}, \bibinfo {author} {\bibfnamefont {P.}~\bibnamefont {Balog}}, \bibinfo {author} {\bibfnamefont {M.}~\bibnamefont {Markham}}, \bibinfo {author} {\bibfnamefont {D.}~\bibnamefont {Twitchen}}, \bibinfo {author} {\bibfnamefont {A.~D.}\ \bibnamefont {Greentree}}, \bibinfo {author} {\bibfnamefont {R.~G.}\ \bibnamefont {Beausoleil}},\ and\ \bibinfo {author} {\bibfnamefont {S.}~\bibnamefont {Prawer}},\ }\href {https://doi.org/10.1063/1.3573768} {\bibfield  {journal} {\bibinfo  {journal} {Journal
  of Applied Physics}\ }\textbf {\bibinfo {volume} {109}},\ \bibinfo {pages} {083530} (\bibinfo {year} {2011})}\BibitemShut {NoStop}%
\bibitem [{\citenamefont {van Dam}\ \emph {et~al.}(2019)\citenamefont {van Dam}, \citenamefont {Walsh}, \citenamefont {Degen}, \citenamefont {Bersin}, \citenamefont {Mouradian}, \citenamefont {Galiullin}, \citenamefont {Ruf}, \citenamefont {IJspeert}, \citenamefont {Taminiau}, \citenamefont {Hanson},\ and\ \citenamefont {Englund}}]{van_dam_optical_2019}%
  \BibitemOpen
  \bibfield  {author} {\bibinfo {author} {\bibfnamefont {S.~B.}\ \bibnamefont {van Dam}}, \bibinfo {author} {\bibfnamefont {M.}~\bibnamefont {Walsh}}, \bibinfo {author} {\bibfnamefont {M.~J.}\ \bibnamefont {Degen}}, \bibinfo {author} {\bibfnamefont {E.}~\bibnamefont {Bersin}}, \bibinfo {author} {\bibfnamefont {S.~L.}\ \bibnamefont {Mouradian}}, \bibinfo {author} {\bibfnamefont {A.}~\bibnamefont {Galiullin}}, \bibinfo {author} {\bibfnamefont {M.}~\bibnamefont {Ruf}}, \bibinfo {author} {\bibfnamefont {M.}~\bibnamefont {IJspeert}}, \bibinfo {author} {\bibfnamefont {T.~H.}\ \bibnamefont {Taminiau}}, \bibinfo {author} {\bibfnamefont {R.}~\bibnamefont {Hanson}},\ and\ \bibinfo {author} {\bibfnamefont {D.~R.}\ \bibnamefont {Englund}},\ }\href {https://doi.org/10.1103/PhysRevB.99.161203} {\bibfield  {journal} {\bibinfo  {journal} {Physical Review B}\ }\textbf {\bibinfo {volume} {99}},\ \bibinfo {pages} {161203} (\bibinfo {year} {2019})},\ \bibinfo {note} {publisher: American Physical Society}\BibitemShut {NoStop}%
\bibitem [{\citenamefont {Chen}\ \emph {et~al.}(2019)\citenamefont {Chen}, \citenamefont {Salter}, \citenamefont {Niethammer}, \citenamefont {Widmann}, \citenamefont {Kaiser}, \citenamefont {Nagy}, \citenamefont {Morioka}, \citenamefont {Babin}, \citenamefont {Erlekampf}, \citenamefont {Berwian}, \citenamefont {Booth},\ and\ \citenamefont {Wrachtrup}}]{chen_laser_2019}%
  \BibitemOpen
  \bibfield  {author} {\bibinfo {author} {\bibfnamefont {Y.-C.}\ \bibnamefont {Chen}}, \bibinfo {author} {\bibfnamefont {P.~S.}\ \bibnamefont {Salter}}, \bibinfo {author} {\bibfnamefont {M.}~\bibnamefont {Niethammer}}, \bibinfo {author} {\bibfnamefont {M.}~\bibnamefont {Widmann}}, \bibinfo {author} {\bibfnamefont {F.}~\bibnamefont {Kaiser}}, \bibinfo {author} {\bibfnamefont {R.}~\bibnamefont {Nagy}}, \bibinfo {author} {\bibfnamefont {N.}~\bibnamefont {Morioka}}, \bibinfo {author} {\bibfnamefont {C.}~\bibnamefont {Babin}}, \bibinfo {author} {\bibfnamefont {J.}~\bibnamefont {Erlekampf}}, \bibinfo {author} {\bibfnamefont {P.}~\bibnamefont {Berwian}}, \bibinfo {author} {\bibfnamefont {M.~J.}\ \bibnamefont {Booth}},\ and\ \bibinfo {author} {\bibfnamefont {J.}~\bibnamefont {Wrachtrup}},\ }\href {https://doi.org/10.1021/acs.nanolett.8b05070} {\bibfield  {journal} {\bibinfo  {journal} {Nano Letters}\ }\textbf {\bibinfo {volume} {19}},\ \bibinfo {pages} {2377} (\bibinfo {year} {2019})}\BibitemShut {NoStop}%
\bibitem [{\citenamefont {Chen}\ \emph {et~al.}(2017)\citenamefont {Chen}, \citenamefont {Salter}, \citenamefont {Knauer}, \citenamefont {Weng}, \citenamefont {Frangeskou}, \citenamefont {Stephen}, \citenamefont {Ishmael}, \citenamefont {Dolan}, \citenamefont {Johnson}, \citenamefont {Green}, \citenamefont {Morley}, \citenamefont {Newton}, \citenamefont {Rarity}, \citenamefont {Booth},\ and\ \citenamefont {Smith}}]{chen_laser_2017-1}%
  \BibitemOpen
  \bibfield  {author} {\bibinfo {author} {\bibfnamefont {Y.-C.}\ \bibnamefont {Chen}}, \bibinfo {author} {\bibfnamefont {P.~S.}\ \bibnamefont {Salter}}, \bibinfo {author} {\bibfnamefont {S.}~\bibnamefont {Knauer}}, \bibinfo {author} {\bibfnamefont {L.}~\bibnamefont {Weng}}, \bibinfo {author} {\bibfnamefont {A.~C.}\ \bibnamefont {Frangeskou}}, \bibinfo {author} {\bibfnamefont {C.~J.}\ \bibnamefont {Stephen}}, \bibinfo {author} {\bibfnamefont {S.~N.}\ \bibnamefont {Ishmael}}, \bibinfo {author} {\bibfnamefont {P.~R.}\ \bibnamefont {Dolan}}, \bibinfo {author} {\bibfnamefont {S.}~\bibnamefont {Johnson}}, \bibinfo {author} {\bibfnamefont {B.~L.}\ \bibnamefont {Green}}, \bibinfo {author} {\bibfnamefont {G.~W.}\ \bibnamefont {Morley}}, \bibinfo {author} {\bibfnamefont {M.~E.}\ \bibnamefont {Newton}}, \bibinfo {author} {\bibfnamefont {J.~G.}\ \bibnamefont {Rarity}}, \bibinfo {author} {\bibfnamefont {M.~J.}\ \bibnamefont {Booth}},\ and\ \bibinfo {author} {\bibfnamefont {J.~M.}\ \bibnamefont {Smith}},\ }\href
  {https://doi.org/10.1038/nphoton.2016.234} {\bibfield  {journal} {\bibinfo  {journal} {Nature Photonics}\ }\textbf {\bibinfo {volume} {11}},\ \bibinfo {pages} {77} (\bibinfo {year} {2017})},\ \bibinfo {note} {publisher: Nature Publishing Group}\BibitemShut {NoStop}%
\bibitem [{\citenamefont {Hao}\ \emph {et~al.}(2024)\citenamefont {Hao}, \citenamefont {He}, \citenamefont {Maksimovic}, \citenamefont {Katkus}, \citenamefont {Xu}, \citenamefont {Juodkazis}, \citenamefont {Li}, \citenamefont {Guo},\ and\ \citenamefont {Castelletto}}]{hao_laser_2024}%
  \BibitemOpen
  \bibfield  {author} {\bibinfo {author} {\bibfnamefont {Z.-H.}\ \bibnamefont {Hao}}, \bibinfo {author} {\bibfnamefont {Z.-X.}\ \bibnamefont {He}}, \bibinfo {author} {\bibfnamefont {J.}~\bibnamefont {Maksimovic}}, \bibinfo {author} {\bibfnamefont {T.}~\bibnamefont {Katkus}}, \bibinfo {author} {\bibfnamefont {J.-S.}\ \bibnamefont {Xu}}, \bibinfo {author} {\bibfnamefont {S.}~\bibnamefont {Juodkazis}}, \bibinfo {author} {\bibfnamefont {C.-F.}\ \bibnamefont {Li}}, \bibinfo {author} {\bibfnamefont {G.-C.}\ \bibnamefont {Guo}},\ and\ \bibinfo {author} {\bibfnamefont {S.}~\bibnamefont {Castelletto}},\ }\href {https://doi.org/10.48550/arXiv.2411.18868} {\bibinfo {title} {Laser writing and spin control of near infrared emitters in silicon carbide}} (\bibinfo {year} {2024}),\ \bibinfo {note} {arXiv:2411.18868 [cond-mat]}\BibitemShut {NoStop}%
\bibitem [{\citenamefont {Liu}\ \emph {et~al.}(2020)\citenamefont {Liu}, \citenamefont {Xu}, \citenamefont {Song}, \citenamefont {Wang}, \citenamefont {Dong}, \citenamefont {Li}, \citenamefont {Ren}, \citenamefont {Li}, \citenamefont {Rommel}, \citenamefont {Gu}, \citenamefont {Liu}, \citenamefont {Hu},\ and\ \citenamefont {Fang}}]{liu_confocal_2020}%
  \BibitemOpen
  \bibfield  {author} {\bibinfo {author} {\bibfnamefont {J.}~\bibnamefont {Liu}}, \bibinfo {author} {\bibfnamefont {Z.}~\bibnamefont {Xu}}, \bibinfo {author} {\bibfnamefont {Y.}~\bibnamefont {Song}}, \bibinfo {author} {\bibfnamefont {H.}~\bibnamefont {Wang}}, \bibinfo {author} {\bibfnamefont {B.}~\bibnamefont {Dong}}, \bibinfo {author} {\bibfnamefont {S.}~\bibnamefont {Li}}, \bibinfo {author} {\bibfnamefont {J.}~\bibnamefont {Ren}}, \bibinfo {author} {\bibfnamefont {Q.}~\bibnamefont {Li}}, \bibinfo {author} {\bibfnamefont {M.}~\bibnamefont {Rommel}}, \bibinfo {author} {\bibfnamefont {X.}~\bibnamefont {Gu}}, \bibinfo {author} {\bibfnamefont {B.}~\bibnamefont {Liu}}, \bibinfo {author} {\bibfnamefont {M.}~\bibnamefont {Hu}},\ and\ \bibinfo {author} {\bibfnamefont {F.}~\bibnamefont {Fang}},\ }\href {https://doi.org/10.1016/j.npe.2020.11.003} {\bibfield  {journal} {\bibinfo  {journal} {Nanotechnology and Precision Engineering}\ }\textbf {\bibinfo {volume} {3}},\ \bibinfo {pages} {218} (\bibinfo {year}
  {2020})}\BibitemShut {NoStop}%
\bibitem [{\citenamefont {Abdedou}\ \emph {et~al.}(2024)\citenamefont {Abdedou}, \citenamefont {Fuchs}, \citenamefont {Fuchs}, \citenamefont {Herrmann}, \citenamefont {Weber}, \citenamefont {Schäfer}, \citenamefont {L'huillier}, \citenamefont {Becher},\ and\ \citenamefont {Neu}}]{abdedou_photoluminescence_2024}%
  \BibitemOpen
  \bibfield  {author} {\bibinfo {author} {\bibfnamefont {Y.}~\bibnamefont {Abdedou}}, \bibinfo {author} {\bibfnamefont {A.}~\bibnamefont {Fuchs}}, \bibinfo {author} {\bibfnamefont {P.}~\bibnamefont {Fuchs}}, \bibinfo {author} {\bibfnamefont {D.}~\bibnamefont {Herrmann}}, \bibinfo {author} {\bibfnamefont {S.}~\bibnamefont {Weber}}, \bibinfo {author} {\bibfnamefont {M.}~\bibnamefont {Schäfer}}, \bibinfo {author} {\bibfnamefont {J.}~\bibnamefont {L'huillier}}, \bibinfo {author} {\bibfnamefont {C.}~\bibnamefont {Becher}},\ and\ \bibinfo {author} {\bibfnamefont {E.}~\bibnamefont {Neu}},\ }\href {https://doi.org/10.48550/arXiv.2404.09906} {\bibinfo {title} {Photoluminescence of {Femtosecond} {Laser}-irradiated {Silicon} {Carbide}}} (\bibinfo {year} {2024}),\ \bibinfo {note} {arXiv:2404.09906 [cond-mat]}\BibitemShut {NoStop}%
\bibitem [{\citenamefont {Day}\ \emph {et~al.}(2023)\citenamefont {Day}, \citenamefont {Dietz}, \citenamefont {Sutula}, \citenamefont {Yeh},\ and\ \citenamefont {Hu}}]{day_laser_2023}%
  \BibitemOpen
  \bibfield  {author} {\bibinfo {author} {\bibfnamefont {A.~M.}\ \bibnamefont {Day}}, \bibinfo {author} {\bibfnamefont {J.~R.}\ \bibnamefont {Dietz}}, \bibinfo {author} {\bibfnamefont {M.}~\bibnamefont {Sutula}}, \bibinfo {author} {\bibfnamefont {M.}~\bibnamefont {Yeh}},\ and\ \bibinfo {author} {\bibfnamefont {E.~L.}\ \bibnamefont {Hu}},\ }\href {https://doi.org/10.1038/s41563-023-01544-x} {\bibfield  {journal} {\bibinfo  {journal} {Nature Materials}\ }\textbf {\bibinfo {volume} {22}},\ \bibinfo {pages} {696} (\bibinfo {year} {2023})}\BibitemShut {NoStop}%
\bibitem [{\citenamefont {Christle}\ \emph {et~al.}(2015)\citenamefont {Christle}, \citenamefont {Falk}, \citenamefont {Andrich}, \citenamefont {Klimov}, \citenamefont {Hassan}, \citenamefont {Son}, \citenamefont {Janzén}, \citenamefont {Ohshima},\ and\ \citenamefont {Awschalom}}]{christle_isolated_2015}%
  \BibitemOpen
  \bibfield  {author} {\bibinfo {author} {\bibfnamefont {D.~J.}\ \bibnamefont {Christle}}, \bibinfo {author} {\bibfnamefont {A.~L.}\ \bibnamefont {Falk}}, \bibinfo {author} {\bibfnamefont {P.}~\bibnamefont {Andrich}}, \bibinfo {author} {\bibfnamefont {P.~V.}\ \bibnamefont {Klimov}}, \bibinfo {author} {\bibfnamefont {J.~U.}\ \bibnamefont {Hassan}}, \bibinfo {author} {\bibfnamefont {N.~T.}\ \bibnamefont {Son}}, \bibinfo {author} {\bibfnamefont {E.}~\bibnamefont {Janzén}}, \bibinfo {author} {\bibfnamefont {T.}~\bibnamefont {Ohshima}},\ and\ \bibinfo {author} {\bibfnamefont {D.~D.}\ \bibnamefont {Awschalom}},\ }\href {https://doi.org/10.1038/nmat4144} {\bibfield  {journal} {\bibinfo  {journal} {Nature Materials}\ }\textbf {\bibinfo {volume} {14}},\ \bibinfo {pages} {160} (\bibinfo {year} {2015})}\BibitemShut {NoStop}%
\bibitem [{\citenamefont {Lukin}\ \emph {et~al.}(2020{\natexlab{a}})\citenamefont {Lukin}, \citenamefont {Dory}, \citenamefont {Guidry}, \citenamefont {Yang}, \citenamefont {Mishra}, \citenamefont {Trivedi}, \citenamefont {Radulaski}, \citenamefont {Sun}, \citenamefont {Vercruysse}, \citenamefont {Ahn},\ and\ \citenamefont {Vučković}}]{lukin_4h-silicon-carbide--insulator_2020}%
  \BibitemOpen
  \bibfield  {author} {\bibinfo {author} {\bibfnamefont {D.~M.}\ \bibnamefont {Lukin}}, \bibinfo {author} {\bibfnamefont {C.}~\bibnamefont {Dory}}, \bibinfo {author} {\bibfnamefont {M.~A.}\ \bibnamefont {Guidry}}, \bibinfo {author} {\bibfnamefont {K.~Y.}\ \bibnamefont {Yang}}, \bibinfo {author} {\bibfnamefont {S.~D.}\ \bibnamefont {Mishra}}, \bibinfo {author} {\bibfnamefont {R.}~\bibnamefont {Trivedi}}, \bibinfo {author} {\bibfnamefont {M.}~\bibnamefont {Radulaski}}, \bibinfo {author} {\bibfnamefont {S.}~\bibnamefont {Sun}}, \bibinfo {author} {\bibfnamefont {D.}~\bibnamefont {Vercruysse}}, \bibinfo {author} {\bibfnamefont {G.~H.}\ \bibnamefont {Ahn}},\ and\ \bibinfo {author} {\bibfnamefont {J.}~\bibnamefont {Vučković}},\ }\href {https://doi.org/10.1038/s41566-019-0556-6} {\bibfield  {journal} {\bibinfo  {journal} {Nature Photonics}\ }\textbf {\bibinfo {volume} {14}},\ \bibinfo {pages} {330} (\bibinfo {year} {2020}{\natexlab{a}})}\BibitemShut {NoStop}%
\bibitem [{\citenamefont {Lukin}\ \emph {et~al.}(2020{\natexlab{b}})\citenamefont {Lukin}, \citenamefont {Guidry},\ and\ \citenamefont {Vučković}}]{lukin_integrated_2020}%
  \BibitemOpen
  \bibfield  {author} {\bibinfo {author} {\bibfnamefont {D.~M.}\ \bibnamefont {Lukin}}, \bibinfo {author} {\bibfnamefont {M.~A.}\ \bibnamefont {Guidry}},\ and\ \bibinfo {author} {\bibfnamefont {J.}~\bibnamefont {Vučković}},\ }\href {https://doi.org/10.1103/PRXQuantum.1.020102} {\bibfield  {journal} {\bibinfo  {journal} {PRX Quantum}\ }\textbf {\bibinfo {volume} {1}},\ \bibinfo {pages} {020102} (\bibinfo {year} {2020}{\natexlab{b}})}\BibitemShut {NoStop}%
\bibitem [{\citenamefont {Castelletto}\ \emph {et~al.}(2022)\citenamefont {Castelletto}, \citenamefont {Peruzzo}, \citenamefont {Bonato}, \citenamefont {Johnson}, \citenamefont {Radulaski}, \citenamefont {Ou}, \citenamefont {Kaiser},\ and\ \citenamefont {Wrachtrup}}]{castelletto_silicon_2022}%
  \BibitemOpen
  \bibfield  {author} {\bibinfo {author} {\bibfnamefont {S.}~\bibnamefont {Castelletto}}, \bibinfo {author} {\bibfnamefont {A.}~\bibnamefont {Peruzzo}}, \bibinfo {author} {\bibfnamefont {C.}~\bibnamefont {Bonato}}, \bibinfo {author} {\bibfnamefont {B.~C.}\ \bibnamefont {Johnson}}, \bibinfo {author} {\bibfnamefont {M.}~\bibnamefont {Radulaski}}, \bibinfo {author} {\bibfnamefont {H.}~\bibnamefont {Ou}}, \bibinfo {author} {\bibfnamefont {F.}~\bibnamefont {Kaiser}},\ and\ \bibinfo {author} {\bibfnamefont {J.}~\bibnamefont {Wrachtrup}},\ }\href {https://doi.org/10.1021/acsphotonics.1c01775} {\bibfield  {journal} {\bibinfo  {journal} {ACS Photonics}\ }\textbf {\bibinfo {volume} {9}},\ \bibinfo {pages} {1434} (\bibinfo {year} {2022})}\BibitemShut {NoStop}%
\bibitem [{\citenamefont {Scheller}\ \emph {et~al.}(2024)\citenamefont {Scheller}, \citenamefont {Hrunski}, \citenamefont {Schwarberg}, \citenamefont {Knolle}, \citenamefont {Soykal}, \citenamefont {Udvarhelyi}, \citenamefont {Narang}, \citenamefont {Weber}, \citenamefont {Hollendonner},\ and\ \citenamefont {Nagy}}]{scheller_quantum_2024}%
  \BibitemOpen
  \bibfield  {author} {\bibinfo {author} {\bibfnamefont {D.}~\bibnamefont {Scheller}}, \bibinfo {author} {\bibfnamefont {F.}~\bibnamefont {Hrunski}}, \bibinfo {author} {\bibfnamefont {J.~H.}\ \bibnamefont {Schwarberg}}, \bibinfo {author} {\bibfnamefont {W.}~\bibnamefont {Knolle}}, \bibinfo {author} {\bibfnamefont {Ã.~O.}\ \bibnamefont {Soykal}}, \bibinfo {author} {\bibfnamefont {P.}~\bibnamefont {Udvarhelyi}}, \bibinfo {author} {\bibfnamefont {P.}~\bibnamefont {Narang}}, \bibinfo {author} {\bibfnamefont {H.~B.}\ \bibnamefont {Weber}}, \bibinfo {author} {\bibfnamefont {M.}~\bibnamefont {Hollendonner}},\ and\ \bibinfo {author} {\bibfnamefont {R.}~\bibnamefont {Nagy}},\ }\href {http://arxiv.org/abs/2410.10750} {\bibinfo {title} {Quantum enhanced electric field mapping within semiconductor devices}} (\bibinfo {year} {2024}),\ \bibinfo {note} {arXiv:2410.10750 [quant-ph]}\BibitemShut {NoStop}%
\bibitem [{\citenamefont {Steidl}\ \emph {et~al.}(2024)\citenamefont {Steidl}, \citenamefont {Kuna}, \citenamefont {Hesselmeier-Hüttmann}, \citenamefont {Liu}, \citenamefont {Stöhr}, \citenamefont {Knolle}, \citenamefont {Ghezellou}, \citenamefont {Ul-Hassan}, \citenamefont {Schober}, \citenamefont {Bockstedte}, \citenamefont {Gali}, \citenamefont {Vorobyov},\ and\ \citenamefont {Wrachtrup}}]{steidl_single_2024}%
  \BibitemOpen
  \bibfield  {author} {\bibinfo {author} {\bibfnamefont {T.}~\bibnamefont {Steidl}}, \bibinfo {author} {\bibfnamefont {P.}~\bibnamefont {Kuna}}, \bibinfo {author} {\bibfnamefont {E.}~\bibnamefont {Hesselmeier-Hüttmann}}, \bibinfo {author} {\bibfnamefont {D.}~\bibnamefont {Liu}}, \bibinfo {author} {\bibfnamefont {R.}~\bibnamefont {Stöhr}}, \bibinfo {author} {\bibfnamefont {W.}~\bibnamefont {Knolle}}, \bibinfo {author} {\bibfnamefont {M.}~\bibnamefont {Ghezellou}}, \bibinfo {author} {\bibfnamefont {J.}~\bibnamefont {Ul-Hassan}}, \bibinfo {author} {\bibfnamefont {M.}~\bibnamefont {Schober}}, \bibinfo {author} {\bibfnamefont {M.}~\bibnamefont {Bockstedte}}, \bibinfo {author} {\bibfnamefont {A.}~\bibnamefont {Gali}}, \bibinfo {author} {\bibfnamefont {V.}~\bibnamefont {Vorobyov}},\ and\ \bibinfo {author} {\bibfnamefont {J.}~\bibnamefont {Wrachtrup}},\ }\href {https://doi.org/10.48550/arXiv.2410.09021} {\bibinfo {title} {Single {V2} defect in {4H} {Silicon} {Carbide} {Schottky} diode at low temperature}} (\bibinfo
  {year} {2024}),\ \bibinfo {note} {arXiv:2410.09021 [quant-ph]}\BibitemShut {NoStop}%
\bibitem [{\citenamefont {Jamali}\ \emph {et~al.}(2014)\citenamefont {Jamali}, \citenamefont {Gerhardt}, \citenamefont {Rezai}, \citenamefont {Frenner}, \citenamefont {Fedder},\ and\ \citenamefont {Wrachtrup}}]{jamali_microscopic_2014}%
  \BibitemOpen
  \bibfield  {author} {\bibinfo {author} {\bibfnamefont {M.}~\bibnamefont {Jamali}}, \bibinfo {author} {\bibfnamefont {I.}~\bibnamefont {Gerhardt}}, \bibinfo {author} {\bibfnamefont {M.}~\bibnamefont {Rezai}}, \bibinfo {author} {\bibfnamefont {K.}~\bibnamefont {Frenner}}, \bibinfo {author} {\bibfnamefont {H.}~\bibnamefont {Fedder}},\ and\ \bibinfo {author} {\bibfnamefont {J.}~\bibnamefont {Wrachtrup}},\ }\href {https://doi.org/10.1063/1.4902818} {\bibfield  {journal} {\bibinfo  {journal} {Review of Scientific Instruments}\ }\textbf {\bibinfo {volume} {85}},\ \bibinfo {pages} {123703} (\bibinfo {year} {2014})}\BibitemShut {NoStop}%
\bibitem [{\citenamefont {Keldysh}(1965)}]{keldysh_ionization_1965}%
  \BibitemOpen
  \bibfield  {author} {\bibinfo {author} {\bibfnamefont {L.~V.}\ \bibnamefont {Keldysh}},\ }\href@noop {} {\bibfield  {journal} {\bibinfo  {journal} {J. Exp. Theor. Phys.}\ }\textbf {\bibinfo {volume} {20}},\ \bibinfo {pages} {1307} (\bibinfo {year} {1965})}\BibitemShut {NoStop}%
\bibitem [{\citenamefont {Chen}(2017)}]{chen_laser_2017}%
  \BibitemOpen
  \bibfield  {author} {\bibinfo {author} {\bibfnamefont {Y.-C.}\ \bibnamefont {Chen}},\ }\emph {\bibinfo {title} {Laser writing of coherent colour centres in diamond}},\ \href@noop {} {Ph.D. thesis} (\bibinfo {year} {2017})\BibitemShut {NoStop}%
\bibitem [{\citenamefont {Castelletto}\ \emph {et~al.}(2008)\citenamefont {Castelletto}, \citenamefont {Johnson}, \citenamefont {Boretti}, \citenamefont {Gattass},\ and\ \citenamefont {Mazur}}]{castelletto_femtosecond_2008}%
  \BibitemOpen
  \bibfield  {author} {\bibinfo {author} {\bibfnamefont {S.}~\bibnamefont {Castelletto}}, \bibinfo {author} {\bibfnamefont {B.~C.}\ \bibnamefont {Johnson}}, \bibinfo {author} {\bibfnamefont {A.}~\bibnamefont {Boretti}}, \bibinfo {author} {\bibfnamefont {R.~R.}\ \bibnamefont {Gattass}},\ and\ \bibinfo {author} {\bibfnamefont {E.}~\bibnamefont {Mazur}},\ }\href {https://doi.org/10.1038/nphoton.2008.47} {\bibfield  {journal} {\bibinfo  {journal} {IOP Conference Series: Materials Science and Engineering}\ }\textbf {\bibinfo {volume} {2}},\ \bibinfo {pages} {219} (\bibinfo {year} {2008})},\ \bibinfo {note} {publisher: IOP Publishing}\BibitemShut {NoStop}%
\bibitem [{\citenamefont {Griffiths}\ \emph {et~al.}(2021)\citenamefont {Griffiths}, \citenamefont {Kirkpatrick}, \citenamefont {Nicley}, \citenamefont {Patel}, \citenamefont {Zajac}, \citenamefont {Morley}, \citenamefont {Booth}, \citenamefont {Salter},\ and\ \citenamefont {Smith}}]{griffiths_microscopic_2021}%
  \BibitemOpen
  \bibfield  {author} {\bibinfo {author} {\bibfnamefont {B.}~\bibnamefont {Griffiths}}, \bibinfo {author} {\bibfnamefont {A.}~\bibnamefont {Kirkpatrick}}, \bibinfo {author} {\bibfnamefont {S.~S.}\ \bibnamefont {Nicley}}, \bibinfo {author} {\bibfnamefont {R.~L.}\ \bibnamefont {Patel}}, \bibinfo {author} {\bibfnamefont {J.~M.}\ \bibnamefont {Zajac}}, \bibinfo {author} {\bibfnamefont {G.~W.}\ \bibnamefont {Morley}}, \bibinfo {author} {\bibfnamefont {M.~J.}\ \bibnamefont {Booth}}, \bibinfo {author} {\bibfnamefont {P.~S.}\ \bibnamefont {Salter}},\ and\ \bibinfo {author} {\bibfnamefont {J.~M.}\ \bibnamefont {Smith}},\ }\href {https://doi.org/10.1103/PhysRevB.104.174303} {\bibfield  {journal} {\bibinfo  {journal} {Physical Review B}\ }\textbf {\bibinfo {volume} {104}},\ \bibinfo {pages} {174303} (\bibinfo {year} {2021})},\ \bibinfo {note} {publisher: American Physical Society}\BibitemShut {NoStop}%
\bibitem [{\citenamefont {Fuchs}\ \emph {et~al.}(2015)\citenamefont {Fuchs}, \citenamefont {Stender}, \citenamefont {Trupke}, \citenamefont {Simin}, \citenamefont {Pflaum}, \citenamefont {Dyakonov},\ and\ \citenamefont {Astakhov}}]{fuchs_engineering_2015}%
  \BibitemOpen
  \bibfield  {author} {\bibinfo {author} {\bibfnamefont {F.}~\bibnamefont {Fuchs}}, \bibinfo {author} {\bibfnamefont {B.}~\bibnamefont {Stender}}, \bibinfo {author} {\bibfnamefont {M.}~\bibnamefont {Trupke}}, \bibinfo {author} {\bibfnamefont {D.}~\bibnamefont {Simin}}, \bibinfo {author} {\bibfnamefont {J.}~\bibnamefont {Pflaum}}, \bibinfo {author} {\bibfnamefont {V.}~\bibnamefont {Dyakonov}},\ and\ \bibinfo {author} {\bibfnamefont {G.~V.}\ \bibnamefont {Astakhov}},\ }\href {https://doi.org/10.1038/ncomms8578} {\bibfield  {journal} {\bibinfo  {journal} {Nature Communications}\ }\textbf {\bibinfo {volume} {6}},\ \bibinfo {pages} {7578} (\bibinfo {year} {2015})},\ \bibinfo {note} {number: 1 Publisher: Nature Publishing Group}\BibitemShut {NoStop}%
\bibitem [{\citenamefont {Bathen}\ \emph {et~al.}(2021)\citenamefont {Bathen}, \citenamefont {Galeckas}, \citenamefont {Karsthof}, \citenamefont {Delteil}, \citenamefont {Sallet}, \citenamefont {Kuznetsov},\ and\ \citenamefont {Vines}}]{bathen_resolving_2021}%
  \BibitemOpen
  \bibfield  {author} {\bibinfo {author} {\bibfnamefont {M.~E.}\ \bibnamefont {Bathen}}, \bibinfo {author} {\bibfnamefont {A.}~\bibnamefont {Galeckas}}, \bibinfo {author} {\bibfnamefont {R.}~\bibnamefont {Karsthof}}, \bibinfo {author} {\bibfnamefont {A.}~\bibnamefont {Delteil}}, \bibinfo {author} {\bibfnamefont {V.}~\bibnamefont {Sallet}}, \bibinfo {author} {\bibfnamefont {A.~Y.}\ \bibnamefont {Kuznetsov}},\ and\ \bibinfo {author} {\bibfnamefont {L.}~\bibnamefont {Vines}},\ }\href {https://doi.org/10.1103/PhysRevB.104.045120} {\bibfield  {journal} {\bibinfo  {journal} {Physical Review B}\ }\textbf {\bibinfo {volume} {104}},\ \bibinfo {pages} {045120} (\bibinfo {year} {2021})},\ \bibinfo {note} {publisher: American Physical Society}\BibitemShut {NoStop}%
\bibitem [{\citenamefont {Rühl}\ \emph {et~al.}(2018)\citenamefont {Rühl}, \citenamefont {Ott}, \citenamefont {Götzinger}, \citenamefont {Krieger},\ and\ \citenamefont {Weber}}]{ruhl_controlled_2018}%
  \BibitemOpen
  \bibfield  {author} {\bibinfo {author} {\bibfnamefont {M.}~\bibnamefont {Rühl}}, \bibinfo {author} {\bibfnamefont {C.}~\bibnamefont {Ott}}, \bibinfo {author} {\bibfnamefont {S.}~\bibnamefont {Götzinger}}, \bibinfo {author} {\bibfnamefont {M.}~\bibnamefont {Krieger}},\ and\ \bibinfo {author} {\bibfnamefont {H.~B.}\ \bibnamefont {Weber}},\ }\href {https://doi.org/10.1063/1.5045859} {\bibfield  {journal} {\bibinfo  {journal} {Applied Physics Letters}\ }\textbf {\bibinfo {volume} {113}},\ \bibinfo {pages} {122102} (\bibinfo {year} {2018})}\BibitemShut {NoStop}%
\bibitem [{\citenamefont {Davidsson}\ \emph {et~al.}(2022)\citenamefont {Davidsson}, \citenamefont {Babar}, \citenamefont {Shafizadeh}, \citenamefont {Ivanov}, \citenamefont {Ivády}, \citenamefont {Armiento},\ and\ \citenamefont {Abrikosov}}]{davidsson_exhaustive_2022}%
  \BibitemOpen
  \bibfield  {author} {\bibinfo {author} {\bibfnamefont {J.}~\bibnamefont {Davidsson}}, \bibinfo {author} {\bibfnamefont {R.}~\bibnamefont {Babar}}, \bibinfo {author} {\bibfnamefont {D.}~\bibnamefont {Shafizadeh}}, \bibinfo {author} {\bibfnamefont {I.~G.}\ \bibnamefont {Ivanov}}, \bibinfo {author} {\bibfnamefont {V.}~\bibnamefont {Ivády}}, \bibinfo {author} {\bibfnamefont {R.}~\bibnamefont {Armiento}},\ and\ \bibinfo {author} {\bibfnamefont {I.~A.}\ \bibnamefont {Abrikosov}},\ }\href {https://doi.org/10.1515/nanoph-2022-0400} {\bibfield  {journal} {\bibinfo  {journal} {Nanophotonics}\ }\textbf {\bibinfo {volume} {11}},\ \bibinfo {pages} {4565} (\bibinfo {year} {2022})},\ \bibinfo {note} {publisher: De Gruyter}\BibitemShut {NoStop}%
\bibitem [{\citenamefont {Heiler}\ \emph {et~al.}(2024)\citenamefont {Heiler}, \citenamefont {Körber}, \citenamefont {Hesselmeier}, \citenamefont {Kuna}, \citenamefont {Stöhr}, \citenamefont {Fuchs}, \citenamefont {Ghezellou}, \citenamefont {Ul-Hassan}, \citenamefont {Knolle}, \citenamefont {Becher}, \citenamefont {Kaiser},\ and\ \citenamefont {Wrachtrup}}]{heiler_spectral_2024}%
  \BibitemOpen
  \bibfield  {author} {\bibinfo {author} {\bibfnamefont {J.}~\bibnamefont {Heiler}}, \bibinfo {author} {\bibfnamefont {J.}~\bibnamefont {Körber}}, \bibinfo {author} {\bibfnamefont {E.}~\bibnamefont {Hesselmeier}}, \bibinfo {author} {\bibfnamefont {P.}~\bibnamefont {Kuna}}, \bibinfo {author} {\bibfnamefont {R.}~\bibnamefont {Stöhr}}, \bibinfo {author} {\bibfnamefont {P.}~\bibnamefont {Fuchs}}, \bibinfo {author} {\bibfnamefont {M.}~\bibnamefont {Ghezellou}}, \bibinfo {author} {\bibfnamefont {J.}~\bibnamefont {Ul-Hassan}}, \bibinfo {author} {\bibfnamefont {W.}~\bibnamefont {Knolle}}, \bibinfo {author} {\bibfnamefont {C.}~\bibnamefont {Becher}}, \bibinfo {author} {\bibfnamefont {F.}~\bibnamefont {Kaiser}},\ and\ \bibinfo {author} {\bibfnamefont {J.}~\bibnamefont {Wrachtrup}},\ }\href {https://doi.org/10.1038/s41535-024-00644-4} {\bibfield  {journal} {\bibinfo  {journal} {npj Quantum Materials}\ }\textbf {\bibinfo {volume} {9}},\ \bibinfo {pages} {34} (\bibinfo {year} {2024})}\BibitemShut {NoStop}%
\bibitem [{\citenamefont {Li}\ \emph {et~al.}(2019)\citenamefont {Li}, \citenamefont {Wang}, \citenamefont {Bian}, \citenamefont {Dong}, \citenamefont {Song}, \citenamefont {Shao}, \citenamefont {Jiang},\ and\ \citenamefont {Guo}}]{li_threshold_2019}%
  \BibitemOpen
  \bibfield  {author} {\bibinfo {author} {\bibfnamefont {W.}~\bibnamefont {Li}}, \bibinfo {author} {\bibfnamefont {L.}~\bibnamefont {Wang}}, \bibinfo {author} {\bibfnamefont {L.}~\bibnamefont {Bian}}, \bibinfo {author} {\bibfnamefont {F.}~\bibnamefont {Dong}}, \bibinfo {author} {\bibfnamefont {M.}~\bibnamefont {Song}}, \bibinfo {author} {\bibfnamefont {J.}~\bibnamefont {Shao}}, \bibinfo {author} {\bibfnamefont {S.}~\bibnamefont {Jiang}},\ and\ \bibinfo {author} {\bibfnamefont {H.}~\bibnamefont {Guo}},\ }\href {https://doi.org/10.1063/1.5093576} {\bibfield  {journal} {\bibinfo  {journal} {AIP Advances}\ }\textbf {\bibinfo {volume} {9}},\ \bibinfo {pages} {055007} (\bibinfo {year} {2019})}\BibitemShut {NoStop}%
\bibitem [{\citenamefont {Storasta}\ \emph {et~al.}(2004)\citenamefont {Storasta}, \citenamefont {Bergman}, \citenamefont {Janzén}, \citenamefont {Henry},\ and\ \citenamefont {Lu}}]{storasta_deep_2004}%
  \BibitemOpen
  \bibfield  {author} {\bibinfo {author} {\bibfnamefont {L.}~\bibnamefont {Storasta}}, \bibinfo {author} {\bibfnamefont {J.~P.}\ \bibnamefont {Bergman}}, \bibinfo {author} {\bibfnamefont {E.}~\bibnamefont {Janzén}}, \bibinfo {author} {\bibfnamefont {A.}~\bibnamefont {Henry}},\ and\ \bibinfo {author} {\bibfnamefont {J.}~\bibnamefont {Lu}},\ }\href {https://doi.org/10.1063/1.1778819} {\bibfield  {journal} {\bibinfo  {journal} {Journal of Applied Physics}\ }\textbf {\bibinfo {volume} {96}},\ \bibinfo {pages} {4909} (\bibinfo {year} {2004})}\BibitemShut {NoStop}%
\bibitem [{\citenamefont {Sullivan}\ and\ \citenamefont {Steeds}(2006)}]{sullivan_investigation_2006}%
  \BibitemOpen
  \bibfield  {author} {\bibinfo {author} {\bibfnamefont {W.}~\bibnamefont {Sullivan}}\ and\ \bibinfo {author} {\bibfnamefont {J.~W.}\ \bibnamefont {Steeds}},\ }\href {https://doi.org/10.4028/www.scientific.net/MSF.527-529.481} {\bibfield  {journal} {\bibinfo  {journal} {Materials Science Forum}\ }\textbf {\bibinfo {volume} {527-529}},\ \bibinfo {pages} {481} (\bibinfo {year} {2006})}\BibitemShut {NoStop}%
\bibitem [{\citenamefont {Kaneko}\ and\ \citenamefont {Kimoto}(2011)}]{kaneko_formation_2011}%
  \BibitemOpen
  \bibfield  {author} {\bibinfo {author} {\bibfnamefont {H.}~\bibnamefont {Kaneko}}\ and\ \bibinfo {author} {\bibfnamefont {T.}~\bibnamefont {Kimoto}},\ }\href {https://doi.org/10.1063/1.3604795} {\bibfield  {journal} {\bibinfo  {journal} {Applied Physics Letters}\ }\textbf {\bibinfo {volume} {98}},\ \bibinfo {pages} {262106} (\bibinfo {year} {2011})}\BibitemShut {NoStop}%
\bibitem [{\citenamefont {Anderson}\ \emph {et~al.}(2019)\citenamefont {Anderson}, \citenamefont {Bourassa}, \citenamefont {Miao}, \citenamefont {Wolfowicz}, \citenamefont {Mintun}, \citenamefont {Crook}, \citenamefont {Abe}, \citenamefont {Ul~Hassan}, \citenamefont {Son}, \citenamefont {Ohshima},\ and\ \citenamefont {Awschalom}}]{anderson_electrical_2019}%
  \BibitemOpen
  \bibfield  {author} {\bibinfo {author} {\bibfnamefont {C.~P.}\ \bibnamefont {Anderson}}, \bibinfo {author} {\bibfnamefont {A.}~\bibnamefont {Bourassa}}, \bibinfo {author} {\bibfnamefont {K.~C.}\ \bibnamefont {Miao}}, \bibinfo {author} {\bibfnamefont {G.}~\bibnamefont {Wolfowicz}}, \bibinfo {author} {\bibfnamefont {P.~J.}\ \bibnamefont {Mintun}}, \bibinfo {author} {\bibfnamefont {A.~L.}\ \bibnamefont {Crook}}, \bibinfo {author} {\bibfnamefont {H.}~\bibnamefont {Abe}}, \bibinfo {author} {\bibfnamefont {J.}~\bibnamefont {Ul~Hassan}}, \bibinfo {author} {\bibfnamefont {N.~T.}\ \bibnamefont {Son}}, \bibinfo {author} {\bibfnamefont {T.}~\bibnamefont {Ohshima}},\ and\ \bibinfo {author} {\bibfnamefont {D.~D.}\ \bibnamefont {Awschalom}},\ }\href {https://doi.org/10.1126/science.aax9406} {\bibfield  {journal} {\bibinfo  {journal} {Science}\ }\textbf {\bibinfo {volume} {366}},\ \bibinfo {pages} {1225} (\bibinfo {year} {2019})}\BibitemShut {NoStop}%
\bibitem [{\citenamefont {Lohrmann}\ \emph {et~al.}(2016)\citenamefont {Lohrmann}, \citenamefont {Castelletto}, \citenamefont {Klein}, \citenamefont {Ohshima}, \citenamefont {Bosi}, \citenamefont {Negri}, \citenamefont {Lau}, \citenamefont {Gibson}, \citenamefont {Prawer}, \citenamefont {McCallum},\ and\ \citenamefont {Johnson}}]{lohrmann_activation_2016}%
  \BibitemOpen
  \bibfield  {author} {\bibinfo {author} {\bibfnamefont {A.}~\bibnamefont {Lohrmann}}, \bibinfo {author} {\bibfnamefont {S.}~\bibnamefont {Castelletto}}, \bibinfo {author} {\bibfnamefont {J.~R.}\ \bibnamefont {Klein}}, \bibinfo {author} {\bibfnamefont {T.}~\bibnamefont {Ohshima}}, \bibinfo {author} {\bibfnamefont {M.}~\bibnamefont {Bosi}}, \bibinfo {author} {\bibfnamefont {M.}~\bibnamefont {Negri}}, \bibinfo {author} {\bibfnamefont {D.~W.~M.}\ \bibnamefont {Lau}}, \bibinfo {author} {\bibfnamefont {B.~C.}\ \bibnamefont {Gibson}}, \bibinfo {author} {\bibfnamefont {S.}~\bibnamefont {Prawer}}, \bibinfo {author} {\bibfnamefont {J.~C.}\ \bibnamefont {McCallum}},\ and\ \bibinfo {author} {\bibfnamefont {B.~C.}\ \bibnamefont {Johnson}},\ }\href {https://doi.org/10.1063/1.4939906} {\bibfield  {journal} {\bibinfo  {journal} {Applied Physics Letters}\ }\textbf {\bibinfo {volume} {108}},\ \bibinfo {pages} {021107} (\bibinfo {year} {2016})}\BibitemShut {NoStop}%
\bibitem [{\citenamefont {Kaneko}\ \emph {et~al.}(2023)\citenamefont {Kaneko}, \citenamefont {Takashima}, \citenamefont {Shimazaki}, \citenamefont {Takeuchi},\ and\ \citenamefont {Kimoto}}]{kaneko_impact_2023}%
  \BibitemOpen
  \bibfield  {author} {\bibinfo {author} {\bibfnamefont {M.}~\bibnamefont {Kaneko}}, \bibinfo {author} {\bibfnamefont {H.}~\bibnamefont {Takashima}}, \bibinfo {author} {\bibfnamefont {K.}~\bibnamefont {Shimazaki}}, \bibinfo {author} {\bibfnamefont {S.}~\bibnamefont {Takeuchi}},\ and\ \bibinfo {author} {\bibfnamefont {T.}~\bibnamefont {Kimoto}},\ }\href {https://doi.org/10.1063/5.0162610} {\bibfield  {journal} {\bibinfo  {journal} {APL Materials}\ }\textbf {\bibinfo {volume} {11}},\ \bibinfo {pages} {091121} (\bibinfo {year} {2023})}\BibitemShut {NoStop}%
\bibitem [{\citenamefont {Onishi}\ \emph {et~al.}(2024)\citenamefont {Onishi}, \citenamefont {Nakanuma}, \citenamefont {Tahara}, \citenamefont {Kutsuki}, \citenamefont {Shimura}, \citenamefont {Watanabe},\ and\ \citenamefont {Kobayashi}}]{onishi_generation_2024}%
  \BibitemOpen
  \bibfield  {author} {\bibinfo {author} {\bibfnamefont {K.}~\bibnamefont {Onishi}}, \bibinfo {author} {\bibfnamefont {T.}~\bibnamefont {Nakanuma}}, \bibinfo {author} {\bibfnamefont {K.}~\bibnamefont {Tahara}}, \bibinfo {author} {\bibfnamefont {K.}~\bibnamefont {Kutsuki}}, \bibinfo {author} {\bibfnamefont {T.}~\bibnamefont {Shimura}}, \bibinfo {author} {\bibfnamefont {H.}~\bibnamefont {Watanabe}},\ and\ \bibinfo {author} {\bibfnamefont {T.}~\bibnamefont {Kobayashi}},\ }\href {https://doi.org/10.35848/1882-0786/ad4449} {\bibfield  {journal} {\bibinfo  {journal} {Applied Physics Express}\ }\textbf {\bibinfo {volume} {17}},\ \bibinfo {pages} {051004} (\bibinfo {year} {2024})},\ \bibinfo {note} {publisher: IOP Publishing}\BibitemShut {NoStop}%
\end{thebibliography}%

\end{document}